# Volatile transport modeling on Triton with new observational constraints

**T. Bertrand[1,2], E. Lellouch[1], B. J. Holler[3], L. A. Young[4], B. Schmitt[5], J. Marques Oliveira[1], B. Sicardy[1], F. Forget[6], W. M. Grundy[7], F. Merlin[1], M. Vangvichith[6], E. Millour[6], P. Schenk[8], C. Hansen[9], O. White[2], J. Moore[2], J. Stansberry[3], A. Oza[10,11], D. Dubois[2], E. Quirico[5], D. Cruikshank[2]**

[1] Laboratoire d'Etudes Spatiales et d'Instrumentation en Astrophysique (LESIA), Observatoire de Paris, Université PSL, CNRS, Sorbonne Université, Univ. Paris Diderot, Sorbonne Paris Cité, 5 place Jules Janssen, 92195 Meudon, France. tanguy.bertrand@obspm.fr

[2] Ames Research Center, Space Science Division, National Aeronautics and Space Administration (NASA), Moffett Field, CA, USA

[3] Space Telescope Science Institute, Baltimore, MD, USA

[4] Southwest Research Institute, Boulder, CO 80302

[5]. Univ. Grenoble Alpes, CNRS, Institut de Planétologie et d'Astrophysique de Grenoble (IPAG), 38000 Grenoble, France

[6] Laboratoire de Météorologie Dynamique, IPSL, Sorbonne Universités, UPMC Université Paris 06, CNRS, Paris, France

[7] Lowell Observatory, 1400 W. Mars Hill Rd., Flagstaff, AZ 86001.

[8] Lunar and Planetary Institute, Houston, TX 77058, USA

[9] Planetary Science Institute, 1700 E. Fort Lowell, Suite 106, Tucson, AZ 85719, United States of America

[10] Jet Propulsion Laboratory, California Institute of Technology, Pasadena, CA, USA

[11] Physikalisches Institut, Universität Bern, Bern, Switzerland 3012

## Abstract

Neptune's moon Triton shares many similarities with Pluto, including volatile cycles of $N_2$, $CH_4$ and CO, and represents a benchmark case for the study of surface-atmosphere interactions on volatile-rich Kuiper Belt objects. The observations of Pluto by New Horizons acquired during the 2015 flyby and their analysis with volatile transport models (VTMs) shed light on how volatile sublimation-condensation cycles control the climate and shape the surface of such objects. Within the context of New Horizons observations as well as recent Earth-based observations of Triton, we adapt a Plutonian VTM to Triton, and test its ability to simulate its volatile cycles, thereby aiding our understanding of its climate.

Here we present numerical VTM simulations exploring the volatile cycles of $N_2$, $CH_4$ and CO on Triton over long-term and seasonal timescales (cap extent, surface temperatures, surface pressure, sublimation rates) for varying model parameters (including the surface ice reservoir, albedo, thermal inertia, and the internal heat flux). We explore what scenarios and model parameters allow for a best match of the available observations. In particular, our set of observational constraints include Voyager 2 observations (surface pressure and cap extent), ground-based near-infrared (0.8 to 2.4 μm) disk-integrated spectra (the relative surface area of volatile vs. non-volatile ice) and the evolution of surface pressure as retrieved from stellar occultations.

Our results show that Triton's poles act as cold traps for volatile ices and favor the formation of polar caps extending to lower latitudes through glacial flow or through the formation of thinner seasonal deposits. As previously evidenced by other VTMs, North-South asymmetries in surface properties can favor the development of one cap over the other. Our best-case simulations are obtained for a bedrock surface albedo of 0.6-0.7, a global reservoir of $N_2$ ice thicker than 200 m, and a bedrock thermal inertia larger than 500 SI (or smaller but with a large internal heat flux). The large $N_2$ ice reservoir implies a permanent $N_2$ southern cap (several 100 m thick) extending to the equatorial regions with higher amounts of volatile ice at the south pole, which is not inconsistent with Voyager 2 images but does not fit well with observed full-disk near-infrared spectra. Our results also suggest that a small permanent polar cap exists in the northern (currently winter) hemisphere if the internal heat flux remains relatively low (e.g. radiogenic, < 3 mW m$^{-2}$). A non-permanent northern polar cap was only obtained in some of our simulations with high internal heat flux (30 mW m$^{-2}$). The northern cap will possibly extend to 30°N in the next decade, thus becoming visible by Earth-based telescopes. On the basis of our model results, we also discuss the composition of several surface units seen by Voyager 2 in 1989, including the bright equatorial fringe and dark surface patches.

Finally, we provide predictions for the evolution of ice distribution, surface pressure and CO and $CH_4$ atmospheric mixing ratios in the next decades. According to our model, the surface pressure should slowly decrease but remain larger than 0.5 Pa by 2060. We also model the thermal lightcurves of Triton for different climate scenarios in 2022, which serve as predictions for future James Webb Space Telescope observations.

# 1. Introduction

The largest satellite of Neptune, Triton, is often described as Pluto's sibling, as the two bodies share similar sizes, densities, ices (non-volatile $H_2O$ and volatile $N_2$, $CH_4$, CO, Cruikshank et al. 1993, 2000, Quirico et al. 1999, De Meo et al., 2010, Merlin et al., 2018), and currently similar heliocentric distances, surface pressures and temperatures, and atmosphere composition ($N_2$ with traces of $CH_4$ and CO). Its climate and albedo patterns are therefore expected to be dominated by long-term and seasonal volatile condensation-sublimation cycles. However, the geological history (and by extension, interior thermal history) of the two bodies differ: the portion of Triton seen by Voyager 2 during the 1989 flyby has extremely flat topography (vs. dramatic topography of Pluto, Schenk et al., 2018, 2021) surface with a globally high Bond albedo (~0.6-0.8, McEwen, 1990, Hillier et al., 1994) without low-albedo red areas, and active geysers (Soderblom et al., 1990). Other differences between the two bodies include the lower amount of atmospheric $CH_4$ and the colder temperatures in the lower (< 200 km) atmosphere of Triton, ultimately related to the ~10x lower abundance of $CH_4$ in the ice on Triton, the presence of $CO_2$ ice on Triton (forming the surface bedrock along with $H_2O$ ice, Quirico et al., 1999, Grundy et al., 2010, Merlin et al. 2018) and the location of the main volatile ice reservoir (Sputnik Planitia on Pluto vs possibly an extended southern cap on Triton, Stern et al., 2015, Smith et al., 1989).

These differences may be related to the fact that Triton is thought to be a captured Kuiper Belt object (McKinnon, 1984; Agnor and Hamilton, 2006, Li et al., 2020) as it has a retrograde, highly inclined orbit, with a differentiated interior possibly heated by tidal braking. This peculiar orbit leads to complex seasons and possibly to strong obliquity tides and subsequent resurfacing processes (e.g., cryovolcanism, Nimmo and Spencer, 2015). Consequently, Triton is a benchmark case of a dwarf planet that is tidally activated by a giant planet and offers new insights into the surface-atmosphere interactions controlling the climate of large volatile-rich objects of the Kuiper Belt (including Pluto, and others that may exhibit a similar atmosphere near perihelion, e.g. Eris and Makemake).

The flyby of Pluto by New Horizons in July 2015 highlighted the extraordinary complexity of the best studied icy body of the Kuiper Belt (Stern et al., 2015). In particular, volatile transport models (VTMs) of Pluto, constrained by the New Horizons observations, shed light on how the volatile sublimation-condensation cycles can shape the surface of such a planetary body (Bertrand et al., 2016, 2018, 2019, 2020a, 2020b, Johnson et al., 2020). In addition, the correlation of volatile ice deposition with the large global topographic variability of Pluto's surface (total relief exceeding 10 km, with a standard deviation of 1.1 km, Schenk et al., 2018) also indicated the importance of elevation in volatile ice transport on icy bodies with thin atmospheres (Trafton et al., 1998, Bertrand and Forget, 2016). Within the context of New Horizons observations as well as recent Earth-based observations of Triton, it is a natural step to adapt these better-constrained VTMs to Triton and test their ability to simulate its volatile cycles, thereby aiding our understanding of its climate.

Many fundamental issues remain unsolved regarding the climate and the volatile cycles on Triton, which may shape Triton's surface as they do on Pluto. For instance, how do seasonal cycles affect the volatile ice surface distribution, and over what temporal and spatial scales? Is the southern cap seasonal or permanent? Does the winter hemisphere act as a sink for volatile ices? If so, what is the extent and the nature (seasonal vs permanent) of the northern cap, considering that Moore and Spencer (1990) suggested a long term net transfer of $N_2$ ice from the north to the south and thus a more extended southern vs northern cap? Where are $N_2$ and $CH_4$ currently sublimating and condensing? What are the properties of the volatile ices (thermal inertia, reservoir, etc.)? How will the surface pressure, and CO and $CH_4$ gas abundances evolve? The cases of Spencer and Moore (1992) that best match occultations had a non-global atmosphere in the 1960's, so does Triton's atmosphere collapse (become non-global)? Near-infrared (NIR) rotational lightcurves suggest the latitudes south of 60°S are bare of $N_2$ (or at least of large-grained $N_2$, e.g. Grundy et al., 2010, Holler et al., 2016), but what could explain a difference in composition between high southern latitudes and southern mid-latitudes? Finally, how do the volatile cycles on Triton compare with those on Pluto?

Here we present numerical simulations designed to model the evolution of Triton's volatiles (surface ice distribution, surface pressure, and atmospheric abundances) over millions of years on the basis of universal physical equations put into practice with some hypotheses. Our main goal is to investigate where the volatile ices tend to accumulate and sublimate on Triton as the seasons change, and compare our results with the available observations. In particular, we investigate if a perennial northern cap of nitrogen ice can form, and whether our simulations can produce results that are consistent with Earth-based spectroscopic and stellar occultation observations, as well as the observations of the surface by Voyager 2.

Section 2 gives an overview of the past observations of Triton relevant to volatile transport. Section 3 describes the model used, the model parameters, the settings of the reference simulations and defines what the best-case simulations would be. Section 4 discusses the sensitivity of bedrock surface temperatures on Triton to several model parameters. Section 5 presents long-term volatile transport simulations performed over several millions of Earth years (Myrs) and explores the impact of North-South asymmetries in internal heat flux, topography and $N_2$ ice albedo on the ice distribution. Section 6 presents short-term volatile transport simulations performed over several seasonal cycles with fixed $N_2$ ice reservoirs and explores the sensitivity of the surface pressure cycle to different model parameters, including the ice distribution. Section 7 presents a large range of simulation results performed with the full volatile transport model and explores what scenarios and model parameters allow for a best match of the available observations. We discuss these results in Section 8 and provide predictions of Triton's climate for the next decades and future observations with the James Webb Space Telescope (JWST).

## 2. Background: observations and characteristics of the volatile cycle on Triton

### 2.1 Triton's orbit and seasons

Triton is moving in a retrograde and circular orbit with an orbital inclination with respect to Neptune of 157°, that is 23° above the planet's equator. It is tidally locked to Neptune with a rotation period of 5.877 days. The combination of this inclination with Neptune's obliquity of 28° leads to complex seasonal cycles, oscillating between low (~5° latitude), moderate (~20° latitude) and extreme (~50° latitude) summer solstices (Trafton, 1984, also see Appendix) over a period of 140-180 years. Each season on Triton lasts ~35-45 terrestrial years. The recent epoch lasting from 1980–2020, during which the highest quality observations of Triton were gathered, corresponds to a period of intense summer in Triton's southern hemisphere (in particular at the south pole), as the subsolar latitude of Triton reached ~50°S in 2000 (Figure 1), and follows a relatively intense winter. At present, the subsolar latitude is rapidly migrating toward the equator. For instance, in 2021, the subsolar latitude will be 36°S and previously unseen regions in the northern hemisphere will come into view (up to 54°N).

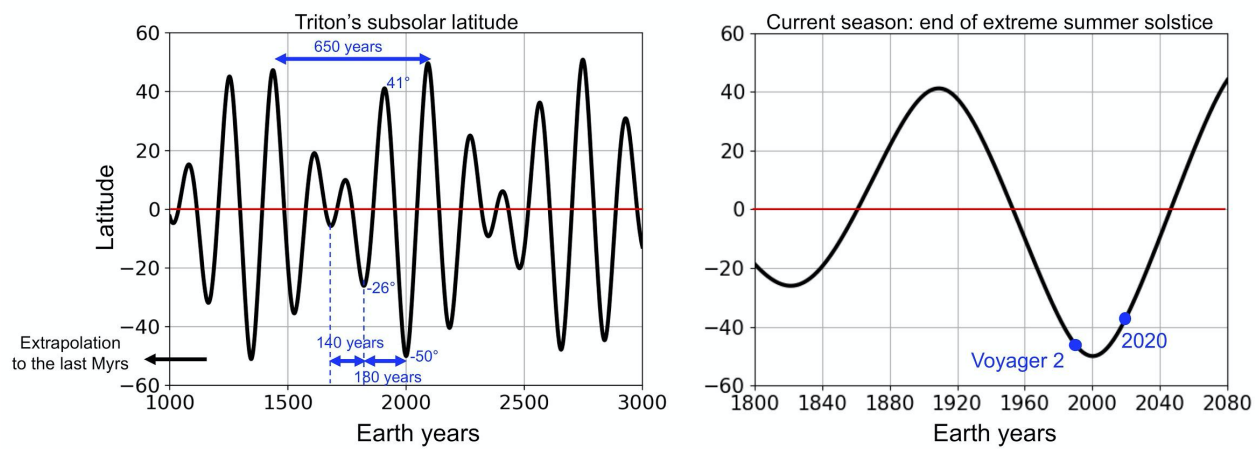

*Figure 1: Variation of the subsolar latitude on Triton over time. Triton experiences periods with low (e.g., years 2300-2400), moderate (e.g., years 1500-1650) and intense seasons (e.g., years 1850-2200).*

## 2.2 Surface albedo, color and texture

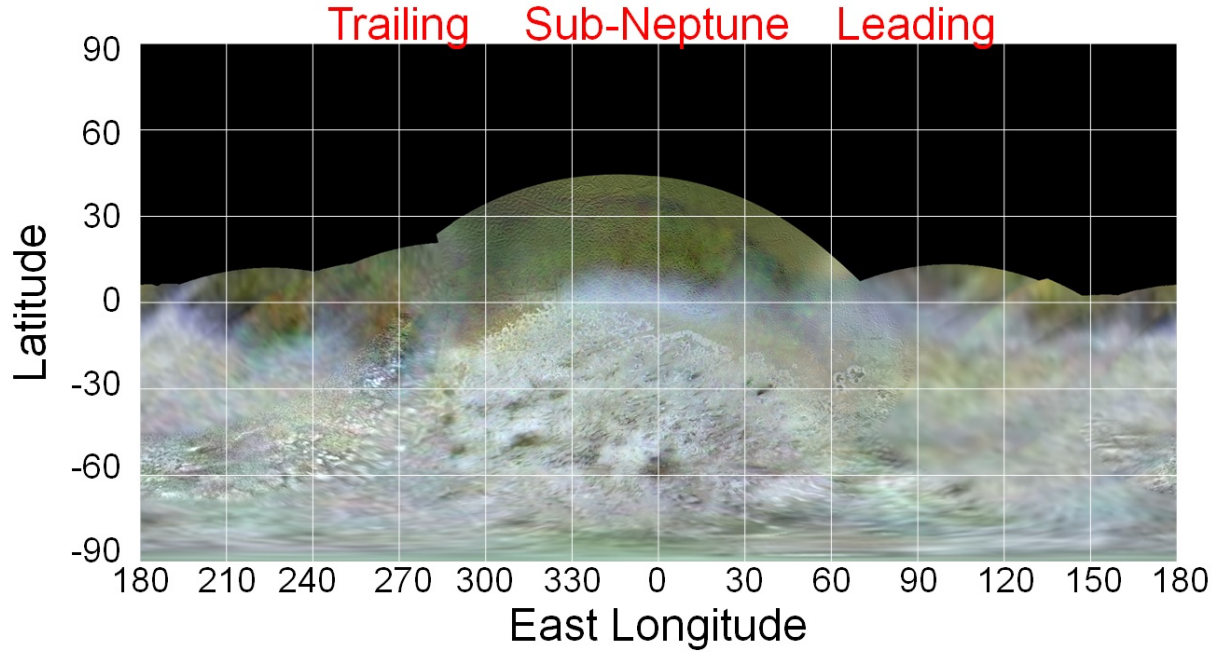

*Figure 2: Global color map mosaic of Triton (simple cylindrical projection) at 0.35 km/pixel, as seen by Voyager 2 (color-composite uses orange, green and UV filters in the red, green and blue colors; Schenk et al., 2021). 0°E, 90°E, 180°E, 270°E correspond to the sub-Neptune, leading, anti-Neptune and trailing longitudes respectively.*

In 1989, the Voyager 2 flyby of Triton revealed an amazing world with a geologically young (<100 Myr), diverse and active surface as suggested by the presence of plumes (evidence of resurfacing processes), tectonic structures, and very few impact craters (Smith et al., 1989, Croft et al., 1995, Stern and McKinnon, 2000, Schenk and Zahnle, 2007). Figure 2 shows a map of Triton as seen by Voyager 2. Only ~60% of the surface has been imaged (including half of that at relatively high resolution), as much of the northern hemisphere was hidden in the polar night. The northern and southern hemispheres display different terrains in color, albedo, and texture, suggestive of distinct compositions (Cruikshank et al., 1993, McEwen, 1990). Unfortunately, Voyager 2 lacked spectral measurements in the near-infrared, and therefore the exact composition of the different regions remains uncertain.

Triton is uniformly very bright, with a normal reflectance of the surface that varies between ~0.7 and ~1 (Hillier et al., 1994) and a Bond albedo of ~0.85 (Hillier et al., 1991), with a higher reflectance in the southern hemisphere. Estimates of surface emissivities from Voyager 2 data range from 0.3 to 0.8 (Hillier et al., 1991, Stansberry et al., 1996a ; see Section 3.4.1). The Voyager 2 observations suggest that the southern hemisphere is covered by a bright cap of volatile ice (likely $N_2$ mixed with $CH_4$ and CO, see Section 2.3), extending to the equator, whereas the observed part of the northern hemisphere may be volatile-free (e.g., Stone and Miner, 1989; Moore and Spencer, 1990). In the absence of spectral measurements, this is supported by the observations of many local dark areas on the bright ice resembling sublimation patterns on Mars (e.g. Mangold, 2011) and on Pluto (e.g. White et al., 2017, Howard et al., 2017) and by the volatile ice longitudinal variability observed from Earth-based near-infrared spectroscopy (see Section 2.3). In this paper, we refer to this geologic terrain as the southern cap (it can include permanent and seasonal volatile deposits), although we note that other types of terrains and ices may be present at the surface in the southern hemisphere. We do not use the term "polar cap" since the caps tend to extend outside the polar regions down to mid-latitudes and to the equator, but both northern and southern caps mentioned in this paper are centered at the pole. In Voyager 2 images, the northern latitudes seem to be depleted in volatile ice, based on the albedo, geology and texture of these terrains, and should therefore consist of Triton's exposed "bedrock" ($CO_2$ and/or $H_2O$). Alternatively, the northern hemisphere could be covered by a thin transparent slab of $N_2$ ice, so that the albedo of the underlying bedrock is not significantly altered. This has been suggested by Lee et al. (1992), who performed photometric analyses of Voyager 2 high-resolution images and revealed unexpected scattering properties of the surface in the northern hemisphere.

Another region of interest is the blue and very bright equatorial fringe of frost seen between 325°E and 30°E in the close encounter images (Smith et al., 1989, McEwen, 1990). A bluer surface is interpreted as a younger, fresher surface where recent deposition of volatiles has occurred. This is due to a shorter period of exposure to radiation and cosmic rays that tends to redden planetary surfaces. More specifically, the blue (or less red) fringe seen by Voyager 2 has been interpreted to be freshly deposited seasonal $N_2$ (e.g. Lunine and Stevenson, 1985; Zent 1989).

Photometric observations of Triton have been performed since the 1950s and revealed changes in surface color with time, possibly due to seasonal volatile transport of $N_2$, CO and $CH_4$. In particular, the surface became bluer (less red) between 1979 and the Voyager 2 flyby in 1989 (Smith et al., 1989; Buratti et al., 1994; Brown et al., 1995). Whether these bluer regions were brought into view by changing viewing geometry or were created by deposition of volatiles on the already visible hemisphere of Triton is unknown. Possible direct evidence for volatile transport between the Voyager 2 flyby and 2005 was reported by Bauer et al. (2010). They used visible images of Triton obtained with the Hubble Space Telescope (HST) in 2005 to construct visible albedo maps and lightcurves to compare with Voyager 2's observations. Rotational and regional changes in surface albedo were identified, with an overall brightening in the equatorial regions, suggesting recent deposition there, and a darker anti-Neptune hemisphere. Photometry of Triton for the 1992–2004 period found that the color of the surface was similar to that of the 1950–1974 period (Buratti et al., 1994, Buratti et al., 2011), thus suggesting that Triton reddened since 1989 (assuming it became bluer before 1989). This may be due to the plume activity occurring over the southern cap which tends to redden the ice (wind streaks appear darker and redder than the surrounding deposits, McEwen, 1990) and therefore the global color of Triton (the observed global reddening of Triton's surface could not be linked to insolation changes, Buratti et al., 2011).

## 2.3 Earth-based surface spectroscopic observations

In addition to the Voyager data, several ground-based spectroscopic observations of Triton's surface have been performed, leading to different scenarios for the surface distribution of the volatile and non-volatile ices. The volatile ices $N_2$, CO, $CH_4$ and non-volatile $CO_2$ and $H_2O$ ices have been detected by Cruikshank et al. (1993, 2000), with the non-volatile ices forming Triton's bedrock being at the estimated mean surface temperature of ~40-50 K (Broadfoot et al., 1989; Gurrola, 1995). The surface temperature of $N_2$ ice was estimated to be 38 ± 1K in 1993 (Tryka et al., 1994), and 37.5 ± 1K for the 2010–2013 epoch (Merlin et al., 2018). It was demonstrated that CO and most of the $CH_4$ are diluted in $N_2$ ice with ice mixing ratios relative to $N_2$ of ~0.05% (Cruikshank et al. 1993; Quirico et al. 1999; Grundy et al. 2010, Merlin et al., 2018), with a small fraction of $CH_4$-rich ice being present elsewhere (Quirico et al., 1999, Merlin et al., 2018). Tegler et al. (2019) also showed that CO and $N_2$ molecules on Triton are intimately mixed in the ice rather than existing as separate regions.

During the 2002-2014 period, ground-based infrared (0.8 to 2.4μm) spectroscopic observations performed using the SpeX instrument at NASA's Infrared Telescope Facility (IRTF) showed a general increase in $N_2$ and $CH_4$ band absorption during this period (Holler et al., 2016). Recent IRTF/SpeX results indicate that they are continuing to increase (Young, L. A.,, personal communication). These trends can be indicative of volatile transport although the monotonic increase in $CH_4$ absorption could be due to a change in viewing geometry only, if southern high latitudes are deficient in $CH_4$.

Observations of Triton with IRTF/SpeX showed variability in spectral band depth suggesting longitudinal variability of the ice distribution (Grundy and Young 2004, Grundy et al., 2010, Holler et al., 2016). $N_2$ and CO spectra displayed a peak absorption at the same longitudes in the sub-Neptune hemisphere, confirming that the two species form a molecular solution. This longitudinal variability for $N_2$ and CO likely results from variations in surface ice coverage, although variations in other physical parameters of the ice (e.g., particle size, vertical segregation, etc.) may also be at work. There is a certain degree of consistency with the Voyager images showing a southern cap (likely $N_2$ + diluted CO and $CH_4$) roughly extending to the equator on the sub-Neptune hemisphere and ~30°S in the opposite hemisphere (Figure 2), which suggests that the longitudinal variation of the spectrum has not changed much since the Voyager 2 flyby (more $N_2$ ice in the sub-Neptune hemisphere). However, the marked rotational variation of the $N_2$ and CO spectra also suggests that the very southernmost latitudes are free of detectable $N_2$ and CO.

The same observations also showed longitudinal variability for $CH_4$, with an offset of 90° compared to that of $N_2$ and CO. All the $CH_4$ bands (weak and strong) were observed in phase, with a maximum absorption near 300°E, but the longitudinal variability decreased from weak to strong $CH_4$ bands (Grundy et al., 2010, Merlin et al., 2018), suggesting localized regions around 300°E with $CH_4$-rich ice (and thus with larger optical path lengths in $CH_4$ and a spectral effect most evident in the weakest $CH_4$ bands).

The $CO_2$ and $H_2O$ spectra showed little longitudinal variability, which means that the equatorial and low latitudes are either $CO_2$ and $H_2O$ free, or have uniform longitudinal coverage of $CO_2$ and $H_2O$. In the former case, it has been suggested that $CO_2$ and $H_2O$ are exposed at the south pole (roughly between 90°S-60°S), which would thus be devoid of volatile ice during the 2002-2014 period (Grundy et al., 2010, Holler et al., 2016). This scenario also has a certain degree of consistency with the observed longitudinal variation of $N_2$ ice with IRTF/SpeX, whose amplitude would be too shallow if $N_2$ ice was present at the south pole, according to the near-IR modeling of Grundy et al. (2010). We discuss further this scenario in Section 8.

New near-IR observations performed with SINFONI on the Very Large Telescope (VLT) for the 2010-2013 period obtain similar variabilities for the band areas and support a spatial configuration involving two main units: one dominated by $N_2$ (with diluted trace amounts of CO and $CH_4$), but also including $CO_2$ ice as small grains, and the other dominated by $H_2O$ and $CO_2$, but also including a few percent of $CH_4$-rich ice (Merlin et al., 2018). Note that Merlin et al. (2018) did not detect $H_2O$ ice directly (the absorption bands are not seen entirely or are hidden by $CH_4$ absorption bands), but their model results remain consistent with little longitudinal variability for $H_2O$ and $CO_2$ ice.

The modeling of the near-IR disk integrated spectrum of Triton's surface also provided constraints on the relative surface area (projected on the visible disk) covered or not covered by $CH_4$ ice (including $CH_4$-rich or diluted in $N_2$ ice, so by extension, the method constrains the fractional area volatile ice/non-volatile ice, e.g. Quirico et al. 1999, Merlin et al. 2018). For instance, Quirico et al. (1999) found for 1995 a volatile/non-volatile fractional area of 55%/45% (see their table IV) whereas Merlin et al. (2018) found for the 2010-2013 period a volatile/non-volatile fractional area of 60-70%/30-40% (see their Table 9), although they note that these results remain model-dependent. The coverage of $CH_4$-rich ice during this period is estimated to represent 2–3% of the observed surface (i.e. the surface projected on the visible disk) and to be larger in the anti-Neptune hemisphere, whereas Quirico et al. (1999) estimated a maximum of 10% for 1995. However, the latitudinal distribution of these units remains uncertain.

## 2.4 Atmospheric surface pressure

Triton's atmosphere is mainly $N_2$ and is in solid-gas equilibrium with the surface $N_2$ ice (Tyler et al., 1989). A comprehensive dataset of Triton's surface pressure is detailed in Marques Oliveira et al. (2021) (see their Table 4, and their Section 5 and 6). On August 25, 1989, the Voyager 2 spacecraft flew by Triton and sent its radio signal (RSS experiment) back to Earth as it passed behind Triton. The analysis of the phases and amplitudes of the recorded radio waves provided the line-of-sight column density of the $N_2$ atmosphere, and an estimation of the surface pressure of ~1.4 ± 0.2 Pa (Broadfoot et al., 1989; Gurrola, 1995).

Different sets of ground-based stellar occultation observations were obtained in the following years. Stellar occultations do not probe the same altitude as RSS, and the methods and assumptions used to retrieve atmospheric pressure vary from one group to another. Consequently, careful comparisons between stellar occultation observations must be made. The first stellar occultation observations performed after Triton's flyby were interpreted as showing that the atmospheric pressure doubled during the 1989–1997 period (Olkin et al., 1997, Elliot et al., 1998). Olkin et al. (1997) derived a surface pressure of 1.7 ± 0.1 Pa for 1995, i.e. a 40% increase since 1989, but at a low 1.8-σ significance level. Elliot et al. (2000) and Marques Oliveira et al. (2021) derived a surface pressure of 2.68 ± 0.34 Pa and $2.28^{+5.4}_{-3.6}$ for the same event (but using different analysis methods) in 1997 (July 18[th]), respectively, i.e. a 60-90% increase since 1989. It is possible that this dramatic increase in surface pressure is related to the sublimation of the $N_2$ southern cap. However, in the analysis from Marques Oliveira et al. (2021), no increase can be claimed between the 1989 and the 1995-1997 epochs at a 3-σ significance level. The 4 November 1997 value of Elliot et al. (2003) has a much lower error bar, and at face value does indicate an increase of pressure by a factor 1.76 between 1989 and 1997 (Ps = 2.11 ± 0.02 Pa). This was a single-chord event but with a central flash (chord close to central). The 21 May 2008 event provided only two grazing chords, making it difficult to infer any change of pressure between 1989 and 2003. Finally, a stellar occultation by Triton was observed on 5 October 2017 and provided 90 chords, 25 of them showing a central flash near the shadow center, from which a surface pressure of 1.41 ± 0.04 Pa was derived (Marques Oliveira et al., 2021). This value for 2017 is similar to that measured during the Voyager 2 flyby, which suggests (although at a small confidence level) a pressure maximum around years 2000-2010 and a recent decrease in pressure related to $N_2$ ice deposition in the northern hemisphere or in the equatorial regions (deposition at the South polar region is ruled out as it remained under continuous daylight). This is also consistent with the $N_2$ ice temperature of 37.5 ± 1K measured for the 2010–2013 epoch from spectral analyses of the $N_2$ band shapes and positions (Merlin et al., 2018), which suggests a $1.6^{+1.6}_{-0.7}$ Pa pressure for this period.

## 2.5 Atmospheric abundances of CO and $CH_4$

Triton's atmosphere contains traces of CO and $CH_4$. In 1989, the Ultraviolet Spectrometer (UVS) on-board Voyager 2 measured a partial pressure of $CH_4$ of $2.45 \times 10^{-4}$ Pa (2.45 nbar) from a solar occultation (Herbert and Sandel, 1991, Strobel & Summers, 1995) and a surface pressure of ~1.4 Pa, which gives a $CH_4$ atmospheric volume mixing ratio of ~0.03% near the surface. This is the only direct measurement of the $CH_4$ mixing ratio on Triton (since both surface pressure and $CH_4$ partial pressure were precisely measured at the same time).

Lellouch et al. (2010) reported in 2009 the first detection of $CH_4$ in the infrared (IR) and the first ever detection of CO in Triton's atmosphere. They found that the partial pressure of $CH_4$ had increased to $9.8 ± 3.7 \times 10^{-4}$ Pa (fourfold increase compared to the 1989 value) and estimated the partial pressure of CO to be about $2.4 \times 10^{-3}$ Pa (24 nbar) with a factor 4 uncertainty. However, they did not measure the surface pressure. If we assume a constant surface pressure of 1.4 Pa for this period (resp. 1.8 Pa), these values correspond to an atmospheric volume mixing ratio for $CH_4$ of ~0.08%-0.17% (resp. 0.06%-0.13%) and for CO of ~0.17% (resp. 0.13%). This corresponds to a fourfold increase in $CH_4$ mixing ratio since 1989.

These values are higher than what was anticipated on the basis of Raoult equilibrium and an ideal $N_2$-$CH_4$-CO mixture ((of the order of $10^{-5}$ %, Trafton et al., 1998). A first explanation for that calls for the presence of relatively pure icy patches on Triton's surface (Stansberry et al., 1996b, Lellouch et al. 2010) which have been indeed identified in small amount at the surface of Triton by near-infrared observations (Quirico et al., 1999, Merlin et al., 2018). It has been demonstrated that it is a plausible mechanism for $CH_4$ ice (Lellouch et al., 2010, Merlin et al., 2018; note that it has clearly been observed on Pluto, e.g. Schmitt et al., 2017), but not for CO, which cannot thermodynamically separate itself from $N_2$ (Prokhvatilov and Yantsevich, 1983, Vetter et al., 2007, Tan and Kargel, 2018). A second explanation, which applies to CO, is that the seasonal volatile cycles lead to the formation of enriched layers of CO on top of the surface according to the "detailed balancing model" (Trafton, 1984, Lellouch et al., 2010). Note however that based on recent (2017) mm-observations with ALMA, Gurwell et al. (2019) reported a ~100 ppm (~0.01%) CO mixing ratio for an assumed 1.7 Pa atmosphere. This leads to a $1.7 \times 10^{-4}$ Pa CO partial pressure, sharply inconsistent (more than a factor 10 lower) with the IR-derived value from Lellouch et al. (2010). These differences remain to be investigated and understood in more detail, especially given the fact that IR and mm-observations of CO in Pluto's atmosphere give very consistent results on the CO abundance (Lellouch et al. 2011, 2017).

## 2.6 Seasonal volatile transport: expectations and modeling

As described in the previous sections, several pre- and post-Voyager 2 datasets show changes in Triton's surface albedo patterns, spectra, color and optical light curve that aren't completely due to changes in viewing geometry, but instead could be due to volatile ice migration driven by seasonal insolation changes. The last decades on Triton's southern hemisphere correspond to a period of intense summer, with a subsolar point poleward of 40˚S (Figure 1). Volatile sublimation in the southern hemisphere is therefore expected and several observations support the fact that the bright southern cap, usually presumed to be mostly made of $N_2$ ice, is currently sublimating away. First, Voyager 2 images of the equatorial regions show local dark areas on the bright ice (on the sub-Neptune hemisphere), which suggest that erosion due to sublimation of $N_2$ ice was already significant in 1989, when the subsolar latitude was 46°S (Smith et al., 1989). Second, spectral observations showed a tentative seasonal reduction (at 2-σ) in Triton's $N_2$ ice absorption during the 2000-2009 period (Grundy et al., 2010). The longitudinal pattern of this decline is consistent with a textural reduction in optical path length in $N_2$ ice, which suggests $N_2$ ice sublimation but not a retreat of the $N_2$ ice southern cap (although this is not a unique interpretation, as complex and poorly known microphysical processes are at play). Third, Triton's global color has become redder during the 1992-2004 period, possibly due to plume activity and deposition of dark reddish material over the southern cap, associated with the rapid sublimation of the cap that also concentrates at the surface non-volatile particles included in the ice (McEwen, 1990, Buratti et al., 2011).

Volatile transport models have been developed to investigate the seasonal variations in surface ice distribution on Triton due to the migration of the subsolar point. Stansberry et al. (1990) and Spencer (1990) (superseded by Moore & Spencer 1990, and again by Spencer and Moore 1992) developed zonally-averaged models of the $N_2$ seasonal cycle and investigated how $N_2$ ice could evolve and be transported. The former model assumed that $N_2$ ice is brighter than the underlying bedrock while the latter assumed that fresh $N_2$ ice is relatively dark and gets brighter with time and insolation, in a similar way than what has been observed on Mars (Paige 1985). However, both models met limited success in reproducing the observed surface albedo. Most importantly, they did not take into account heat conduction in the subsurface, which has a significant impact on the behavior of the seasonal ice on Mars and on Pluto (Wood and Paige 1992, Bertrand and Forget, 2016, Bertrand et al., 2018, 2019).

Hansen and Paige (1992) investigated both scenarios of bright and dark $N_2$ ice with a thermal model of the $N_2$ cycle which included conduction in the subsurface. Their best agreement between the model results and the available observations at that time was obtained with a relatively dark or transparent frost. However, in light of the occultation observations in the 1990s and in 2017, it seems that their scenario with a bright frost now gives a better agreement. They also found that the southern cap is likely to be a large permanent deposit of $N_2$ ice, otherwise it could not have been extended to the equator in 1989 as seen by Voyager 2. Brown and Kirk (1994) also developed a volatile transport model for Triton, and coupled it with an internal heat flow. They showed that an anisotropic internal heat flow could produce permanent caps of considerable latitudinal extent. Other volatile transport modeling studies showed that a bright permanent $N_2$ cap (with a reduced $N_2$ ice reservoir) can also be produced and/or maintained by ( a permanent albedo difference between the northern (bedrock) and southern hemispheres ($N_2$ ice caps) which affects the radiative balance (the "Koyaanismuuyaw" hypothesis, Moore and Spencer 1990, Spencer and Moore, 1992), or by changes in radiative properties of $N_2$ ice as it goes through the alpha-beta phase transition (this would not produce an asymmetry but would instead help maintain one, Eluszkiewicz, 1991, Duxbury and Brown, 1993, Tryka et al, 1993).

A difference in bedrock topography could also trigger an asymmetry between northern and southern $N_2$ deposits. On Pluto, $N_2$ ice tends to accumulate and be more stable in topographic basins, where the surface pressure and therefore the equilibrium ice temperature is higher. This is because the warmer $N_2$ ice deposits at lower altitudes radiate more heat to space, which is balanced by latent heat of sublimation through increased $N_2$ deposition rates in order to maintain local conservation of energy, vapor-pressure equilibrium, and hydrostatic equilibrium (Trafton et al. 1998, Trafton & Stansberry 2015, Bertrand & Forget 2016, Bertrand et al. 2018, 2019, and Young et al 2017). This atmospheric-topographic process could also apply to Triton, although the surface (at least the portion imaged by Voyager 2) seems to be much flatter than Pluto's (Schenk et al., 2021).

In addition to the volatile cycles, Triton's surface activity may be (globally or locally) impacted by the formation of active geysers, such as those observed at the south pole by Voyager 2. Such processes could act to darken the ice, increase the ice sublimation rates, and thus affect volatile transport. The formation of complex chemical molecules (whose spectral properties in the visible range are fairly reproduced by tholin materials formed in the laboratory, Materese et al., 2015, Auge et al., 2016, Jovanovic et al., 2020), through volatile photochemistry occurring in the atmosphere (Krasnopolsky and Cruikshank, 1999), or through direct volatile irradiation at the surface (Moore and Hudson, 2003), could play a similar role at a more global scale.

Finally, complex evolution of $N_2$:CO:$CH_4$ mixtures that result in compositional stratification and formation of a bright lag deposit of $CH_4$ can occur on Triton, in particular where the $N_2$ sublimation is more intense (Grundy and Fink, 1991, Cruikshank et al 1991, Quirico et al., 1999). For instance, on Pluto, the northern edge of the $N_2$ ice sheet Sputnik Planitia is enriched in $CH_4$ due to intense $N_2$ sublimation at these latitudes (Protopapa et al., 2017, Schmitt et al., 2017, Bertrand et al., 2018). This could affect the ice and atmospheric mixing ratio of the volatile species.

## 2.7 Reservoirs of volatile ice and internal heat flow

Triton's orbital elements likely indicate that it is a captured Trans-neptunian Object (e.g., McCord, 1966; McKinnon, 1984; Agnor and Hamilton, 2006, Li et al., 2020), sharing a common volatile origin as the Kuiper Belt Objects (Johnson et al. 2015). Tidal interactions associated with its capture and subsequent circularization around Neptune should have heated its interior significantly (McKinnon et al., 1995, Correia, 2009; Nogueira et al., 2011). Remnant heat stemming from these interactions may have persisted to the present day, as suggested by recent studies showing that the heat flow on Triton should be around 10-100 mW m$^{-2}$ (Ruiz, 2003, Martin-Herrero et al., 2018). These values are much larger than those obtained by assuming only radiogenic production and tidal dissipation for fixed orbital eccentricities (2-4 mW m$^{-2}$, Gaeman et al. 2012, Brown et al. 1991, Hussmann et al. 2006). The large heat flow could also be produced by internal ocean tidal heating due to Triton's orbit obliquity (Chen et al., 2014; Nimmo and Spencer, 2015; Dubois et al., 2017).

# 3. The Triton Volatile Transport Model: description and simulation settings

## 3.1 Model description: the Pluto VTM legacy

We use the latest version of the Triton volatile transport model (VTM) of the Laboratoire de Météorologie Dynamique (LMD). The Triton VTM is a 2D surface thermal model derived from the LMD Pluto VTM, taking into account the volatile cycles of N$_2$, CH$_4$, and CO (insolation, surface thermal balance, condensation-sublimation, Bertrand and Forget, 2016), a glacial flow scheme for N$_2$ ice (Bertrand et al., 2018; based on the equations presented in Umurhan et al., 2017), and the seasonal variation of the subsolar point specific to Triton. The calculations of this complex variation is detailed in the Appendix.

As in the Pluto VTM, we consider that Triton's atmosphere is very thin, almost transparent and thus has a negligible influence on the surface thermal balance aside from the condensation, sublimation and exchanges of latent heat with the surface (this is even more true than for Pluto, given the 10x lower CH$_4$ abundance, and more tenuous haze). We parametrize the atmospheric transport using a simple global mixing function for N$_2$, CH$_4$, and CO in place of 3D atmospheric transport and dynamics, with a characteristic time τ for the redistribution of the surface pressure and trace species, based on reference 3D global climate model simulations of Triton. Tests done with the 3D model determined the timescales for atmospheric transport of CH$_4$ ($10^7$ s, i.e., about 4 Earth months), N$_2$ (1 s, instantaneous mixing) and CO (1 s; CO is well mixed in the atmosphere in 3D global climate model simulations) used in the VTM (the same values were used on Pluto, Bertrand et al., 2019). The topography in the model is controlled by the amount of volatile ice on the surface, but we use a flat topography for the non-volatile bedrock.

## 3.2 General simulation settings and paleoclimate algorithm

The simulations of this paper are performed on a horizontal grid of 32×24 points, which corresponds to a grid-point spacing of 11.25° in longitude and 7.5° in latitude (about 270 km and 180 km at the equator, respectively). We use 96 timesteps per Triton day. At each time step the local solar insolation is calculated taking into account the variation of the subsolar point, the distance Triton-Sun, and the diurnal cycle. In Sections 5 and 7, we perform simulations over several million years (Myrs) using the paleoclimate and ice equilibration algorithm described in detail in Bertrand et al. (2018):

Typically, the model is first run over ~4,000 Earth years to capture several seasonal cycles of Triton so that the ice distribution and surface pressure reach an equilibrium. We consider that the first ~2,000 Earth years correspond to a spin up time and we use the last seasonal cycles covering ~2,000 Earth years to estimate the mean sublimation-condensation rates over that period (which is also largely enough to capture several seasonal cycles). These mean sublimation-condensation rates are used to calculate the new amounts of ice over a paleo-timestep of $\Delta t$=20,000 Earth years, and finally the topography is updated according to the new amounts of volatile ice. These steps are repeated so that the simulation covers at least 2 Myrs, which is enough for large and small $N_2$ ice reservoirs, glacial flow, and surface and subsurface temperatures to reach a steady state insensitive to the initial state. Consequently, in this paper, the distribution of surface ices and subsurface temperatures of all the simulations (unless stated otherwise) are the outcome of several Myrs of evolution.

Note that here we neglect any change over time in obliquity (with respect to Triton's orbit around Neptune) and orbital parameters, which is valid over the relatively short timescales of a few Myrs (the obliquity of Neptune is suggested to be primordial, Laskar and Robutel, 1993).

Also note that the numerical model only applies to global atmospheres (we neglect the effects of local atmospheres). For Pluto, the approximate limit at which point the atmosphere is non-global is 0.006 Pa (Johnson et al., 2021, Spencer et al., 1997). For Triton, the rescaled threshold is 0.009 Pa (owing to the larger size and gravity constant). In this paper, most of our simulations do not reach that threshold. A few of them do, but over an extremely short range of time, and therefore we do not expect this process to impact the results significantly (it involves very slow sublimation and condensation rates).

## 3.3 Hypothesis and initial state of the different simulation cases

| Section | Simulation type | Initial State | Model parameters |
|---|---|---|---|
| 4 | Bedrock surface thermal balance only<br><br>Several seasonal cycles (~10 000 Earth years) | Bedrock only, no volatile ice | **Fixed:** $A_{bed}$=0.6, $\varepsilon_{bed}$=0.8<br>**Variable:** TI=500-2000 SI, HF = 0-30 mW m$^{-2}$ |
| 5 | Full volatile transport<br><br>Long term (~7 Myrs) with viscous flow<br>North-South asymmetries in heat flux (HF), topography (TP) and $N_2$ ice albedo ($A_{N2}$) | Global and uniform cover of 300 m of $N_2$ ice | **Fixed:** $A_{bed}$=0.7, $\varepsilon_{bed}$=0.8, $A_{N2}$=0.7, $\varepsilon_{N2}$=0.8<br>**Asymmetry HF:** $\Delta F$=5-45 mW m$^{-2}$, TI=1000 SI<br>**Asymmetry TP:** $\Delta h$=2-8 km, TI=500-2000 SI<br>**Asymmetry A:** $\Delta A_{N2}$=0.04-0.1, TI=500-2000 SI |
| 6 | Volatile transport limited to fixed $N_2$ reservoirs<br>Different polar cap extensions<br>Several seasonal cycles (~10 000 Earth years)<br>Including current seasonal cycle (1900-2100) | Fixed and infinite $N_2$ reservoirs (no glacial flow, flat topography)<br><br>$N_2$ albedo adjusted by model | **Fixed:** $\varepsilon_{N2}$=0.8<br>**Variable:** TI=500-2000 SI<br>**S. PC size:** 90°S-30°S, 90°S-0°, 80°S-0°<br>**N. PC size:** No cap, 90°S-75°N, 60°N, 45°N |
| 7 | Full volatile transport<br>Long term with viscous flow<br>Including current seasonal cycle (1900-2100)<br><br>North-South asymmetries in $N_2$ ice albedo | $N_2$ ice initially confined to the south pole (90°S-50°S)<br>Northern hemisphere initially volatile-free and warm (T~100 K)<br>$N_2$ albedo adjusted by model | **Fixed:** $\varepsilon_{bed}$=0.8, $\varepsilon_{N2}$=0.8, $TI_{N2}$=1000 SI<br>**Variable:** $A_{bed}$=0.1-0.9, $TI_{bed}$=100-2000 SI,<br>$\Delta F$=0-30 mW m$^{-2}$, $R_{N2}$=0.3-650 m<br>**Asymmetry in $N_2$ ice albedo:** $\Delta A_{N2}$ = 0-0.1 |

*Table 1: Summary of the simulation types, initial states and parameters presented in this paper (see text): albedo (A), emissivity (ε), seasonal thermal inertia (TI) in J $s^{-0.5}$ $m^{-2}$ $K^{-1}$ (SI), internal heat flux (HF), topography (TP), pole-to-pole internal flux, topography gradient and $N_2$ albedo difference (ΔF, Δh, $\Delta A_{N2}$). S. and N. PC are the southern and northern polar caps.*

In this paper, we postulate that the bright terrains observed by Voyager 2 in the southern hemisphere correspond to a $N_2$ ice southern cap (mixed with $CH_4$ and CO ices), based on Voyager images and Earth-based spectroscopic observations. As described in Section 2.3, some Earth-based spectroscopic observations suggest that $N_2$ is not exposed or not detectable at the very southernmost latitudes (i.e. the south pole, Grundy & Young 2004, Grundy et al. 2010, Holler et al. 2016). Note that this does not necessarily means that $N_2$ ice is not present at the south pole (see further discussions on this topic in section 8.3). In general, we leave the model calculate the surface distribution, but we also added simulation cases in which we force the south pole to be depleted in $N_2$ ice.

Since 1989, the southern cap has been under constant illumination and is likely dominated by $N_2$ ice sublimation. Under these conditions, the evolution of surface pressure depends on the $N_2$ condensation in the northern polar night. Table 1 summarizes the different simulations performed in this paper.

In Section 4, we run the subsurface and surface thermal balance model without volatile ice in order to estimate the mean surface temperatures of the bedrock, for different values of the thermal inertia and internal heat flow. The bedrock albedo is fixed to 0.6. Note that throughout this paper, the "albedo" in the VTM refers to a Bond albedo.

In Section 5, we run long term simulations of the $N_2$ cycle to explore how North-South asymmetries in internal heat flux, surface $N_2$ ice albedo and topography lead to asymmetries in the northern and southern cap extents. All simulations start with a global and uniform cover of 300 m of $N_2$ ice.

In Section 6, we run the model over the last seasonal cycles with artificially-prescribed fixed $N_2$ ice distributions in the northern and southern hemispheres. We use a flat topography, no glacial flow, and volatile transport is limited to these artificially-prescribed $N_2$-covered regions (no formation of seasonal frost outside these regions). We analyze the evolution of the surface pressure that results from these specific latitudinal distributions of $N_2$ ice and compare to observations.

In Section 7, we simulate the full volatile transport over the last 4 Myrs, exploring a large range of parameters (bedrock albedo, thermal inertia, $N_2$ ice reservoir, internal heat flux, and North-South asymmetry). In these simulations, the ice is initially placed southward of 50°S and the northern hemisphere is warm and volatile-free with its surface and subsurface temperatures initialized to an extremal value of 100 K. This is to demonstrate that the end state of the simulation and any formation of a northern (polar) cap is independent of the initial ice distribution and surface and subsurface temperatures (see discussions in Section 5 and Section 7). The bedrock topography is flat, so the surface topography is equal to the thickness of $N_2$ ice lying on top.

Note that in Section 5, Section 6, and Section 7, the $N_2$ ice albedo in the simulations is not a free model parameter but instead is constrained by the 1989's surface pressure: it is automatically adjusted by the model during the first seasonal cycles of the simulation (spin-up time) so that the calculated surface pressure converges toward ~1.4 Pa in 1989, as observed by Voyager 2 (see Section 3.4.1).

## 3.4 Surface properties

### 3.4.1. Albedos and emissivities

As on Pluto, the nitrogen cycle on Triton is very sensitive to the nitrogen ice Bond albedo $A_{N2}$ and emissivity $\varepsilon_{N2}$. The local energy balance on a $N_2$-covered surface on Triton can be written, to first order (assuming a spatially uniform and isothermal ice with a flat topography, an efficient global transport of $N_2$, and neglecting thermal inertia and latent heat exchanges) by the classical equation $\varepsilon_{N2}\sigma T^4 \approx (1 - A_{N2})$ F/4, where F is the solar constant at Triton, $\sigma$ is the Stefan-Boltzmann constant, and the factor ¼ comes from global averaging. As a result, the $N_2$ ice equilibrium temperature and therefore the $N_2$ surface pressure, depend on $(1 − A_{N2})/\varepsilon_{N2}$ (Spencer, 1990). On Pluto, the surface pressure dataset inferred from stellar occultations give a strong constraint on the combination of $A_{N2}$ and $\varepsilon_{N2}$ ; only a small range for these parameters allows for a satisfactory match to the observations (Bertrand et al., 2016, Meza et al., 2018, Johnson et al., 2020). In this paper, we intend to apply this study to Triton. We assume a fixed and relatively high emissivity of $\varepsilon_{N2}$=0.8, as in the Pluto VTM (it would correspond to large cm-sized $N_2$ ice grains, Stansberry et al., 1996a). This value is consistent with the values estimated for Triton from Voyager 2 data (0.7 < $\varepsilon_{N2}$ < 0.77, Stansberry et al., 1992), and remains within 10% of the upper value estimated on Pluto from a surface energy balance model (0.47 < $\varepsilon_{N2}$ < 0.72, Lewis et al., 2021). However, we note that Hillier et al. (1991) derived a lower surface emissivity of 0.46 ± 0.16, which could be consistent with smaller grain sizes of ~1 mm (Stansberry et al., 1996a). Consequently, in Section 7.3.2, we briefly explore the model sensitivity to lower emissivities than 0.8 (fixed $\varepsilon_{N2}$=0.3 and $\varepsilon_{N2}$=0.5). In addition, since the emissivity of $N_2$ ice is also thought to vary with the ice temperature, being lower in its α-phase than that in its β-phase (Stansberry and Yelle, 1999; Lellouch et al., 2011b), we also explore the case of a temperature dependent emissivity.

For practical reasons, $A_{N2}$ is automatically calculated by the model so that the surface pressure is close to 1.4 Pa in 1989, as measured by Voyager 2 (tuning manually $A_{N2}$ would be extremely expensive in time and computing resources). In all simulations of this paper, $A_{N2}$ is initially fixed to 0.75, and then the model increments or decrements $A_{N2}$ by steps of 0.005 during each extreme southern summer (at subsolar latitude 45°S corresponding to that of Voyager 2's flyby) so that the surface pressure converges towards 1.4 Pa at this season. Typically, it takes less than 10,000 Earth years for $A_{N2}$ to reach a stable value in the model and thus for the pressure cycle to be consistent with the observed surface pressure in 1989. Since all our simulations of the present-day Triton climate are the results of Myrs of simulations, this means that the convergence occurs quickly at the beginning of the simulations (spin-up time) We ensured that our model results are not sensitive to the initial $N_2$ ice albedo value and that all our simulations reach a steady state for $A_{N2}$ and surface pressure. By contrast, the albedo of the bedrock $A_{bed}$ is a free parameter of the model and we tested the sensitivity of the results to several values ranging from 0.1 to 0.9.

### 3.4.2. Thermal inertia (TI)

We assume a high seasonal thermal inertia in the sub-surface for $N_2$ ice, fixed to 800 J s$^{-1/2}$ m$^{-2}$ K$^{-1}$ (or SI) as has been suggested on Pluto (Bertrand et al., 2016, Johnson et al., 2020). A high thermal inertia was also inferred by Spencer and Moore (1992), and, in retrospect, their high thermal inertia models are more in agreement with the later observed pressure evolution. Note that to first order, this parameter does not significantly impact the $N_2$ cycle (see Section 2.2 in Bertrand et al., 2018). The diurnal thermal inertia (for all ices) is set to 20 SI by analogy to Pluto (Lellouch et al. 2011, 2016). Note that Pluto has large areas covered by tholins-like dark materials, which are expected to have low diurnal thermal inertias as they form very porous layers of very small grains. They may have played a significant role in the retrieval of thermal inertia values on Pluto (the lightcurves are mostly sensitive to the thermal inertia of the tholins), so the values on Triton could be higher. More thermal observations of Triton's surface are needed to constrain this parameter. For the bedrock (assumed to be water ice or $CO_2$ ice, but here in this paper it is only characterized by its albedo, emissivity and thermal inertia), we explored the sensitivity of the VTM results to low and high values of seasonal thermal inertia ranging from 200 to 2000 SI.

The thermal skin depth is defined as δ = TI/C x $\sqrt{(P/\pi)}$ with TI the thermal inertia, C the ground volumetric specific heat (assumed to be 10$^6$ J m−3 K−1) and P the period (s) of the thermal wave. With this definition, the diurnal skin depth for 20 SI of ~8 mm and a seasonal skin depth for 200 SI and 2000 SI of ~8 m and 80 m respectively. As in the Pluto VTM, the subsurface is divided into 24 discrete layers, with a geometrically stretched distribution of layers with higher resolution near the surface to capture the short period diurnal thermal waves (the depth of the first layer is $z_1 = 1.4 \times 10^{-4}$ m) and a coarser grid for deeper layers and long seasonal thermal waves (the deepest layer depth is near 1000 m). Note that when a $N_2$ ice deposit (e.g., 300 m thick) is present at the surface, we assume a subsurface seasonal thermal inertia of 800 SI in the subsurface levels that correspond to the thickness of the deposits (e.g., down to 300 m) and then the seasonal thermal inertia of the bedrock down to the deepest level.

### 3.4.3. $N_2$ ice reservoir and internal heat flux

Previous volatile transport modeling on Triton only used low $N_2$ ice reservoirs ($R_{N2}$ = 0.2-2 m, Moore and Spencer, 1990, Hansen and Paige, 1992, Spencer and Moore, 1992) and thus were limited to the simulation of seasonal frost. Spencer and Moore (1992) also tested simulations in which a permanent southern cap is artificially and indefinitely maintained to a large size, with the aim of reproducing to first order the effect of large reservoir and viscous spreading of $N_2$ ice from the pole to the equatorial regions. In retrospect, these simulation cases (e.g. case "L" in Spencer and Moore, 1992) are in relatively better agreement with the later observed pressure evolution. Here, on the basis of these results, we used different $N_2$ ice reservoirs $R_{N2}$ ranging from 1 m to 650 m in global surface coverage, thus exploring both small and large reservoirs. With large $N_2$ ice reservoirs, the model is able to self-consistently simulate the formation of thick perennial deposits and their viscous flow.

At the deepest subsurface level of the model, there may be a positive heat flow, which is balanced by upward thermal conduction from a negative thermal gradient (-k dT/dz) as in Bertrand et al. (2018, 2019). The reference simulations are performed assuming no internal heat flux, but we also explored the effect of assuming an internal heat flux of 30 mW m$^{-2}$.

### 3.4.4. Assumptions on the state of $N_2$ and $CH_4$ ice in the model

$N_2$, $CH_4$, and CO ices easily mix together and are not expected to exist in perfectly pure states on Triton. Instead, they should form non-ideal solid solutions whose phases follow ternary phase equilibria (Trafton, 2015; Tan and Kargel, 2018). Complex mixtures have been revealed on Pluto's surface by the analyses of New Horizons observations, with $N_2$:$CH_4$ ($N_2$-rich mixtures, e.g., Sputnik Planitia) and $CH_4$:$N_2$ ($CH_4$-rich mixtures, e.g., the north pole) solid solutions involving different molecular mixing ratios (Grundy et al., 2016; Protopapa et al., 2017; Schmitt et al., 2017). Observations also suggest mixtures of both $N_2$-rich + $CH_4$-rich ice phases at some locations, although the exact 'organization' (i.e. intimately mixed and/or vertically stratified) of these deposits remains uncertain (Protopapa et al., 2017; Schmitt et al., 2017).

We note that sophisticated equations of state exist for the $N_2$–$CH_4$ and $N_2$–$CH_4$–CO systems under surface conditions (surface pressure and temperature) similar to that of Triton (CRYOCHEM, Tan and Kargel, 2018). Although these ternary and binary systems, when applied to Pluto, give results relatively consistent with the diversity of phases seen on Pluto's surface, it remains unclear how they would apply to the case of Triton. In fact, on Triton, despite a surface pressure and temperatures similar to that of Pluto, the observed mole fraction of $CH_4$ in $N_2$-rich ice is ~0.05-0.11% (Quirico et al., 1999, Merlin et al., 2018), which is 5-10 times less than on Pluto (Douté et al. 1999; Merlin, 2015). In addition, the mechanisms controlling the formation and evolution of such mixtures remain largely unknown.

In this context, and given a certain lack of data on the ice mixtures on Triton, the model presented in this paper is rather simple and sticks to the available observations. As in previous VTM studies (Bertrand and Forget, 2016, Bertrand et al., 2018, 2019, 2020a), for simplicity in coding with a VTM, the model does not compute any evolution of ice mixing ratio and adopts Raoult's law to describe the solid-gas equilibria. In the simulations, the surface is either volatile-free, covered by pure $CH_4$ ice or by $N_2:CH_4$ ice. When both $CH_4$ and $N_2$ ices are present on the surface, we assume that $CH_4$ is diluted in a solid solution $N_2:CH_4$ with 0.05% of $CH_4$, as retrieved from the last telescopic observations (Merlin et al., 2018), but we also explored the addition of small areas of $CH_4$-rich ice (see Section 7.4.2). For CO concentration, we used 0.04% (close to the 0.05% reported by Quirico et al., 1999) and also explored 0.08% to bracket the values 0.04-0.08% suggested by Merlin et al., 2018 (see Section 7.4.1). The modeled $N_2:CH_4$ ice sublimates by conserving the 0.05% of diluted $CH_4$. We make the approximation that $CH_4$-rich ice behaves almost like pure $CH_4$ ice (Tan and Kargel, 2018), in terms of temperature and vapor pressure at saturation of $CH_4$, as $CH_4$-rich ice can contain only up to 3% $N_2$ around 40K (Prokhvatilov and Yantsevich 1983). It can form after sublimation of $N_2$ ice (in which $CH_4$ was trapped before) or directly on a volatile-free surface. In the next sections of this paper, we refer to this phase as $CH_4$ ice. The formation and evolution of other types of binary or ternary phases is out of the scope of this paper, and we neglect their effect in the model (although we acknowledge that it could lead to some unevaluated uncertainties in $N_2$-rich and $CH_4$ solid-phase stability).

## 3.5 Observational constraints for the definition of best-case simulations

As detailed in Section 3.4.1, the simulations are constrained by one unique observation: the surface pressure of ~1.4 Pa in 1989. The $N_2$ ice albedo is automatically adjusted by the model during the spin up time to match this constraint.

The other available observational constraints, presented in this section, are not explicitly included in the numerical calculations but are considered when interpreting the simulation results and when looking for the best-case simulations. We search for simulations including (1) an ice distribution consistent with Voyager 2 observations and ground-based near-infrared hemispheric spectra (the relative surface area of volatile vs. non-volatile ice), and (2) an evolution of surface pressure consistent with that retrieved from stellar occultations.

### 3.5.1. Constraints from the ground-based near-IR spectroscopy

As detailed in Section 2.3, near-IR ground-based spectroscopy of Triton's surface provides a relatively good constraint on the relative surface area (projected on the visible disk) covered or not covered by volatile ice. Based on the results from Quirico et al. 1999 and Merlin et al. 2018 for 1995 and 2010-2013, respectively, we assume with some margin that the volatile/non-volatile fractional area is 45-65%/35-55% in 1995 and 55-75%/25-45% in 2010. To take these constraints into account, we used the calculation in the appendix of Holler et al. (2016) to project our VTM-modeled surface on the visible disk at the time of observation (subsolar latitude 49°S and 46°S for 1995 and 2010, respectively). Note, that, in 2010, 60-70% of the disk-projected surface covered by $N_2$ would correspond to a southern cap extending from 90°S to ~20°S, or 80°S to ~15°S, or 60°S to ~0°.

### 3.5.2. Constraints from Voyager 2 images and infrared surface emission measurements

Based on the Voyager 2 images, we assume that the bright southern cap is made of $N_2$-rich ice, and that its northern edge extends to 30°S-0° in 1989. Note that the volatile/non-volatile fractional area derived from near-IR surface spectroscopy, detailed in Section 3.5.1, is a stronger constraint and is given more weight when interpreting the results because it is a more direct observation (Voyager 2 did not carry any IR spectrometer). We do not make any assumption regarding whether the southernmost latitudes are covered by $N_2$-rich ice or not. Our model self-consistently calculates the surface ice distribution and we consider that the best-case simulations should have a $N_2$-rich southern cap (with or without $N_2$-rich ice at the very pole) with a northern edge extending to 30°S-0° in 1989. In practice, as shown in Section 5 and Section 7, the south pole is always the most efficient cold trap for $N_2$ ice in seasonal average and all our simulations present a perennial $N_2$-rich deposit at the south pole (except in the cases where $N_2$ is artificially removed from the south pole). This point is further discussed in Section 8.3.

We assume that the northern (polar) cap, if it exists, was in winter night and therefore not seen by Voyager 2 in 1989, implying that it did not extend to latitudes southward of 45°N. Based on the analysis of Voyager 2 images (McEwen, 1990), we also assume that the albedo of all terrains is higher than 0.6.

The Infrared Interferometer Spectrometer (IRIS) instrument on-board Voyager 2 measured the infrared radiation emitted by Triton's surface. The spectra were relatively noisy and the surface emission was mostly detected at the longest wavelengths of the IRIS spectral range (i.e., ~40-50 μm). From these measurements, Conrath et al. (1989) and Stansberry et al. (2015) derived a full-disk averaged surface temperature of Triton in 1989 ranging from 37-44 K, depending on the emissivity of the different surface units. We use this range of surface temperature to constrain the best-case simulations. In fact, most simulations satisfying the previous albedo constraint of > 0.6 fall in this range.

### 3.5.3. Constraints from the surface pressure from Voyager 2 and stellar occultations

Section 2.4 summarizes the different observations that provide estimates of Triton's surface pressure. Table 2 gives the surface pressure data points used in this paper. Based on the analysis of an extremely high-quality occultation dataset from 2017, and the re-analysis of earlier occultation curves, Marques Oliveira et al. (2021) conclude that the increase in surface pressure reported in 1995-1997 (compared to the Voyager 2 value in 1989) remains elusive (as the data are not available for reanalysis using an approach consistent with theirs), but that the 2017 value has been obtained at a high significance level and is fully compatible with that measured by the Voyager 2 RSS experiment.

Consequently, in this paper, we consider the values reanalyzed by Marques Oliveira et al. (2021) as the strongest observational constraints of surface pressure for our volatile transport simulations, but we will also assess if the model can predict an increase in surface pressure maximum in the 1990s, as suggested by the other values. We also note that the 2017 event and the Voyager pressure are the best quality datasets among the analyzed events.

| Date | Surf. Pressure (Pa) | Reference |
|------|---------------------|-----------|
| 25 Aug. 1989 | $1.4 \pm 0.2$ | Gurrola et al., 1995 |
| 14 Aug. 1995 | $1.7 \pm 0.1$ | Olkin et al., 1997 |
| 18 Jul. 1997 | $2.28^{+0.54}_{-0.36}$<br>$2.68 \pm 0.34$ | Marques Oliveira et al., 2021<br>Elliot et al., 2000 |
| 4 Nov. 1997 | $2.11 \pm 0.02$ | Elliot et al., 2003 |
| 21 May 2008 | $1.38^{+1.24}_{-0.44}$ | Marques Oliveira et al., 2021 |
| 5 Oct. 2017 | $1.41 \pm 0.04$ | Marques Oliveira et al., 2021 |

*Table 2: Surface pressures on Triton derived from different observations and detailed in Marques Oliveira et al. (2021).*

## 4. Bedrock surface temperatures on Triton

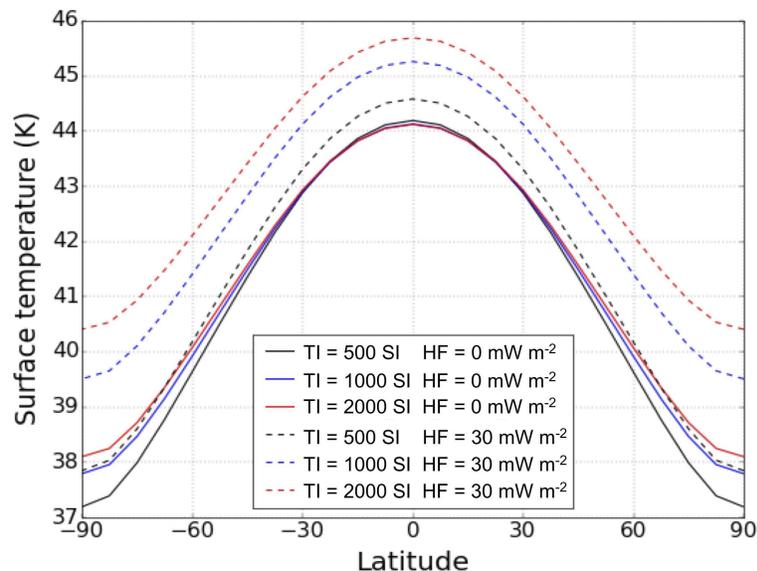

*Figure 3: Zonal mean surface temperatures of the bedrock of Triton (assumed to be water ice, no volatile ice) averaged over several seasonal cycles (~10,000 Earth years), with subsurface thermal inertias within 500-2000 SI and internal heat fluxes 0 and 30 mW $m^{-2}$. The bedrock albedo is fixed to 0.6 and its emissivity to 0.8.*

On average over one or several seasonal cycles the poles on Triton receive less insolation flux than the equator. As a result, the poles are colder than the equatorial regions, as shown by Figure 3. High subsurface thermal inertia allows the subsurface to store more heat accumulated during extreme summer and release it during winter, thus dampening and delaying the response of the surface temperatures to insolation (Spencer, 1989) and raising the mean temperatures (the poles are ~1 K warmer with high thermal inertia than with low thermal inertia, on average). While on Pluto, high subsurface thermal inertia allows the poles to be as warm or warmer than the equator on annual or multi-annual average (see Fig. 3 and 4 in Bertrand et al., 2018), on Triton, the poles remain colder than the equator by 5-6 K on average because they receive much less flux (~0.25 W $m^{-2}$) than the equator (~0.44 W $m^{-2}$) on average over several seasonal cycles (the subsolar point never reaches high enough latitudes to make the poles warmer than the equator on average, and extreme summers are not occurring every seasonal cycle). For a subsurface (i.e., $H_2O$ ice) TI of 1000 SI, a bedrock surface albedo of 0.6 and no internal flux, maximum and minimum bedrock surface temperatures are respectively ~49 K and ~33 K at the poles (during extreme seasons) and respectively ~45 K and ~43 K at the equator.

Volatile ice would therefore accumulate at the poles on Triton, and form a cold trap. Due to the very low eccentricity of Neptune's orbit and the circular orbit of Triton around Neptune, the northern and southern latitudes of Triton undergo the same seasons over several seasonal cycles (with the same insolation and heliocentric distances). Figure 3 shows that both poles on Triton are symmetric in terms of insolation and surface temperatures, on average over several seasonal cycles. Under these conditions, the volatile transport model would simulate the formation of symmetric southern and northern caps if all ice and surface properties were constant and uniform.

However, in reality, it is likely that one cap "wins" over the other (see Section 5). For instance, Figure 3 shows that an internal heat flux of 30 mW m$^{-2}$ would raise the bedrock surface temperatures by up to ~2 K (for high subsurface thermal inertias), and could locally prevent N$_2$ condensation at the locations where such a flux reaches the surface.

# 5. Long-term VTM simulation of the N$_2$ cycle with North-South asymmetries

As discussed in Section 2.6, the Voyager 2 observations suggest that a southern cap (presumably made of N$_2$ ice) extends to the equator. It is not clear whether a north (polar) cap exists on Triton, as Voyager 2 did not detect it outside the polar night (southward of 45°N) in 1989. If it does exist, it must therefore be smaller than the southern cap. In Section 2.6, we listed several mechanisms that have been proposed to explain this asymmetry between the northern and southern caps, including the asymmetry in internal heat flux and in surface ice albedo, explored in detail by Brown and Kirk (1994) and Moore and Spencer (1990), respectively.

In this section, we use our volatile transport model to test how the N$_2$ ice deposits evolve when we set a North-South asymmetry in internal heat flux, surface N$_2$ ice albedo, and topography. All simulations start with a global and uniform cover of 300 m of N$_2$ ice, take into account glacial viscous flow of N$_2$ ice (Umurhan et al., 2017), and are run over 9 Myrs, which is long enough so that the surface and subsurface reach a steady state. Other simulation parameters and settings are summarized in Table 1.

## 5.1 North-South asymmetry in internal heat flux

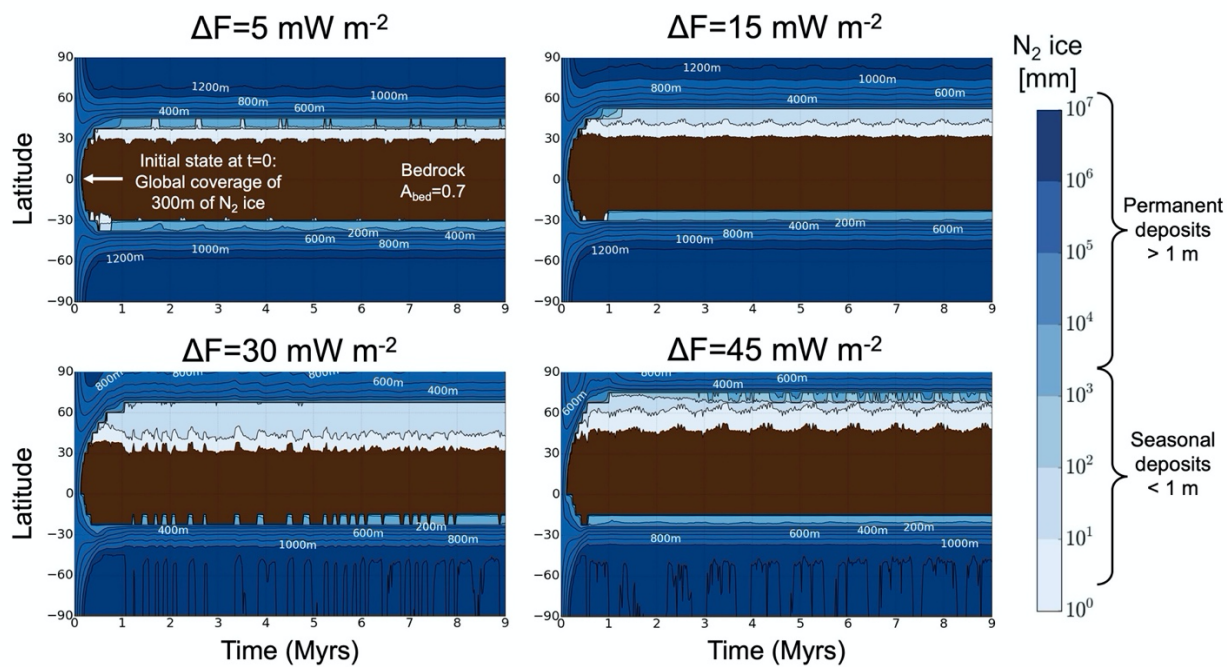

*Figure 4: Annual mean evolution of nitrogen ice thickness (in zonal mean) throughout entire VTM simulations (from t=0 to t=9 Myrs), assuming an internal heat flux in the northern hemisphere only*

*(uniform from 0°N to 90°N) of 5, 15, 30 and 45 mW m$^{-2}$, and a subsurface bedrock thermal inertia of 1000 SI. Deposits thinner than 1 m tend to be seasonal. The N$_2$ ice albedo obtained at equilibrium is 0.755, 0.785, 0.790, and 0.795, respectively.*

Brown and Kirk (1994) showed that Triton's internal heat source could significantly affect volatile transport and that asymmetries in its latitudinal distribution (possibly driven by volcanic activity or internal convection) could result in permanent caps of unequal latitudinal extent, including the case with only one permanent cap. As expected, we find similar results when we test this scenario in our volatile transport model.

Figure 4 shows the annual mean evolution of N$_2$ ice thickness when we assume an internal heat flux of 5, 15, 30 and 45 mW m$^{-2}$ in the northern hemisphere only (the internal heat flux is set to 0 mW m$^{-2}$ in the southern hemisphere). In the northern hemisphere, the internal heat flux transferred to the surface N$_2$ ice is consumed through the latent heat of sublimation of N$_2$ ice and the maintenance of vapor pressure equilibrium. This leads to enhanced N$_2$ sublimation rates and reduced N$_2$ condensation rates and thus favors a larger southern N$_2$ ice cap.

Our results show that a permanent northern cap, centered at the pole with a thickness of at least a few hundred meters, forms in all cases. In all cases there is also a southern permanent cap, that is at least a kilometer thick at the pole and hundreds of meters thick at ~30°S, in agreement with our interpretation of Voyager 2 observations.

Figure 4 shows that a heat flux difference of at least +15 mW m$^{-2}$ between both hemispheres is necessary for permanent N$_2$ ice deposits (>1 m) in the northern cap to remain confined poleward of 45°N (and therefore be hidden in the polar night during the Voyager 2 flyby). For example, if the difference reaches +45 mW m$^{-2}$, then the permanent northern cap is very small and only extends down to 80°N, according to our model. We note that in all cases, mm-to-m thick seasonal deposits extend to 30°N-45°N.

## 5.2 North-South asymmetry in topography

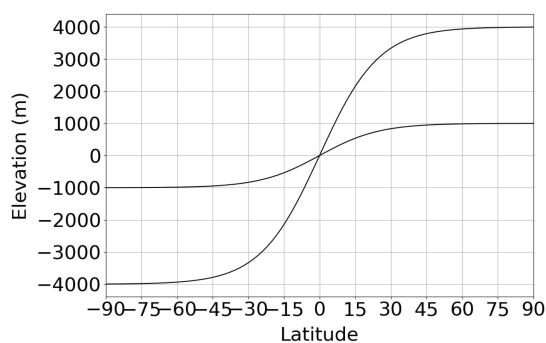

*Figure 5: Topography of the bedrock for the simulations described in Section 5.2, with a North-South asymmetry of ± 1 and ± 4 km.*

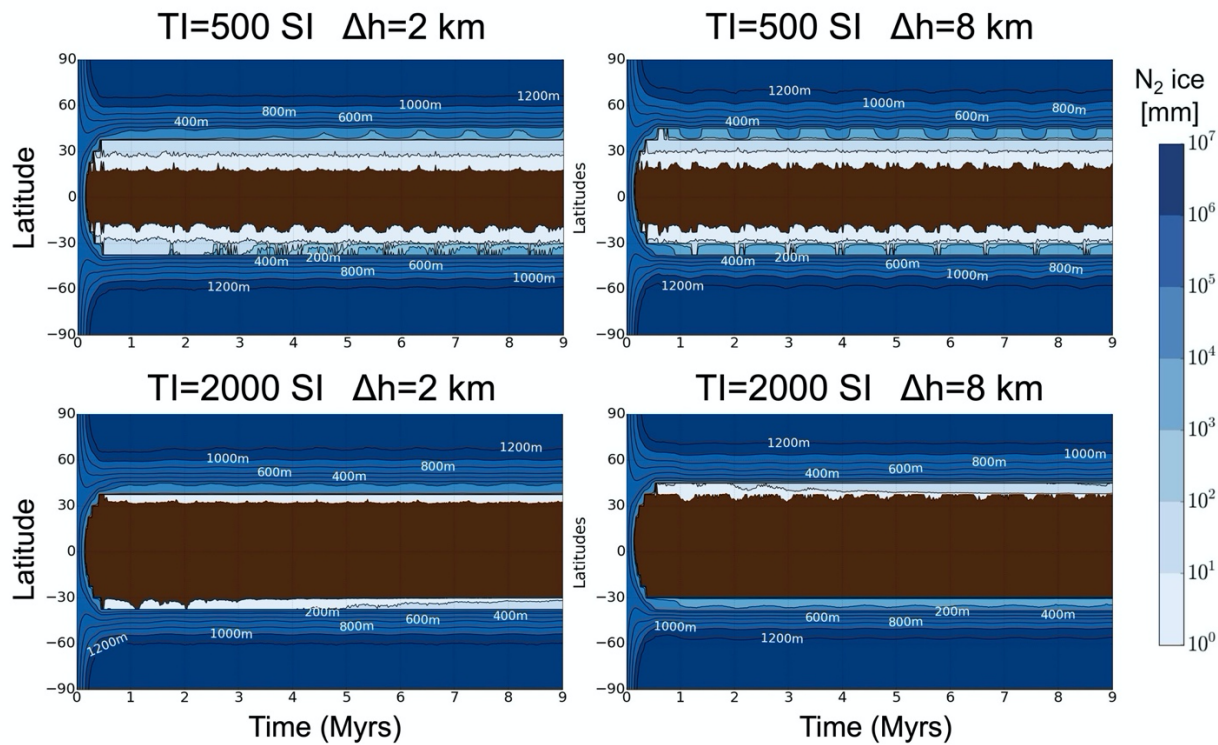

*Figure 6: Annual mean evolution of nitrogen ice thickness (in zonal mean) assuming a North-South asymmetry in topography (as shown on Figure 5, with a pole-to-pole difference of 2 and 8 km), and a subsurface thermal inertia of 500-2000 SI. The $N_2$ ice albedo obtained at equilibrium is 0.740 and 0.750 (for TI=500 SI), and 0.740, and 0.745 (for TI=2000 SI), respectively.*

As detailed in Section 2.6, the main reservoir of $N_2$ ice on Pluto is confined within the massive topographic basin of Sputnik Planitia (estimated to be as much as ~10 km deep, McKinnon et al., 2016) due to (1) its location in the equatorial regions (long-term cold traps induced by the high obliquity cycles) and (2) higher condensation rates in the basin induced by higher surface pressure and infrared cooling of the ice. This atmospheric-topographic process is expected to apply on Triton too, and one could imagine that a North-South topographic asymmetry would favor the formation of a permanent cap of $N_2$ ice in the lower-elevation hemisphere. Voyager stereo and limb observations of Triton's topography (Schenk et al., 2021) are much more limited than for Pluto, but indicate that topographic amplitudes of the areas observed (<25% of the surface) are only ~1 km or less, including the bright deposits of the southern hemisphere. Although topographic data were lacking over large areas, Schenk et al. (2021) concluded from the lack of discrete large bright or dark patches similar to Sputnik Planitia that Triton also lacked deep basins or high plateaus of similar scale. Unresolved basins and plateaus and local extremes of ~1 km scale are possible in these areas, however.

We tested the above scenario by performing simulations with a North-South asymmetry in bedrock topography, as shown by Figure 5, with a high-standing northern hemisphere and low-standing southern hemisphere of ± 1 and ± 4 km (the former case is well within the current limited constraints while the latter case is extremely unrealistic but provides insights into the magnitudes and extents of the processes involved). Figure 6 shows the annual mean evolution of $N_2$ ice thickness resulting from this model configuration and for different subsurface thermal inertias.

Our results show that the topography asymmetry does not have a strong effect on reducing the extent of the northern cap, even in the extreme case with a pole-to-pole dichotomy of 8 km. In this scenario, the low-elevated $N_2$ ice in the southern hemisphere is ~1 K warmer than the high-elevated $N_2$ ice in the northern hemisphere. As a result, the $N_2$ condensation rates should always be larger during southern winter than during northern winter. However, as $N_2$ ice extends to lower latitudes in the southern hemisphere (and because this is maintained by glacial flow in the model), more $N_2$ ice is available for sublimation during southern summer, which counteracts this difference. This allows for a permanent northern cap of $N_2$ ice to form in all cases and to be stable with an expansion of $N_2$ deposits to relatively low latitudes (40°N-50°N) while the permanent southern cap extends to 30°S. We also note that low TI allows for more seasonal deposits forming at the edge of the permanent deposits.

## 5.3 North-South asymmetry in nitrogen ice albedo

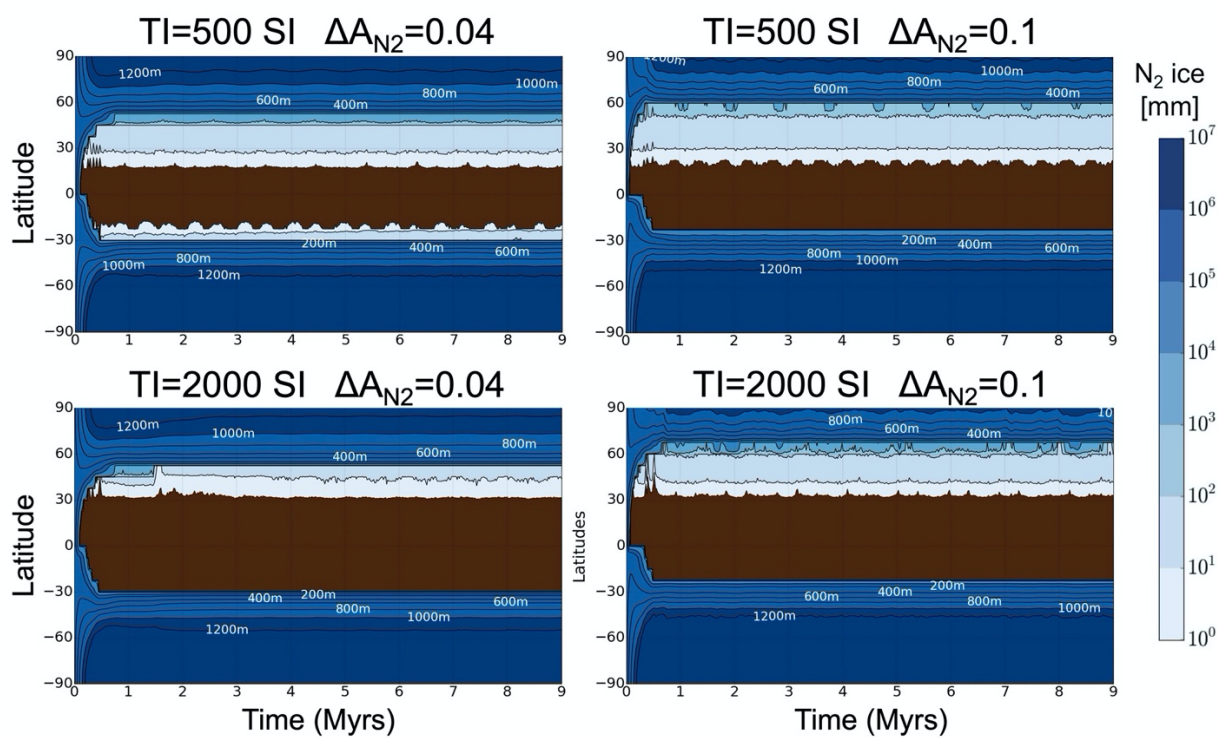

*Figure 7: Annual mean evolution of nitrogen ice thickness (in zonal mean) assuming a North-South asymmetry in $N_2$ ice albedo and a subsurface thermal inertia of 500-2000 SI. The $N_2$ ice albedo in the southern hemisphere is automatically changed by the model so that it quickly converges toward a value allowing a surface pressure of ~1.4 Pa during the season of the Voyager 2 flyby (see Section 3.4.1). The $N_2$ ice albedo in the northern hemisphere remains always lower than that in the southern hemisphere by 0.04 (left panels) and 0.1 (right panels). The $N_2$ ice albedo obtained at equilibrium is 0.765 and 0.775 (for TI=500 SI), and 0.795, and 0.775 (for TI=2000 SI), respectively.*

Volatile transport modeling performed by Moore and Spencer (1990) showed that a permanent North-South albedo dichotomy should result in a net, long-term transfer of $N_2$ from one cap to the other.

Here we tested this scenario with our volatile transport model by performing simulations with a North-South asymmetry in nitrogen ice albedo and we obtained similar results. Figure 7 shows the annual mean evolution of $N_2$ ice thickness predicted by our model when the $N_2$ ice albedo in the northern hemisphere is lower by 0.04 (left panels) and 0.1 (right panels) than in the southern hemisphere (where the albedo is calculated by the model so that P1989~1.4 Pa), for thermal inertia TI=500 SI (top) and 2000 SI (bottom).

A permanent northern cap forms in all cases, but extends to 45-55°N (with mm-to-m thick seasonal deposits extending to 30°N if TI is low) while the permanent southern cap extends to 30°S.

## 5.4 Discussions about the North-South asymmetry

We tested the scenarios of a North-South asymmetry in heat flux, topography, and $N_2$ ice albedo, and our results are consistent with previous work (Brown and Kirk, 1994, Moore and Spencer, 1990).

We note that:

(1) We used a large global $N_2$ reservoir of 300 m, which is required for the expansion of the southern cap to 30°S-0°, as observed by Voyager 2.

(2) As described in Brown and Kirk (1994), this $N_2$ inventory is significantly larger than needed to supply volatile transport. Viscous flow of $N_2$ ice from the poles (where $N_2$ condensation is more intense) toward low latitudes balances the net sublimation-condensation flow. On average over several seasonal cycles, net $N_2$ condensation occurs at the poles and net sublimation occurs in the warm tropical regions, where viscous flow from the pole to the equator ensures that $N_2$ ice remains available for sublimation. This favors the formation of a permanent northern cap and its expansion to relatively low latitudes.

(3) North-South asymmetries in internal heat flux and in surface ice albedo are efficient means to limit the extent of the permanent northern cap to latitudes northward of 45°N. On average over several seasonal cycles, the $N_2$ condensation rates at the south pole are larger than those at the north pole by a factor ~8 if the surface $N_2$ ice albedo is 0.6 in the northern hemisphere vs 0.7 in the southern hemisphere (Figure 7), or if the internal heat flux is 45 W $m^{-2}$ in the northern hemisphere vs 0 W $m^{-2}$ in the southern hemisphere (Figure 4). The asymmetry in topography does not produce a significant asymmetry in cap extents. Even in the case of a topography gradient of 8 km between both poles (Figure 6), the surface temperature gradient is less than 1 K (the ice is warmer in the modeled depression at the south pole) and the condensation rates are larger at the south pole by only a factor of ~1.4. In other words, our results remain quantitatively sensitive to the ice properties (albedo, thermal inertia, $N_2$ reservoir), and insensitive to topography differences, even those that are larger than expected based on Voyager 2 imagery (although this is true for large $N_2$ reservoirs only, see below).

(4) The northern and southern permanent caps simulated with our model quickly reach a steady state after ~1 Myrs and then remain relatively stable over time. If we assume an initially warm volatile-free northern hemisphere, it could take a few Myrs longer, but the end result would remain the same. Note that the timescale associated with viscous flow relaxation is ~1 Myrs for a relatively flat bedrock, a layer of 100 m of $N_2$ ice, and a characteristic length scale for the southern cap of 1000 km (calculated for Triton as detailed in Umurhan et al., 2017, see their Fig. 9b).

(5) Figure 8 shows a few simulation results performed with lower reservoirs (30 m and 100 m). Due to the lower reservoirs, the southern and northern caps are smaller in size and thickness, but the general tendencies observed for the simulations with 300 m remain valid. As in Spencer (1990), Stansberry et al. (1990), Hansen and Paige (1992) and Spencer and Moore (1992), simulations performed with very low global $N_2$ reservoirs (e.g. 1 m, not shown) show that most of Triton's $N_2$ inventory accumulates at the south pole (favored by a North-South asymmetry), and after several seasonal cycles, forms a small southern permanent deposit confined to the south pole and a few meters thick (referred as the problem of ever-shrinking permanent polar caps in Spencer and Moore, 1992). Due to the low amounts of $N_2$ ice involved, the ice does not flow, and the polar cap never extends to the mid-to-equatorial regions. Thin seasonal deposits can form at the north pole during winter. The topography, albedo and internal flux differences between both poles becomes a more significant driver for $N_2$ ice migration in these low reservoir cases.

(6) Brown and Kirk (1994)  and Moore and Spencer (1990) state that an albedo asymmetry between the northern and southern hemispheres is unlikely to be maintained over a long-term period due to (1) seasonal deposition of meter-thick frost layers each winter in the low-albedo hemisphere, possibly brightening the surface except if the layers are transparent enough to have no significant effect on the albedo of the substrate (which remains unlikely), and (2) an increase in albedo as the ice is fractured by passage through the $\alpha$-$\beta$ phase transition (Scott, 1976, Duxbury and Brown, 1993). On the other hand, pole-to-pole difference in ice contamination by dark material, sensible heat flux from the atmosphere or positive feedbacks on the surface (shown to have a non-significant effect on Pluto's ice, Bertrand et al., 2020a) may help maintaining an albedo asymmetry, but these processes would need to be investigated in more detail on Triton.

(7) Amongst the simulations explored here, we do not get any case with $N_2$ ice at southern low- and mid-latitude but no $N_2$ ice at the south pole, as suggested by Earth-based spectroscopic observations (see Section 2.3).

(8) Many other processes could explain the asymmetry between the caps. The reader is referred to Section 2.6 and previous work by Brown and Kirk (1994) and Moore and Spencer (1990) for more detail about this topic. Constraining the composition of the ices on both hemispheres will be key to distinguishing between these alternatives.

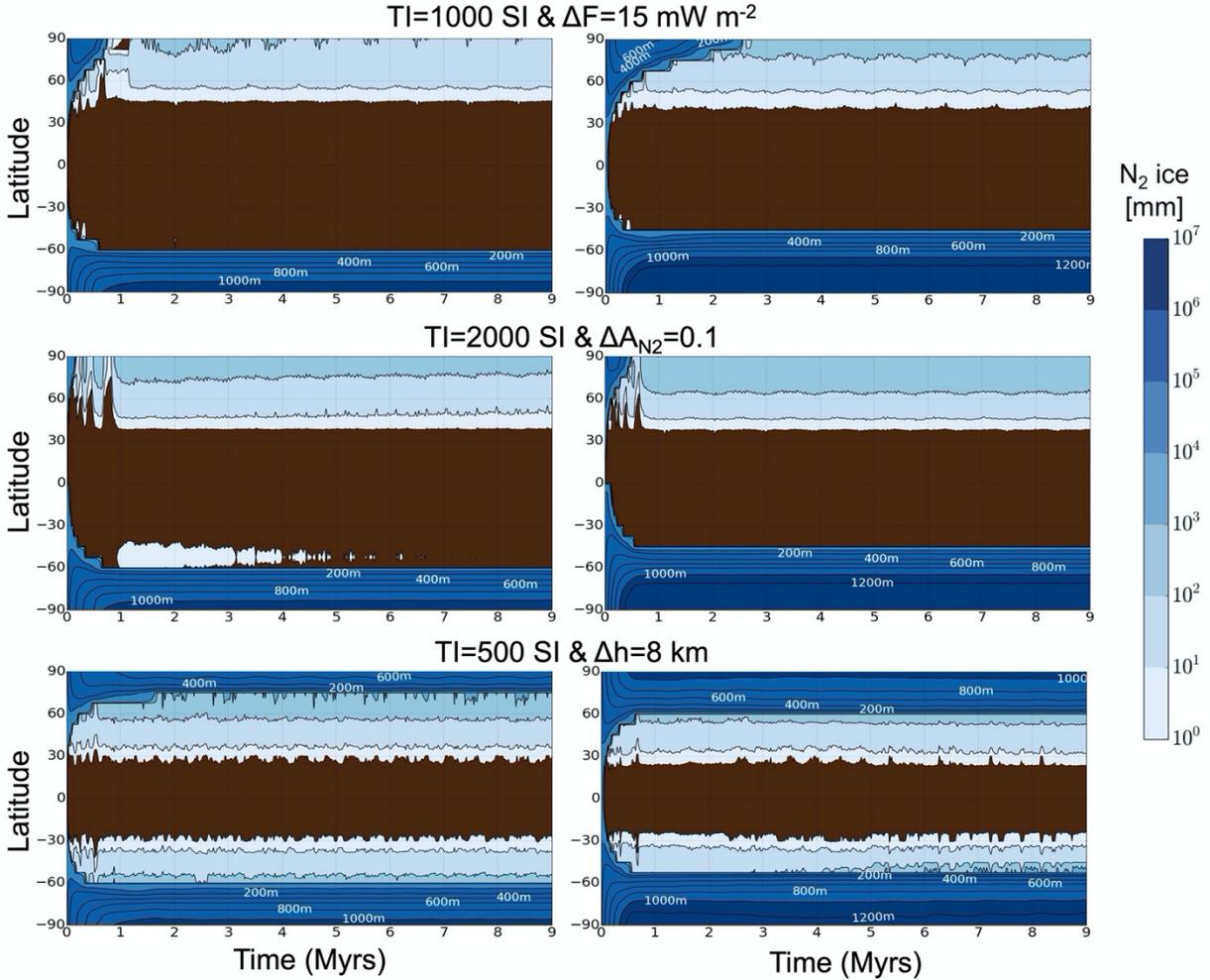

*Figure 8: As for previous figures but for a global $N_2$ reservoir R=30 m (left) and R=100 m (right). The $N_2$ ice albedo obtained at equilibrium is 0.775, 0.750 and 0.720 (for R=30 m), and 0.790, 0.765 and 0.745 (for R=100 m), respectively.*

# 6. Surface pressure cycle assuming fixed $N_2$ ice distribution

As shown in Section 5, the permanent (i.e., non seasonal) northern and southern caps remain relatively stable over time (if we assume the seasonal changes in insolation as described in Section 2.1) but their latitudinal extents depend on the model parameters such as albedo, thermal inertia, and $N_2$ ice reservoir.

In this section, we explore how fixed $N_2$ ice distributions on Triton and surface properties affect the current surface pressure cycle. For sake of simplicity, and in order to understand the impact of each parameter one after another, we neglect the impact of seasonal deposits and therefore we only use fixed volatile ice distribution (more realistic simulations with full, i.e., self-consistent, volatile transport including seasonal deposits are presented in Section 7). We use the stellar occultation datasets to constrain the properties of the northern and southern caps.

6.1 Initial state of the simulations with fixed $N_2$ ice distribution

Table 1 summarizes the settings of the volatile transport simulations presented in this section. We place fixed and infinite $N_2$ reservoirs (no glacial flow, flat topography) confined to 90°S-30°S, 90°S-0°, or 80°S-0° in the southern hemisphere, and to 90°N-75°N, 90°N-60°N or 90°N-45°N in the northern hemisphere (including the case without a northern cap). We limit the volatile transport to these reservoirs (no seasonal frost can form). Consequently, the simulation results are independent of the bedrock surface properties (albedo and emissivity). The simulations are performed over several seasonal cycles (covering ~10 000 Earth years) in order to reach a steady state. The $N_2$ ice albedo is automatically adjusted by the model so that the surface pressure reaches ~1.4 Pa in 1989, as observed (see Section 3.4.1).

6.2 Simulation results with fixed $N_2$ ice distribution

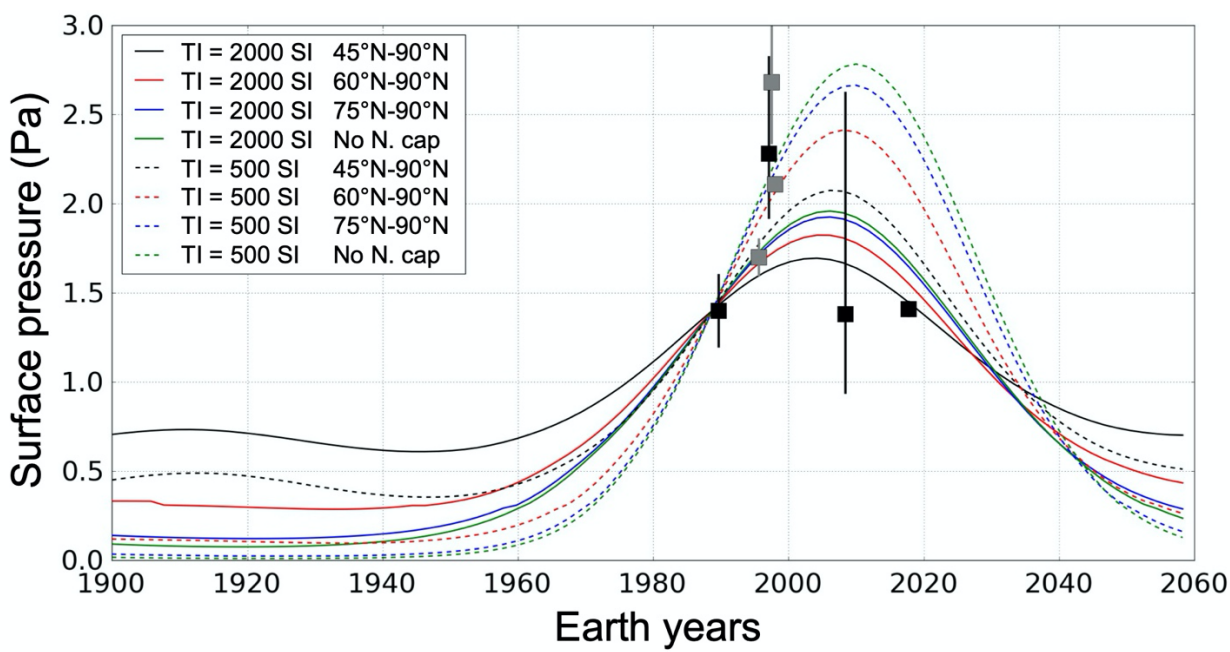

*Figure 9: Surface pressure on Triton for the period 1900-2060 as predicted by the model when we assume a fixed $N_2$ ice distribution with a southern cap placed between 90°S-30°S and a northern cap placed between 90°N-45°N (black), 90°N-60°N (red), 90°N-75°N (blue) or no northern cap (green). The seasonal thermal inertia of $N_2$ ice is set to 2000 SI (solid lines) or 500 SI (dotted lines). Black and grey data points and 3-σ error bars represent the pressure observations as presented in Table 2 (black are for the data points that we consider are the strongest observational constraints).*

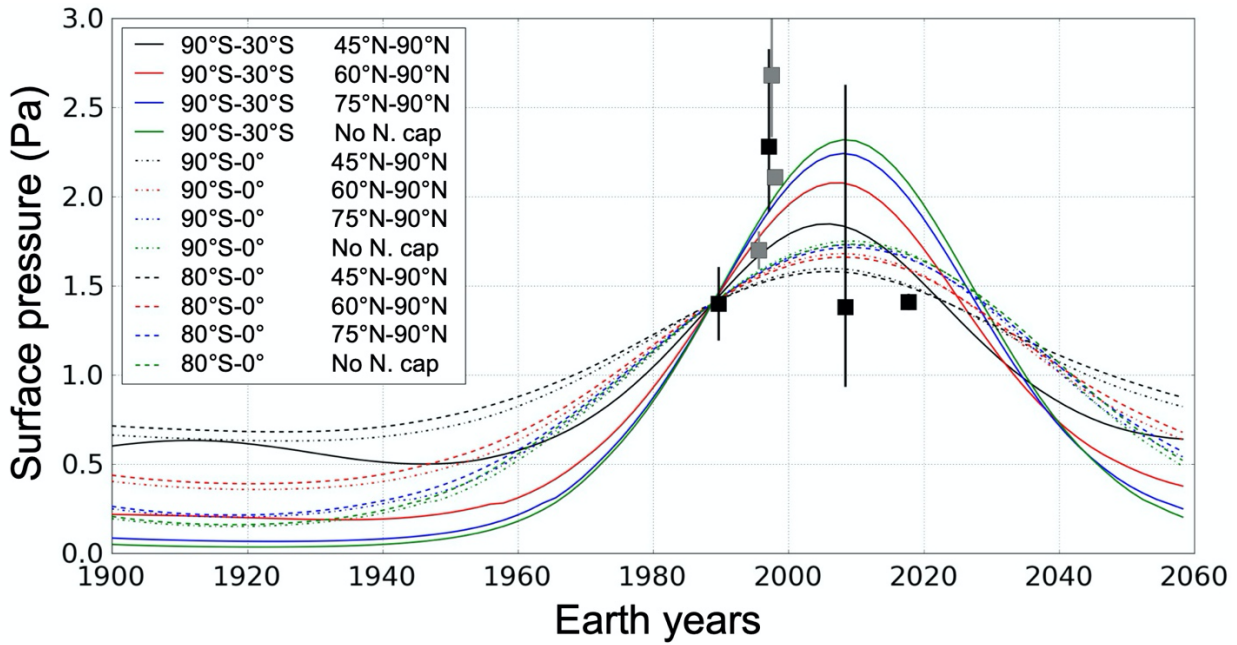

*Figure 10: As Figure 9 but for different fixed N₂ ice distributions in the southern and northern hemisphere. The seasonal thermal inertia is set to 1000 SI.*

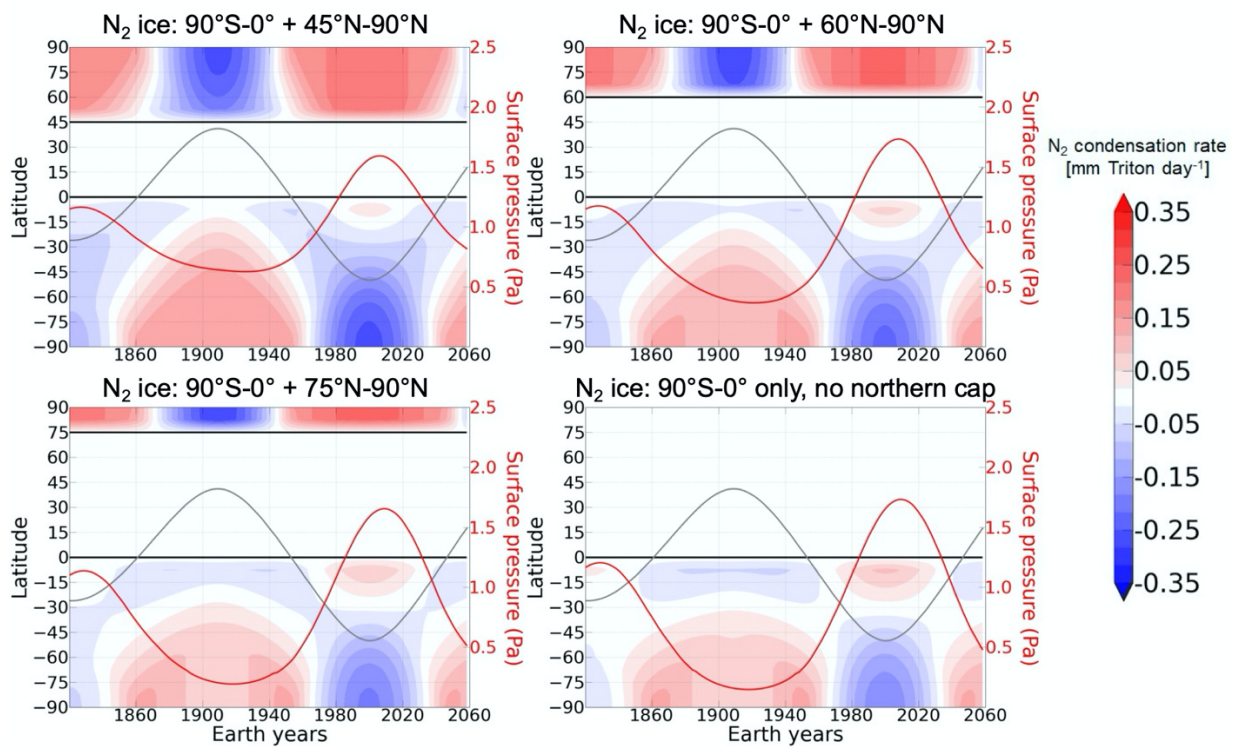

*Figure 11: Zonal and diurnal mean N₂ condensation-sublimation rate (mm per Triton day) during the period 1820-2060 as obtained in the model for the cases of fixed N₂ ice distributions: 90°S-0° + 45°N-90°N (top left), 90°S-0° + 60°N-90°N (top right), 90°S-0° + 70°N-90°N (bottom left), 90°S-0° and no northern cap (bottom right). The seasonal N₂ ice thermal inertia is set to 1000 SI. The*

*thin black contour indicates the extent of the caps. The red line indicates the surface pressure (right y-axis) and the grey line indicates the subsolar latitude.*

Figures fixedscap and fixedti show the surface pressure for the period 1900-2060, obtained with the model for different $N_2$ ice distributions and $N_2$ ice thermal inertia, while Figure 11 shows the latitudinal condensation-sublimation flux of $N_2$ during the period 1820-2060 for the four different extents of the northern cap explored.

The surface pressure is at its minimum during the 1880-1940 period (northern summer), when the subsolar point is above 15°N, because the northern cap is always smaller than the southern cap and therefore the condensation-dominated areas (most of the southern $N_2$ ice deposits) overcome the sublimation-dominated areas (the northern deposits). During the opposite season (southern summer), the pressure is at its maximum as the sublimation-dominated areas (most of the southern deposits) overcome the condensation-dominated areas (northern deposits), as shown by Figure 11 (all panels). In these simulations, the surface pressure peak occurs slightly after the southern summer solstice (2000) between year 2000-2010. The surface pressure evolutions obtained here are similar to those obtained by Spencer and Moore (1992) when they artificially maintained a permanent large southern cap of bright $N_2$ (see their Fig. 7).

As shown by Figure 9, the larger the northern cap, the more it can serve as a condensation area and buffer $N_2$ sublimation in the southern hemisphere, which results in a surface pressure peak that is lower and occurs sooner. In addition, the higher the thermal inertia of $N_2$ ice, the lower the amplitude of surface pressure over a seasonal cycle (higher minimum and lower maximum), and the earlier the surface pressure peaks.

We also tested different $N_2$ ice distributions in the southern hemisphere. Figure 10 shows that the amplitude of the surface pressure peak is strongly attenuated if $N_2$ ice remains between 30°S-0° (non-solid lines). This is because these latitudes are dominated by condensation rather than sublimation between the years 1980-2020, thus dampening the pressure peak (Figure 11, all panels). We also tested a scenario without $N_2$ ice between 90°S-80°S, leading to a pressure peak slightly lower as less $N_2$ is available for sublimation in southern summer (Figure 10, dotted lines).

These results suggest that a mid-to-high thermal inertia of $N_2$, coupled with a northern cap extending down to 45°N-60°N, is needed so that the surface pressure is ~1.4 Pa (back to Voyager 2 levels) in 2017, as observed from stellar occultations. If we assume no northern cap, then the modeled surface pressure remains higher than 1.6 Pa in 2017, which becomes inconsistent with the observations. A strong increase in surface pressure before 2000 cannot be obtained if $N_2$ ice is present between 30°S-0°. We also note that the surface pressure remains greater than 0.5 Pa even during the opposite season (southern winter) when a permanent northern cap extending down to 45°N is assumed. The presence of permanent southern and northern caps prevents Triton's atmosphere from collapsing.

Finally, as discussed in Section 2.3, IRTF/SpeX observations of Triton's near-IR spectrum in 2002 suggest that $N_2$ is undetectable or absent at high southern latitudes. Here, Figure 11 shows that about 30-50 cm of $N_2$ ice is removed at the south pole by sublimation during the period 1970-2000. It is not clear whether the removal of 30-50 cm of $N_2$ ice could have altered its detectability, even if $N_2$ is not completely removed. We discuss more on this point in Section 8.3.

# 7. VTM Simulations of the $N_2$ cycle

In this Section, we simulate the full volatile transport across Triton over the last 4 Myrs and explore a large range of parameters (bedrock albedo, thermal inertia, $N_2$ ice reservoir, internal flux and North-South asymmetry). The results (e.g., cap extents in 1989 and surface pressure in 2017) are compared to the available observations in order to better constrain these model parameters.

## 7.1 Initial state of the simulations and main model parameters

Section 3.3 and Table 1 summarize the initial state of the simulations. Note that all the results of this section are the outcome of 4-Myrs simulations in which the end state does not depend on the initial ice distribution and surface and subsurface temperatures, and in which the $N_2$ ice albedo is calculated by the model and constrained by the surface pressure in 1989 (~1.4 Pa). The $N_2$ ice albedo obtained in the model in 1989 usually ranges within 0.7-0.8, depending on the cap extents and on the internal heat flux. We first perform simulations with a spatially uniform $N_2$ ice albedo (Section 7.2) and then with a lower albedo in the northern hemisphere (Section 7.3, where the albedo of the northern cap is always lower than that of the southern cap by 0.1).

The bedrock and $N_2$ ice emissivities are fixed to $\varepsilon$=0.8, and the $N_2$ ice thermal inertia to $TI_{N2}$=1000 SI. We explore different values for the bedrock's thermal inertia $TI_{bed}$ (Low: 200 SI; Moderate: 500 SI, 1000 SI; High: 2000 SI) and albedo $A_{bed}$ (from 0.1 to 0.9), for the global $N_2$ ice reservoir $R_{N2}$ (from 1 m to 650 m), and with and without a globally uniform internal heat flux of 30 mW m$^{-2}$.

## 7.2 Full volatile transport simulations with spatially uniform ice properties (no North-South asymmetries)

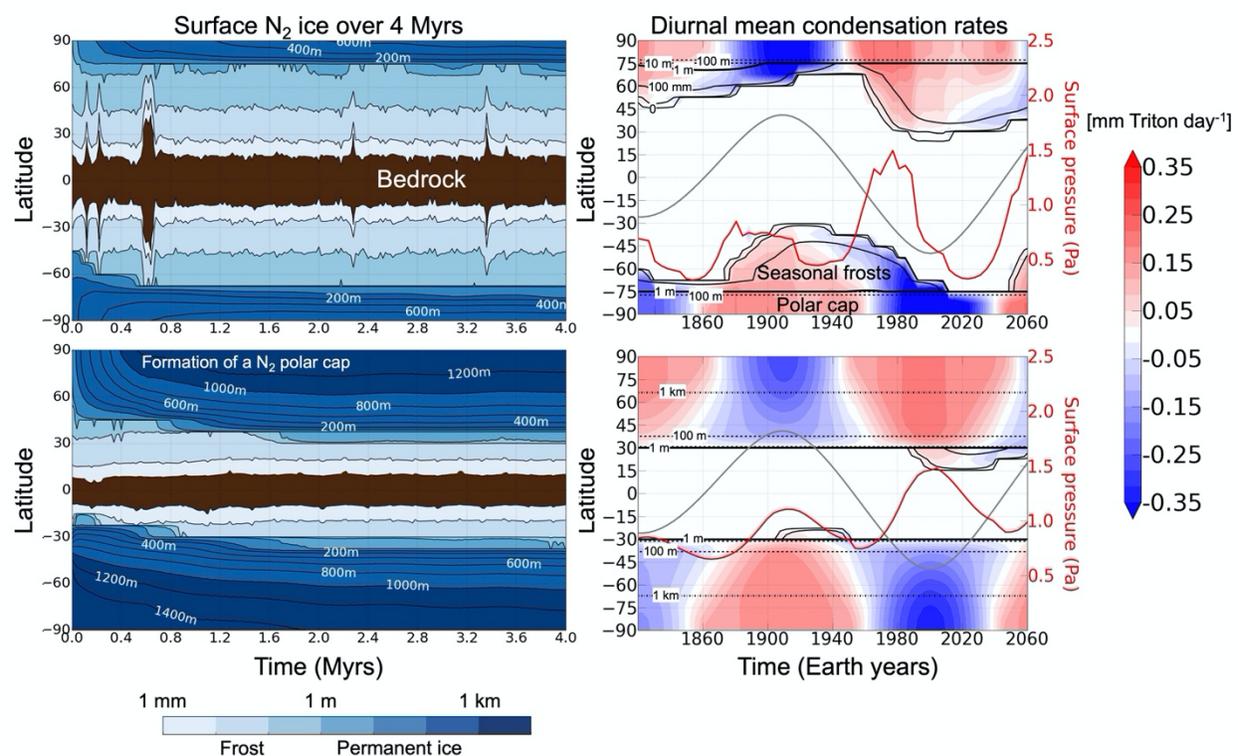

*Figure 12: Results from a VTM simulation performed with spatially uniform $N_2$ ice properties. Top: Simulation with $TI_{bed}$=1000 SI, $A_{bed}$=0.8, $R_{N2}$=12 m. Bottom: $TI_{bed}$=500 SI, $A_{bed}$=0.7, $R_{N2}$=350 m.*

*Left: annual mean evolution of $N_2$ ice thickness (in zonal mean) over 4 Myrs. Right: Zonal and diurnal mean $N_2$ condensation-sublimation rate (mm per Triton day) during the period 1820-2060. The black contours indicate the extent of the caps (0-line, 10 mm, 100mm, solid), and where the ice is 1 m (thick solid line) and 1 km thick (dash-dotted line). The red line indicates the surface pressure (right y-axis) and the grey line indicates the subsolar latitude. Deposits thinner than 1 m tend to be seasonal.*

Figure 12 shows examples of model results when a spatially uniform $N_2$ ice albedo is assumed, primarily in two formats: the long-term evolution of $N_2$ ice thickness over 4 Myrs (left panel) and the recent seasonal evolution of the diurnal mean $N_2$ condensation rates obtained during the 1820-2060 period (as the outcome of the 4-Myrs simulation).

First, in the long-term evolution, the northern hemisphere, initially warm and volatile-free, quickly cools down and allows the formation of a permanent northern cap, which stabilizes along with a symmetric southern cap after ~2 Myrs. The thickness and the extent of the caps depend on the global reservoir of $N_2$ ice. In the tropical regions, mm-to-m thick seasonal deposits can form and extend to the equatorial regions.

Second, the seasonal evolution (right panel) shows that the southern cap sublimates from ~1960 to ~2040, with $N_2$ condensing in the northern hemisphere. In the case of Figure 12 (top), the low amount of $N_2$ ice in the system reduces $N_2$ deposits to seasonal layers (mm-to-m thick) at latitudes equatorward of 70°N or of 75°S, and the southern cap retreats from 40°S to 75°S while the northern cap expands from 70°N to 20°N between 1960 and 2020. This leads to a strong decrease of surface pressure from 1989 as $N_2$ condensation-dominated areas increase in the north and as sublimation is limited in the south. In the case of Figure panel_uniform1 (bottom), there is enough $N_2$ ice on the surface so that the permanent northern and southern caps expand to 30°N and 30°S, respectively. The peak of surface pressure is reached around year 2000, when the sources of $N_2$ in the south balance the sinks in the north (~2-3 mm Triton day$^{-1}$).

In general, all simulations that assume spatially uniform subsurface and ice properties produce symmetric permanent caps. As shown by Figure 12, this results in either a northern cap that extends to latitudes southward of 45°N in 1989 (inconsistent with Voyager 2 observations) and/or a strong decrease of surface pressure between 1989 and 2017 (with values largely below the reported value of 1.41 Pa in 2017, inconsistent with stellar occultation observations).

In the next section, we impose a North-South asymmetry in $N_2$ ice albedo in order to force a smaller northern cap.

## 7.3 Full volatile transport simulations with a North-South asymmetry in $N_2$ ice albedo

In this section, we perform volatile transport simulations following the *Koyaanismuuyaw* model (Moore and Spencer, 1990): we assume that the $N_2$ ice albedo in the northern hemisphere is lower by 0.1 than that in the southern hemisphere, which leads to a larger permanent southern cap.

### 7.3.1. Comparison to the case without North-South asymmetry

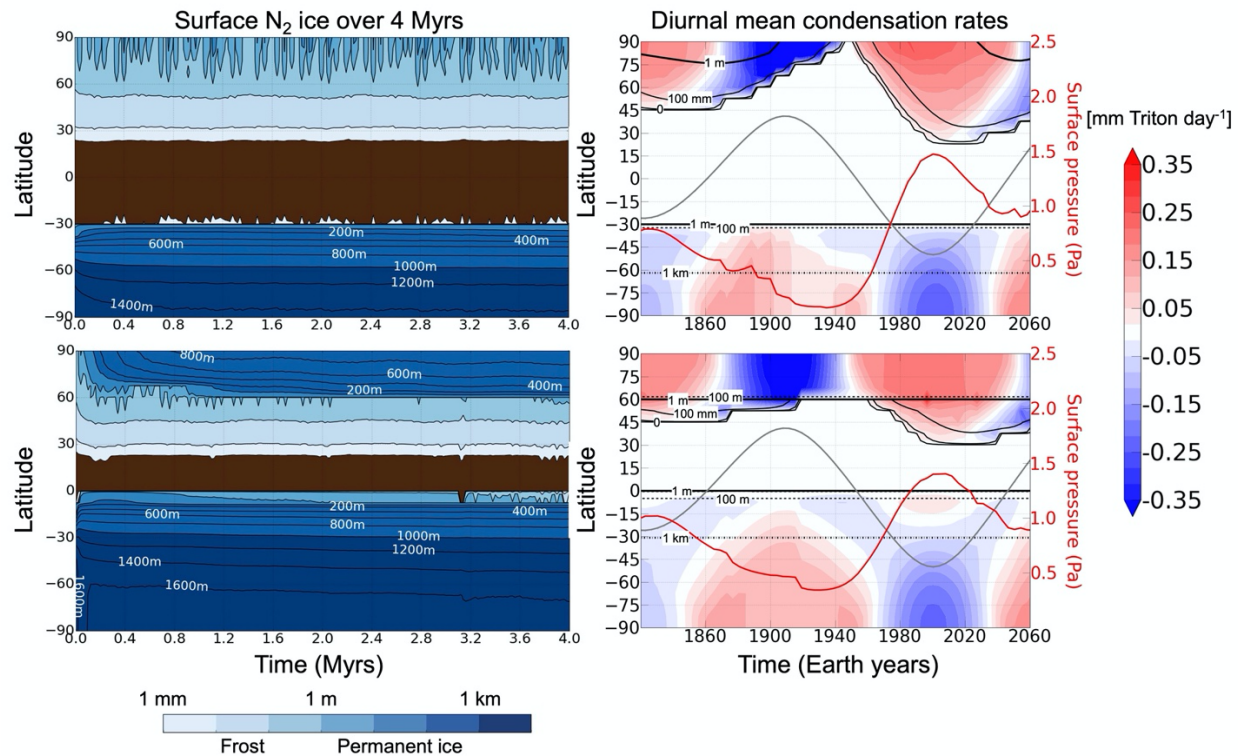

*Figure 13: As Figure 12 but from a VTM simulation performed with a North-South asymmetry in N$_2$ ice albedo ($\Delta A_{N2}$=0.1). Top: Simulation with TI$_{bed}$=1000 SI, A$_{bed}$=0.7, R$_{N2}$=200 m. Bottom: TI$_{bed}$=2000 SI, A$_{bed}$=0.7, R$_{N2}$=550 m.*

Figure 13 (top) shows an example of a simulation performed with the North-South asymmetry in N$_2$ ice albedo and a global N$_2$ reservoir of 200 m. A permanent, 1-km thick southern cap forms and stabilizes with an extent to 30°S, due to the relatively high ice albedo and subsurface thermal inertia. In the northern hemisphere, only seasonal mm-to-m thick deposits form (Figure 13, top left). During the northern summer (1890-1940), these northern deposits entirely sublimate and disappear (Figure 13, top right). In this simulation, N$_2$ re-condenses in the northern hemisphere in the current southern summer, with a maximum extent of the frost to ~30°N, reached in 2000. The surface pressure peaks in ~2000, and is back to Voyager levels in 2017, consistent with the observations.

Figure 13 (bottom) describes a similar simulation but with a larger global N$_2$ ice reservoir of 550 m. In this simulation, the permanent southern cap extends to the equator and a smaller northern cap forms and extends to 60°N (Figure 13, bottom left). The seasonal frosts in the Northern hemisphere extend to ~30°N in 2000, and the pressure cycle is still consistent with the observations with a maximum reached in 2005 and a value close to that of 1989 in 2017. Note that N$_2$ re-condenses in the equatorial regions at the edge of the southern cap between 1980-2020 (Figure 13, bottom right), as also shown in Section 6.2. This may be related to the formation of the bright blue fringe observed in these regions by Voyager 2 (see discussion on this topic in Section 8.4).

7.3.2. Sensitivity of the results to surface properties, internal heat flux and ice reservoir

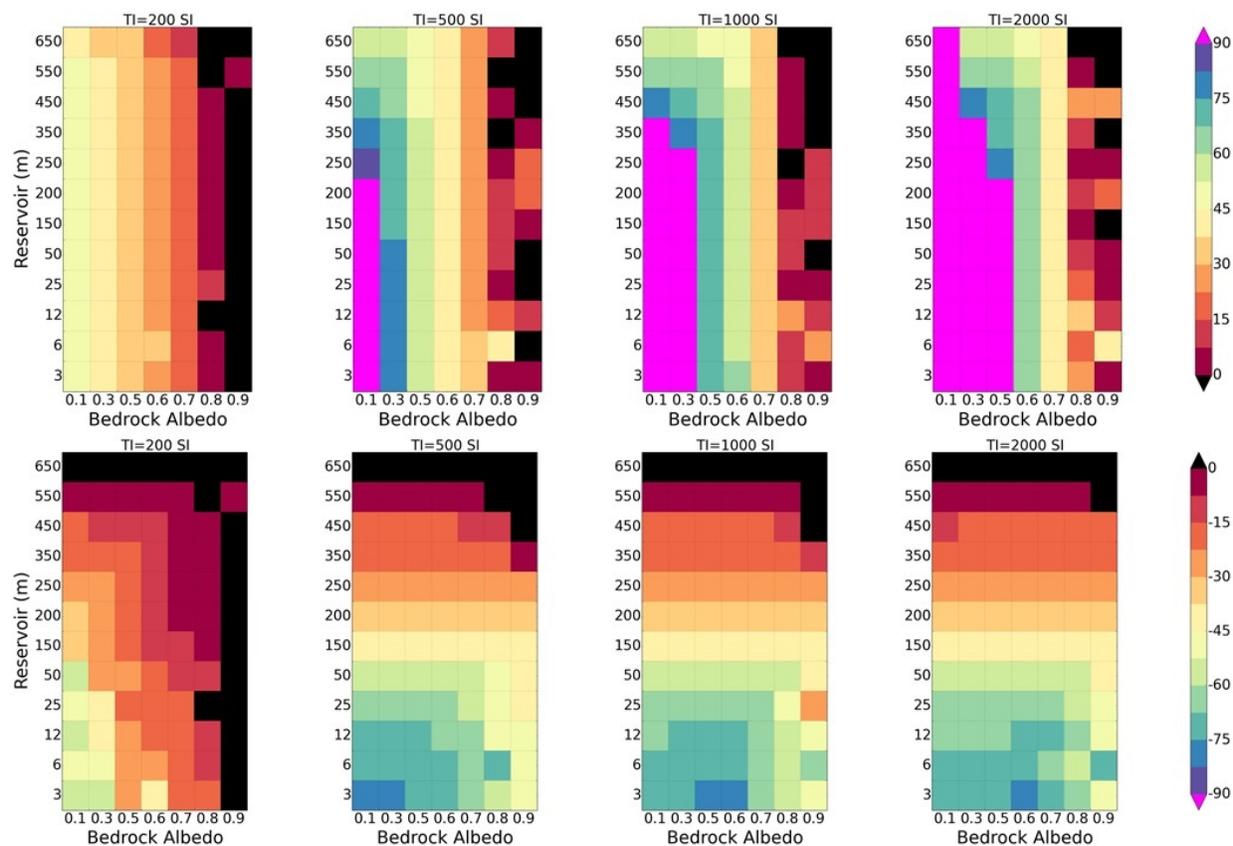

*Figure 14: Latitude to which the northern (top) and southern (bottom) caps expand (from the pole) in 1989 as modeled by simulations with varying subsurface thermal inertia (columns, TI = 200, 500, 1000, and 2000 SI from left to right), $N_2$ global reservoir (y-axis) and bedrock albedo (x-axis). The simulations that best match Voyager 2 observations in 1989 are those that show a northern cap confined poleward of 45°N and a southern cap that extends to 30°S at least.*

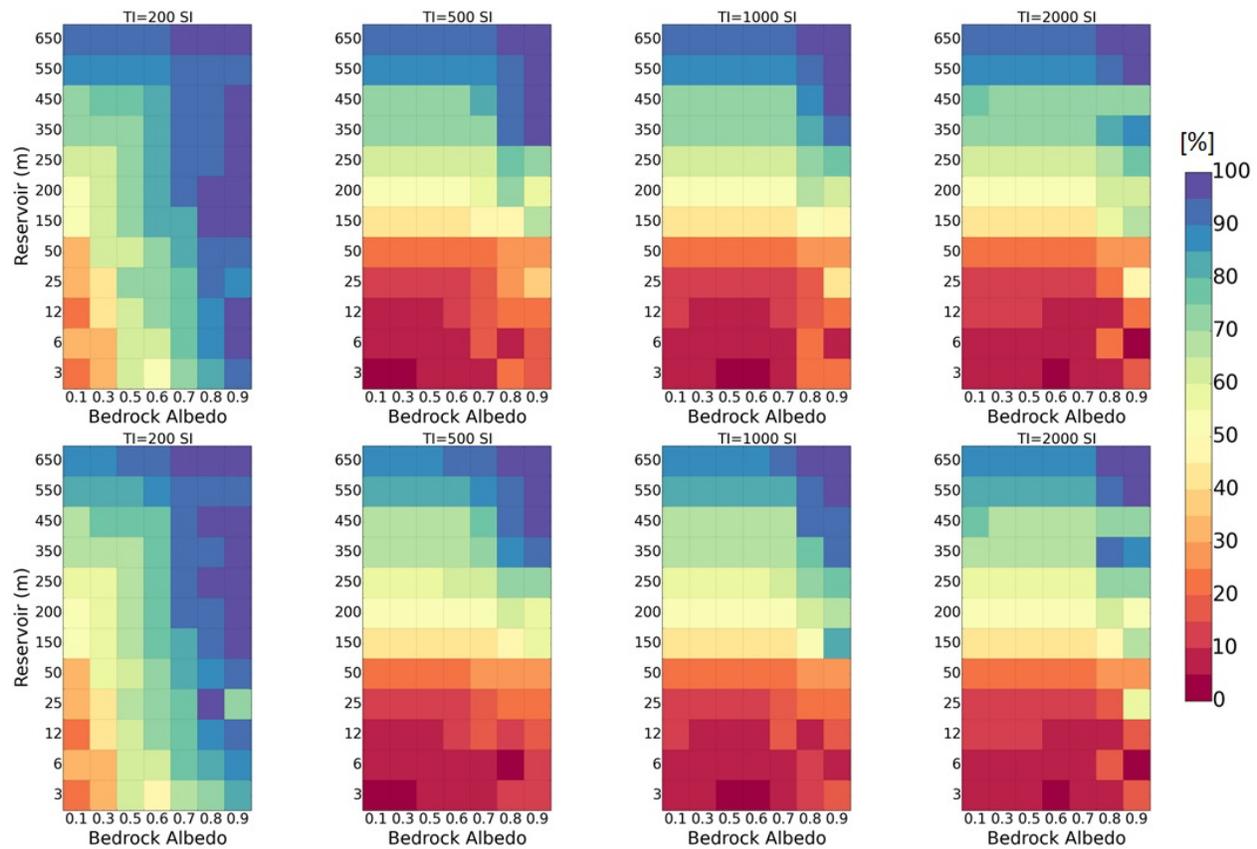

*Figure 15: Modeled volatile fractional area (%) in 1995 (top) and 2010 (bottom), projected on the visible disk at these dates, with varying subsurface thermal inertia (columns, TI = 200, 500, 1000, and 2000 SI from left to right), $N_2$ global reservoir (y-axis) and bedrock albedo (x-axis). The simulations that best match Earth-based spectroscopic observations are those that show a volatile fractional area of 50-60% in 1995 (Quirico et al., 1995) and 60-70% in 2010 (Merlin et al., 2018). Note that the extent of the southern cap and the latitude of the subsolar point do not change much between 1995 and 2010, hence the relatively unchanged volatile fractional area during this period of time.*

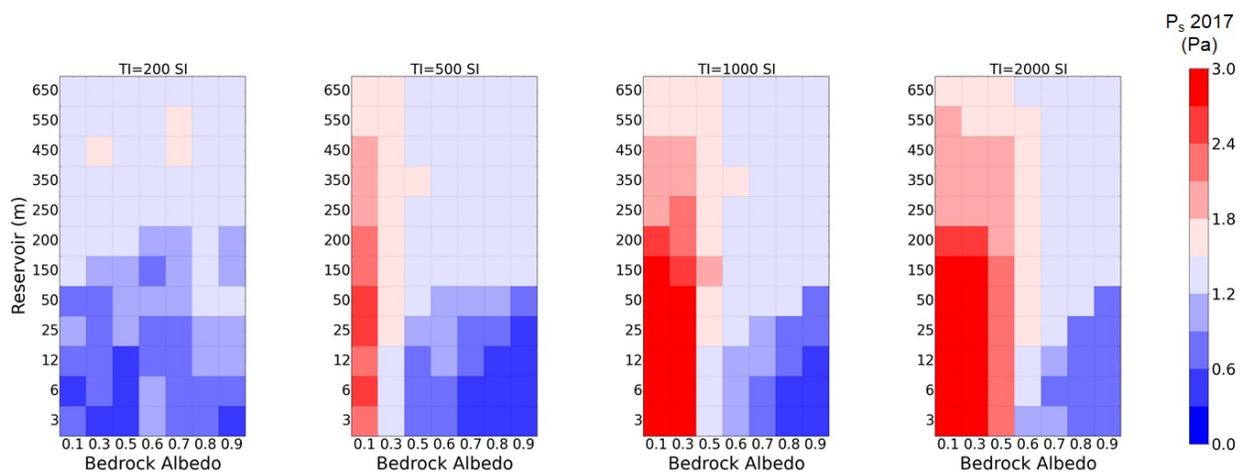

*Figure 16: Surface pressure on Triton in 2017 as modeled by simulations with varying subsurface*

*thermal inertia (columns, TI = 200, 500, 1000, and 2000 SI from left to right), $N_2$ global reservoir (y-axis) and bedrock albedo (x-axis). The simulations that best match the 2017 stellar occultation data are those that show a surface pressure of ~1.41 Pa (Marques Oliveira et al., 2021).*

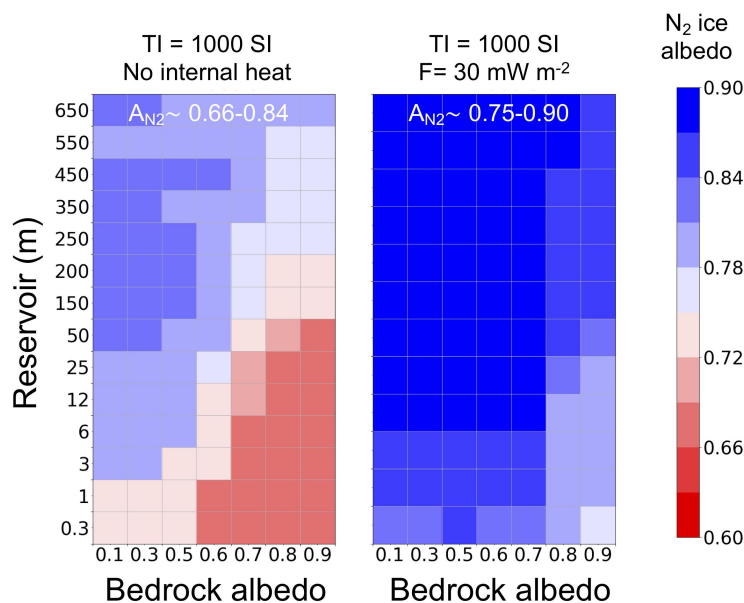

*Figure 17: $N_2$ ice albedo in the southern hemisphere in 1989 as modeled by simulations without (left) and with internal heat flux (right), and for different $N_2$ global reservoir (y-axis) and bedrock albedo (x-axis). Albedos in the model are Bond albedos. The subsurface thermal inertia is set to 1000 SI.*

We performed similar simulations with varying combinations of bedrock subsurface thermal inertias (TI = 200, 500, 1000, and 2000 SI), $N_2$ global reservoirs (from 0.3 m to 650 m) and bedrock albedos (from 0.1 to 0.9). We also explored how results are impacted by a uniform internal heat flux of 30 mW m$^{-2}$. The model sensitivity to these parameters is described by (1) Figure 14, which shows the extent of the caps in 1989, (2) Figure 15, which shows the modeled volatile fractional area as seen from Earth in 1995 and 2010, (3) Figure 16, which shows the surface pressure in 2017, and (4) Figure 17, which shows the value of the $N_2$ ice albedo in 1989 calculated by the model so that the surface pressure reached the observed value (~1.4 Pa). We also tested how the volatile cycle responds to a lower $N_2$ ice emissivity ($\varepsilon_{N2}$=0.3, 0.5, and taking into account the α-β phase transition) and higher values of thermal inertias (TI = 4000, 8000 SI).

**Sensitivity to the $N_2$ reservoir**

$N_2$ ice accumulation is always favored in the southern hemisphere, due to the north-south $N_2$ ice albedo asymmetry. As a result, the extent of the southern cap (and thus the volatile fractional area) strongly depends on the global $N_2$ reservoir and the viscous flow of $N_2$ (Figure 14, bottom). This is illustrated by Figure 18. The more $N_2$ in the system, the thicker the cap, the more it flows toward the equator, and the more the ice subsists through an extreme summer (larger perennial cap). On the other hand, a low reservoir would lead to a small and thin southern cap, with seasonal deposits that disappear during the recent extreme southern summer. As a result, simulations with low reservoirs also lead to a low volatile fraction area in 1995-2010 (except for the low thermal inertia case) and to surface pressures much lower in 2017 than in 1989 as condensation in the north then dominates sublimation in the south. Reservoirs lower than 150 m coupled with moderate-to-high bedrock TI lead to a southern cap extending to latitudes poleward of 45°S in 1989, a volatile fraction area less than 40% in 1995-2010 (Figure 15) and surface pressure lower than 1.2 Pa in 2017 (Figure 16s), which is inconsistent with Earth-based spectroscopic observations and the 2017 stellar occultation. Very large reservoirs (> 350 m) tend to lead to a higher volatile fraction area than observed (>70%). They also lead to a higher seasonal minimum and a lower seasonal maximum in surface pressure, due to the fact that the caps extend to lower latitudes and are more perennial (limited amounts of seasonal deposits), which better balances the sources and sinks of $N_2$ (see Figure 9). A maximal surface pressure of ~2 Pa is reached during southern extreme summer for intermediate $N_2$ reservoirs of 12 m - 450 m coupled with intermediate thermal inertia (not shown).

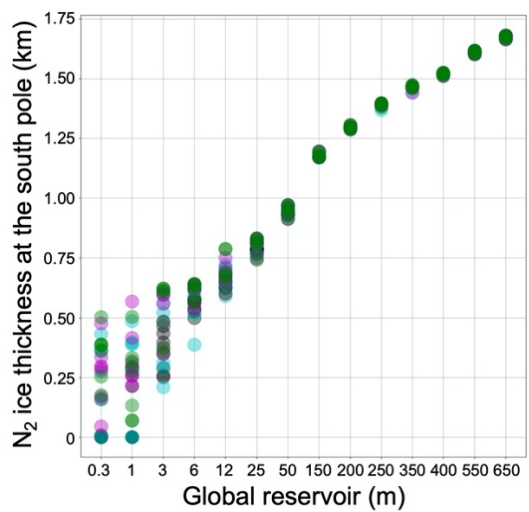

*Figure 18: Modeled maximum $N_2$ ice thickness at the south pole in 1989 vs initial global $N_2$ ice reservoirs for all simulations performed with north-south $N_2$ ice albedo asymmetry, no internal heat flux. Colors indicate the thermal inertia (black: 200 SI; cyan: 500 SI; magenta: 1000 SI; green: 2000 SI). For large reservoirs greater than 25 m in global average, the maximum thickness of the ice at the south pole mostly depend on the reservoir (and not on the thermal inertia or albedo of the bedrock).*

**Sensitivity to the bedrock albedo**

The extent of the northern cap does not strongly depend on the global $N_2$ reservoir but rather on the bedrock albedo (Figure 14). This is because $N_2$ ice is more stable within the southern cap and only allows a cap to form in the northern hemisphere if the surface is cold enough there (which, in our model, is driven by the bedrock albedo). As described in Section 7.1, our simulations are initialized with a warm northern hemisphere depleted of volatile ice. After several Myrs required to reach a steady state, most of these simulations present a permanent and/or seasonal northern cap in 1989, except when we assume a bedrock albedo much lower than 0.6 and a mid-to-high subsurface thermal inertia (Figure 14, top). Such a low albedo value for Triton's bedrock is unrealistic, as the surface was observed to be relatively bright in 1989 (with a bolometric Bond albedo at least greater than 0.5, McEwen, 1990). Furthermore, Figure 16 shows that a low bedrock albedo coupled with a mid-to-high thermal inertia tends to lead to a high surface pressure (as this would limit $N_2$ condensation in the northern hemisphere), greater than 1.8 Pa in 2017 (Figure 16), which is not consistent with the 2017 stellar occultation. On the other hand, a very high bedrock albedo would favor more $N_2$ condensation in the north during southern summer, leading to a northern cap extending to the equator in 1989 and a low surface pressure in 2017, which is also inconsistent with observations (Figure 14, top, Figure 16). For instance, if the $N_2$ reservoir is lower than 150 m, the southern cap considerably shrinks during this season, leading to surface pressures lower than 1.2 Pa in 2017, inconsistent with the 2017 stellar occultation (Figure 16).

### Sensitivity to the bedrock thermal inertia

The lower the bedrock thermal inertia, the higher the amplitude of bedrock surface and subsurface temperatures across the seasons (higher temperatures during summer, lower temperatures during winter), and, on average over seasonal cycles, the lower the temperatures, in particular at the poles. As a result, a lower bedrock thermal inertia leads to the formation of more mm-to-m seasonal frosts that extend to lower latitudes. In our simulations performed with the low TI = 200 SI and relatively high bedrock albedo (>0.5), the northern cap extends to 15°N in 1989 (Figure 14, top), which is inconsistent with Voyager 2 observations. However, the volatile fractional area is consistent with Earth-based spectroscopic observations for low thermal inertia coupled with low to moderate reservoirs and a surface albedo of ~0.6 (Figure 15). The increased presence of seasonal $N_2$ frosts forming in the northern hemisphere during winter in the case of low thermal inertia prevents an increase of surface pressure after 1989. As a result, low thermal inertia tends to lead to slightly low surface pressure (< 1.2 Pa) in 2017, especially when the $N_2$ reservoir is low (frosts would disappear in the southern hemisphere shortly after 1989). Moderate-to-large thermal inertia (500-2000 SI) limits the decrease of surface and subsurface temperature in the northern hemisphere during winter and thus allows for a limited extent of the northern cap in 1989. Surface pressure tends to increase from 1989 to ~2005-2010 and then decrease with a 2017 value close to that of 1989. Finally, we show on Figure 19.D two simulation cases performed with extremely (and possibly unrealistically; Ferrari and Lucas, 2016) high thermal inertias of 4000 SI and 8000 SI. Changes to the $N_2$ cycle are not significant in the case with TI=4000 SI. With TI=8000 SI, thicker deposits accumulate in the northern hemisphere and the seasonal changes of surface pressure are reduced, with a higher pressure minimum at 0.5 Pa.

### Sensitivity to the internal heat flux

A higher internal heat flux (uniformly applied across the globe) increases the surface and subsurface temperature for a given insolation. In order to keep a surface pressure of ~1.4 Pa in 1989, the $N_2$ ice must thus remain at the same vapor-pressure equilibrium temperature and must therefore be brighter. In general, an internal heat flux of 30 mW m$^{-2}$ leads to an increase in $N_2$ ice albedo of 0.1 compared to the case without internal heat flux (Figure 17) and as a result to slightly lower condensation and sublimation rates. The evolution of the surface pressure and the extent of the southern and northern caps remain relatively similar to the cases without internal heat flux.

### Sensitivity to the $N_2$ ice emissivity

As detailed in Section 3.4.1, the $N_2$ ice emissivity on Triton may be lower than what is assumed in the model ($\varepsilon_{N2}$=0.8). Figure 19.A-B shows how the volatile cycle is impacted by a lower $N_2$ ice emissivity by comparing a reference simulation using a fixed emissivity $\varepsilon_{N2}$=0.8 with the same simulation but using $\varepsilon_{N2}$=0.3 and $\varepsilon_{N2}$=0.5. In addition, Figure 19.C shows the case of a temperature dependent $N_2$ ice emissivity, with an $\alpha$-$\beta$ phase transition for $N_2$ ice at $T_{\alpha-\beta}$=35.6 K, as implemented in the model and tested for Pluto in Bertrand et al., 2019 (see their equation 1). In all these "low emissivity cases" (and as for all the other simulations of this section), the model had to adjust (increase by steps of 0.005, typically from 0.75 to 0.8) the $N_2$ ice albedo during the spin-up time so that the surface pressure remained close to 1.4 Pa in 1989.

Overall, the fixed lower emissivity induce higher pressure minima, which remain above 0.4 Pa in the case with $\varepsilon_{N2}$=0.3. Together with the subsequent increase of $N_2$ ice albedo needed to match the Voyager 2 pressure constraint, it induces lower condensation-sublimation rates and a smaller northern cap.

The cases with the $\alpha$-transition are slightly different. The change of emissivity forces the ice surface temperature to remain at the transition temperature $T_{\alpha-\beta}$=35.6 K during the periods of low pressure (i.e., northern summers) and therefore the surface pressure during these periods also remains constant at ~0.5 Pa. However, $\varepsilon_{N2}$ remains at 0.8 during most of time, and therefore the condensation-sublimation rates and the extension and thickness of the northern cap do not change much compared to the reference case.

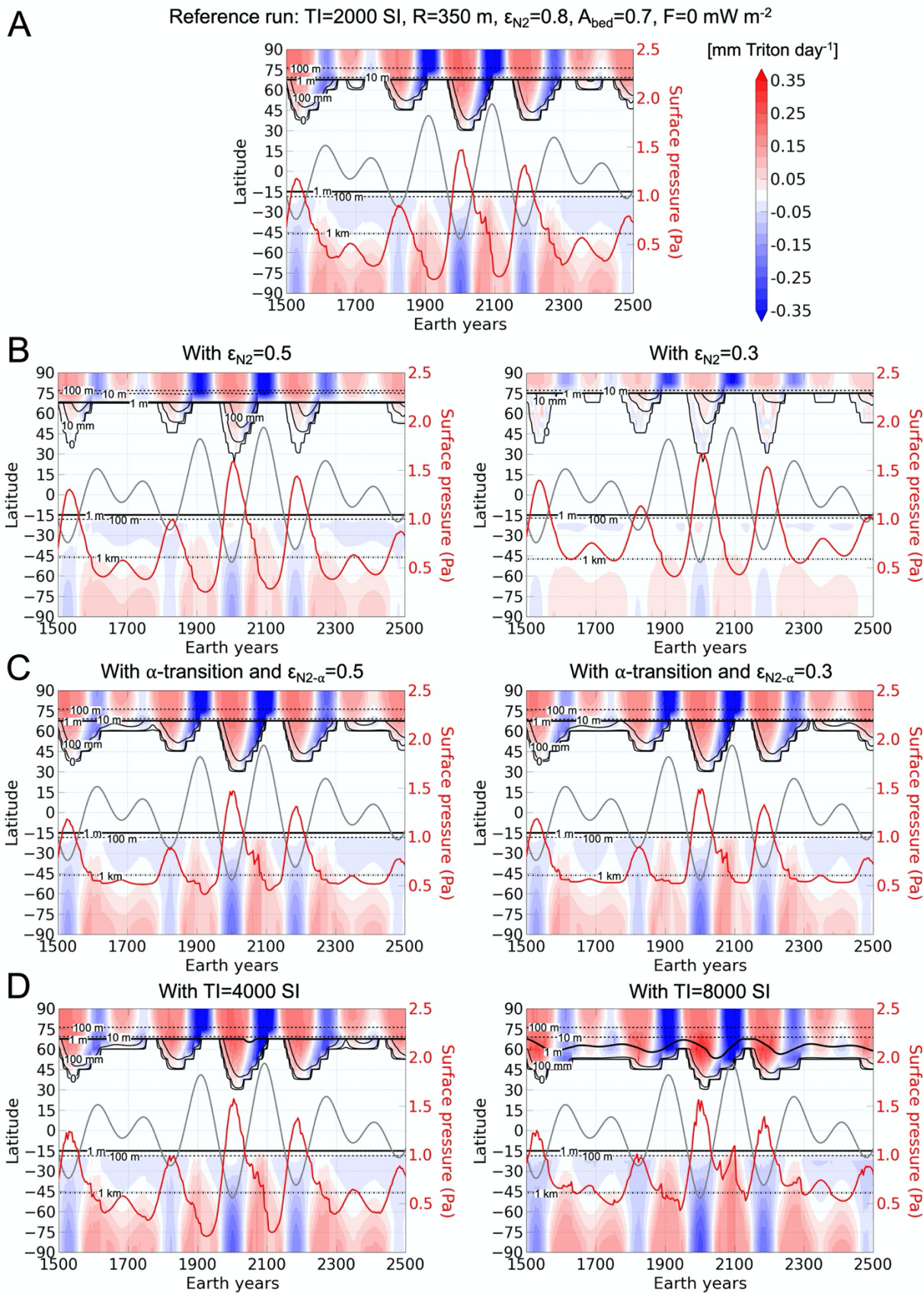

*Figure 19: Sensitivity of the volatile cycle to low $N_2$ ice emissivity and high thermal inertia. (A) Reference simulation among best-case runs, with the North-South asymmetry in $N_2$ ice albedo, TI=2000 SI, $\varepsilon_{N2}$=0.8, showing the zonal and diurnal mean $N_2$ condensation-sublimation rate (mm Triton day$^{-1}$) during the period 1500-2500 (about 7 seasonal cycles). The black contours indicate the extent of the caps (0-line, 10 mm, 100mm, solid), and where the ice is 1 m (thick solid line) and 1 km thick (dash-dotted line). The red line indicates the surface pressure (right y-axis) and the grey line indicates the subsolar latitude. (B) As in (A) but with $\varepsilon_{N2}$=0.5 (left) and $\varepsilon_{N2}$=0.3 (right). (C) As in (A) but with the $\alpha$-$\beta$ phase transition, $\varepsilon_{N2-\beta}$=0.8, $\varepsilon_{N2-\alpha}$=0.5 (left) and $\varepsilon_{N2-\alpha}$=0.3 (left). (D) As in (A) but with TI=4000 SI (left) and TI=8000 SI (right).*

### 7.3.3. Best simulations matching observations

As defined in Section 3.5, realistic simulations must be consistent with the volatile fractional area retrieved from Earth-based spectroscopic observations, the relatively bright surface and the extent of the caps seen by Voyager 2 in 1989, and the surface pressure of ~1.41 Pa retrieved by stellar occultation in 2017 (all simulations match by construction the 1.4 Pa surface pressure measured by Voyager 2 in 1989).

Here we define four classes of simulations, summarized in Table 3. Simulations of Class #1 are consistent with the observed surface pressure in 2017 (1.41 Pa, with 30% margin, i.e. ±0.4 Pa) and with a relatively bright surface Bond albedo ($A_{bed}$ >= 0.6). Note that all simulations of Class #1 satisfy the constraint of having a full-disk averaged surface temperature of 37-44 K in 1989 (although this constraint is not stringent enough to distinguish between best-case simulations). Simulations of Class #2.1 are of Class #1 and are in addition consistent with the volatile fractional area observed in 1995 and 2010, with relatively large margins (we assume 45-75% for both 1995 and 2010). Simulations of Class #2.2 are also of Class #1, but are consistent with the extent of the caps observed in 1989 by Voyager 2 (poleward of 45°N in the northern hemisphere and about 15°S in the southern hemisphere, with some margins on these values). Both Class #2.1 and Class #2.2 require a more constrained surface pressure in 2017 of 1.41 Pa ± 0.2 Pa. Finally, simulations of Class #3 are the most realistic ones. They are of Class #1, #2.1 and #2.2. Here we only analyze the simulations performed with the North-South $N_2$ albedo asymmetry (those without the asymmetry can match Class #1 criteria but do not match any of the Class #2 criteria, as shown in Section 7.2).

| |
|---|
| **Class 1: Matching 2017 stellar occultations and surface albedo** |
| Surface pressure in 2017: 1.41 ± 0.4 Pa |
| Surface albedo (all ices, including bedrock) ≥ 0.6 (relatively bright surface) |
| **Class 2.1: Matching Earth-based spectroscopic observations** |
| Must be of Class 1 |
| Volatile fractional area within 45-75% in 1995 and in 2010 |
| Surface pressure in 2017: 1.41 ± 0.2 Pa |
| **Class 2.2: Matching cap extents seen by Voyager 2** |
| Must be of Class 1 |
| Southern cap extent within 35°S-5°N in 1989 |
| Northern cap extent poleward of 35° in 1989 |
| Surface pressure in 2017: 1.41 ± 0.2 Pa |
| **Class 3: Best cases** |
| Must be of Class 1, 2.1 and 2.2 |

*Table 3: Classification of the simulations, based on the available observations and with some margins. Most realistic simulations are of Class #3.*

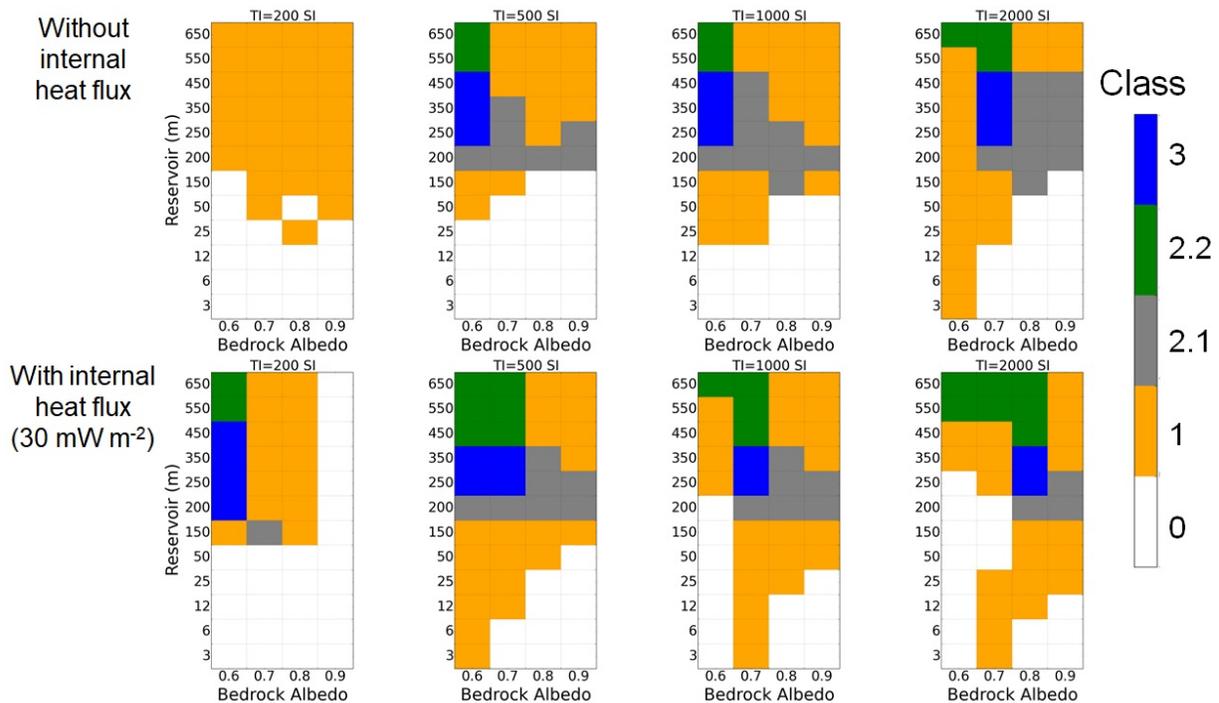

*Figure 20: Classification of our simulations (with the North-South $N_2$ albedo asymmetry) without (top) and with (bottom) 30 mW $m^{-2}$ internal flux as defined by Table 3, with varying subsurface thermal inertia (columns, TI = 200, 500, 1000, and 2000 SI from left to right), $N_2$ global reservoir (y-axis) and bedrock albedo (x-axis). Best case simulations (Class #3) are shown in blue. Class 0 does not meet any constraint.*

Figure 20 shows the classification for all the simulations performed with the North-South $N_2$ albedo asymmetry. We note that:

- All simulations performed with a low TI of 200 SI are generally of Class #0 or #1 because they are inconsistent with the extent of the caps seen in 1989 and with the observed volatile fractional area (too large), except for the cases with an internal heat flux, a large reservoir and a bedrock surface albedo of 0.6 (Classe #3).

- The most realistic simulations in terms of surface pressure and $N_2$ ice distribution (extent in 1989 and fractional area in 1995-2010) have a bedrock albedo within 0.6-0.8 and a large $N_2$ ice reservoir greater than 200 m. In particular, the northern cap is confined to poleward of 45˚N in 1989 for moderate bedrock albedo (< 0.8) and mid-to-high bedrock thermal inertia, while the southern cap extends to at least 30˚S in 1989 for large $N_2$ reservoirs (> 200 m). As a result, these simulations are of Class #2 (2.1 or 2.2) at least.

- The fact that best case simulations are obtained for a bedrock albedo within 0.6-0.8 is consistent with Voyager 2 observations.

- Simulations of Class #3 that closely match all observations listed in Table 3 are generally obtained for mid-to-large reservoir (200-450m) coupled with (1) high thermal inertia (2000 SI) and a bedrock albedo of 0.7-0.8, or (2) intermediate thermal inertia (500-1000 SI) and a bedrock albedo of 0.6-0.7, or (3) low thermal inertia (200 SI), a bedrock albedo of 0.6 and high internal heat flux.

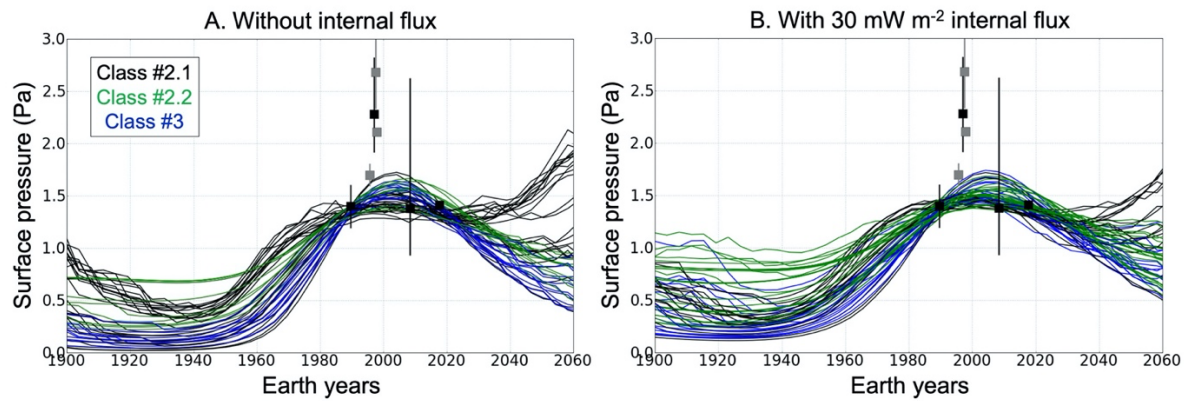

*Figure 21: Surface pressure from 1900 to 2060 as obtained in the best case simulations, Classes #2.1 (black), #2.2 (green) and #3 (blue), without (A) and with internal heat flux (B). Black and grey data points and error bars represent the pressure observations as presented in Table 2 (black are for the data points that we consider are the strongest observational constraints).*

Figure 21 shows the evolution of surface pressure in the best case simulations. The evolution is relatively similar for all these simulations: the surface pressure increases by a factor of 1.5-2 during the period 1980-2010, reaches a maximum of 1.5-1.8 Pa in ~2005-2010 and then decreases to reach 1-1.5 Pa in the period 2020-2040 and 0.5-1 Pa by 2060.

We note an increase in surface pressure between 2040-2060 in some Class #2.1 simulations. This is due to the fact that the seasonal northern cap is strongly extended (equatorward) in these simulations around year 2000, following a large accumulation of deposits at low latitude during southern summer. As a result, when the subsolar point moves northward (close to 0° in 2040-2060), there are large amounts of $N_2$ ice available for sublimation (at the equator and mid-latitudes) and sublimation tends to dominate condensation (at the poles) in global average, causing an increase in surface pressure.

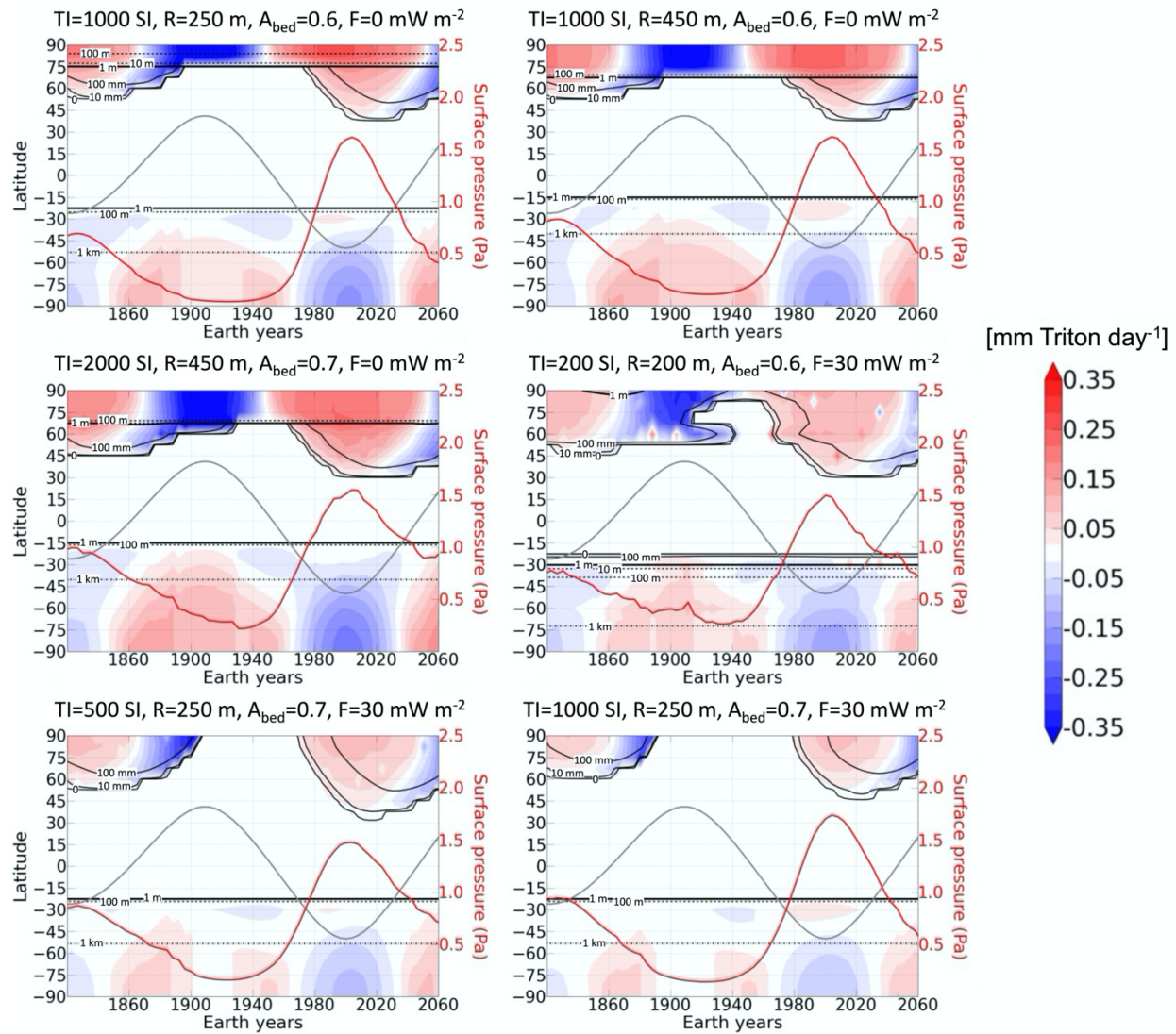

*Figure 22: Best case simulations with the North-South asymmetry in $N_2$ ice albedo. Zonal and diurnal mean $N_2$ condensation-sublimation rate (mm Triton day$^{-1}$) during the period 1820-2060. The black contours indicate the extent of the caps (0-line, 10 mm, 100mm, solid), and where the ice is 1 m (thick solid line) and 1 km thick (dash-dotted line). The red line indicates the surface pressure (right y-axis) and the grey line indicates the subsolar latitude. Deposits thinner than 1 m tend to be seasonal.*

Figure 22 shows the seasonal evolution of the $N_2$ condensation-sublimation rates for six of the best-case (class #3) simulations, while Figure 23 shows the latitudinal distribution of ice and surface temperatures as obtained for 1989 for the same simulations. The modeled southern cap is permanent and extends to ~15°S, with low latitudes being dominated by sublimation, but replenished in $N_2$ ice by viscous glacial flow. All best simulations suggest a southern cap that is >1 km thick poleward of 60°S. A small permanent northern polar cap is suggested for best-case simulations without internal heat flux, while only a seasonal northern cap is suggested for best-case simulations with 30 mW m$^{-2}$ internal heat flux. In all simulations, the northern cap extends at least to 45°N in the period 2020-2040, and to ~30°N in some cases. At the southern summer solstice, the sublimation and condensation rates are about ~0.1-0.2 mm per Triton day (i.e. 2-4x10^-7 mm s-1), and about 0.13-0.26 m if applied over 20 Earth years. Simulations performed with a high internal heat flux (F=30 mW m⁻²) show reduced $N_2$ condensation-sublimation rates due to the higher $N_2$ ice albedo (~0.05-0.15 mm per Triton day), and smaller permanent caps. In all cases, $N_2$ recondenses in the low latitudes of the southern hemisphere during the period 1980-2020. Sublimation from the polar southern cap and condensation at the northern edge was also suggested by the thermal balance model of Grundy et al. (2010) and could help explain the longitudinal variation of the $N_2$ 2.15 μm feature.

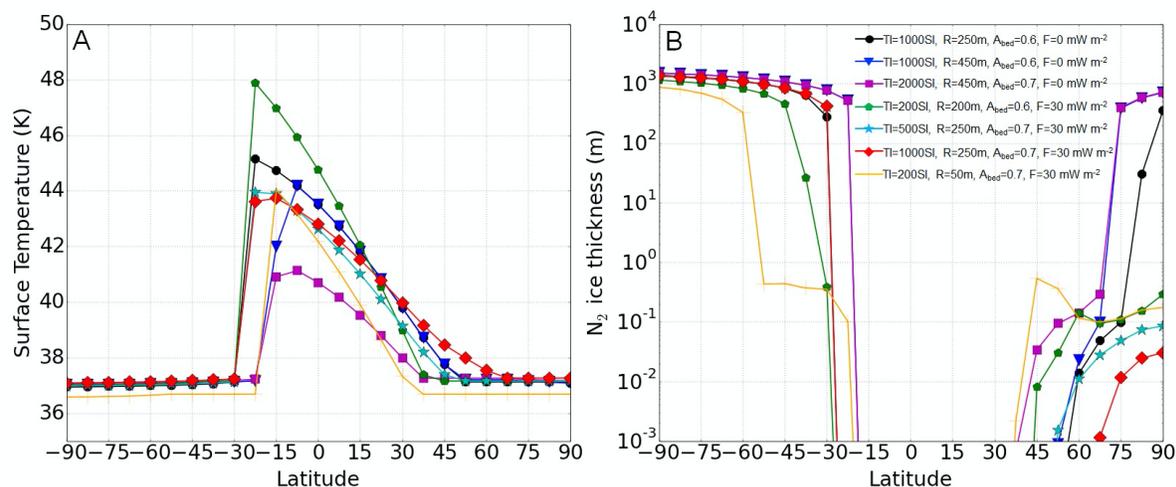

*Figure 23: (A) Zonal mean surface temperature vs latitude for the best-case simulations in 1989, as shown by Figure best_flux1: Without internal heat flux : TI=1000 SI - R=250 m - $A_{bed}$=0.6 (black circle), TI=1000 SI - R=450 m - $A_{bed}$=0.6 (blue triangle), TI=2000 SI - R=450 m - $A_{bed}$=0.7 (purple square). With internal heat flux: TI=200 SI - R=200 m - $A_{bed}$=0.6 (green polygon), TI=500 SI - R=250 m - $A_{bed}$=0.7 (cyan star), TI=1000 SI - R=250 m - $A_{bed}$=0.7 (red losange). The orange solid line shows the cases of a low reservoir and low thermal inertia simulation (TI=200 SI - R=50 m - $A_{bed}$=0.7 - F=30 W m⁻²) that is consistent with observed caps extent and volatile fractional area but inconsistent with the observed pressure in 2017 (too low, see text in Section 8.3). (B). Same as A but showing the surface $N_2$ ice thickness.*

## 7.4 On the seasonal cycle of CO and CH₄

Our results show that CO and $CH_4$ remain mixed with $N_2$ and never form pure deposits (except for pure $CH_4$ frost residuals forming at the cap edge when $N_2$ ice sublimes and disappears, but these frosts quickly disappear within one Earth years or less). This contrasts with Pluto, where $CH_4$-rich ice deposits cover a large part of the surface (thick and permanent deposits in the equatorial regions, seasonal frosts or permanent mantle at mid-to-polar latitudes). This may be due to a difference in the global reservoir of $CH_4$ ice (which may be too small on Triton to form permanent $CH_4$-rich deposits). Another basic understanding of the absence of broad permanent $CH_4$-rich deposits on Triton is the following: on Pluto, $N_2$ and $CH_4$ ices tend to accumulate in the equatorial regions, which are colder than the mid-to-high latitudes, in average over a Pluto year and over several astronomical cycles, due to the relatively high obliquity of the spin axis (e.g., Bertrand et al., 2019). Whereas $N_2$ ice preferably accumulates in depressions, $CH_4$ ice preferably accumulates at high altitude (e.g., Bertrand and Forget, 2016, Bertrand et al., 2020b). The fact that the equatorial regions on Pluto correspond to a large area displaying a variegated topography, including deep depressions (such as the ~6-10 km deep Sputnik Planitia impact crater) and tall mountains (e.g. Pigafetta Montes, or the region of Tartarus dorsa) allows the formation of both permanent $N_2$-rich (Sputnik Planitia ice sheet) and $CH_4$-rich (bladed terrain) deposits at different locations. On Triton, however, the coldest regions are the poles (maybe one pole in particular if one assumes an asymmetry in internal heat flux for instance), which correspond to a relatively small area where only small variations in topography may be displayed, thus not allowing the formation of both deposits (instead, the ice mix together at the poles).

Although we do not expect broad and thick $CH_4$-rich ice deposits to form on Triton, the formation of small and transitory $CH_4$-rich ice patches in our model may be limited by the spatial resolution used or the different assumptions made regarding the behavior of ice mixtures. Here we tested different scenarios for CO and $CH_4$ and describe the model predictions of CO and $CH_4$ atmosphere mixing ratios, and compare to observations. The various observations of CO and $CH_4$ surface ice and atmosphere mixing ratios are summarized in Section 2.3 and Section 2.5, respectively.

### 7.4.1. The CO seasonal cycle

In our model, when we impose a constant CO ice mixing ratio of 0.04% (respectively 0.08%) into $N_2$ ice and we assume that the vapour pressure equilibrium is controlled by Raoult's law, we obtain a mean CO gas volume mixing ratio of 0.006% (respectively 0.012%) during the period 2000-2020, in good agreement with the 2017 ALMA mm-observations (Gurwell et al. 2019), and therefore inconsistent with the 10x higher IR-derived value from Lellouch et al. (2010). The limited seasonal variation is due to the volatility of CO being close to that of $N_2$ and the absence of CO-rich ice deposits (Tegler et al., 2019, and absent on Pluto's surface as well, Bertrand and Forget 2016, Schmitt et al., 2017). In the model, CO tends to follow $N_2$ and condenses where $N_2$-rich deposits are already present. This is consistent with ground-based spectroscopic data of Triton (CO follows $N_2$ in longitude) and with observations of Pluto's surface by New Horizons. According to our model, the CO gas volume mixing ratio should remain very close to these values in the next decades (with a slight decrease in 2040 to 0.005% and 0.01%, respectively), although small variations in the CO gas volume mixing ratio are possible if the surface of some areas becomes enriched or depleted in CO ice as $N_2$ sublimates or condenses.

### 7.4.2. The $CH_4$ seasonal cycle

As detailed in Section 2.5, the observed $CH_4$ atmospheric mixing ratio is much larger (by three orders of magnitude) than expected for an ideal mixture, i.e. if $CH_4$ ice only exists on Triton as $CH_4$ diluted in a $N_2$-rich solid solution (Stansberry et al. 1996b), or for an intimate mixture. Instead, using the $CH_4$ patch model of Stansberry et al. (1996b), Merlin et al., 2018 (respectively, Quirico et al., 1999) estimated that $CH_4$-rich ice (at ~40 K) covered 2-3% in 2010-2013 (respectively, a maximal value of 10% in 1995) of the visible disk of Triton, which could be sufficient to maintain a $CH_4$ atmospheric volume mixing ratio around 0.03%, as observed by Voyager 2.

We tested this scenario in one of our reference simulations (TI=2000 SI, R=400 m, $A_{bed}$=0.6) by imposing pure $CH_4$ ice deposits on Triton's surface (with a fractional area consistent to the observations of Quirico et al., 1999 and Merlin et al., 2018), assuming that they follow Raoult's law and are a proxy for $CH_4$-rich ice (this assumption is good enough, as described in Young et al., 2021). We tested two different locations for these deposits:

(1) At the edges of the southern cap, thus forming a latitudinal band of $CH_4$ ice near the equator, with a fractional area of ~5% of the projected disk as seen from Earth at the time (~2010) of the observations of Merlin et al. (2018): these regions should contain seasonal volatile frosts, which would favor the formation of $CH_4$-rich ice by segregation as $N_2$ ice sublimates and disappears. Voyager 2 images showed features in these regions suggestive of intense $N_2$ ice sublimation, which could lead to an enrichment in $CH_4$-rich ice on the surface.

(2) At the south pole (~2% of the disk surface seen from Earth in 2010): our simulations show that the $N_2$ ice sublimation rates are strongest at the south pole, which could favor the formation of $CH_4$-rich deposits on top of the surface.

Figure 24 shows the seasonal evolution of the global mean mixing ratio of atmospheric $CH_4$ over time for both scenarios, and assuming different albedos for the pure $CH_4$ ice. Note that the evolution of the surface pressure remains relatively unchanged for all cases compared to that in the reference simulation (without $CH_4$-rich ice).

The simulations with $CH_4$-rich ice at the cap edge produce a $CH_4$ partial pressure of $1\text{-}5\times10^{-4}$ Pa in 1989, which is consistent with the $2.45\times10^{-4}$ Pa observed by Voyager 2 (Figure 24, solid lines). Slightly larger amounts can be obtained with a lower $CH_4$ ice albedo and a higher surface $CH_4$ ice coverage (solid grey lines). The simulations with $CH_4$-rich ice at the south pole produce an increase in $CH_4$ partial pressure until 2005, followed by a decrease as the south pole exits polar day (Figure 24, dashed lines). A small fraction of the south pole covered by $CH_4$-rich ice is sufficient to produce a $CH_4$ partial pressure of a few nbar, as observed. The simulation including a $CH_4$-rich patch at the south pole with albedo of 0.7, is particularly promising to explain the factor of 4 (2.45 to $9.8\times10^{-4}$ Pa) increase in the $CH_4$ partial pressure from 1989 to 2009, as reported by Lellouch et al. (2010). In contrast, the models in which $CH_4$-rich ice is concentrated at the $N_2$ cap edge do not reproduce this increase in the $CH_4$ atmospheric abundance over 1989-2009. However, a decrease in $CH_4$ ice albedo or increase in $CH_4$ ice coverage during this period can also explain the observed increase in $CH_4$ partial pressure. It is possible that both the pole and cap edge become enriched in $CH_4$ ice, with different amounts and timescales.

Note that here we assume that these modeled $CH_4$-rich deposits, exposed at the surface once $N_2$ ice locally disappeared, are permanent during southern summer. If $CH_4$ ice also disappears from the surface, the $CH_4$ atmospheric mixing ratio would slowly fall back to the levels expected for an ideal mixture (a few $10^{-5}$ %).

Also note that the relatively low $CH_4$ atmospheric mixing ratio observed (~0.03%) and simulated (>0.01%), would still be enough to block most of the incoming Lyman-α radiation (121.6 nm, 10.19 eV) from direct sunlight and from backscattering from the interplanetary medium, which has implication for the direct photolysis and subsequent darkening of the ices on the surface.

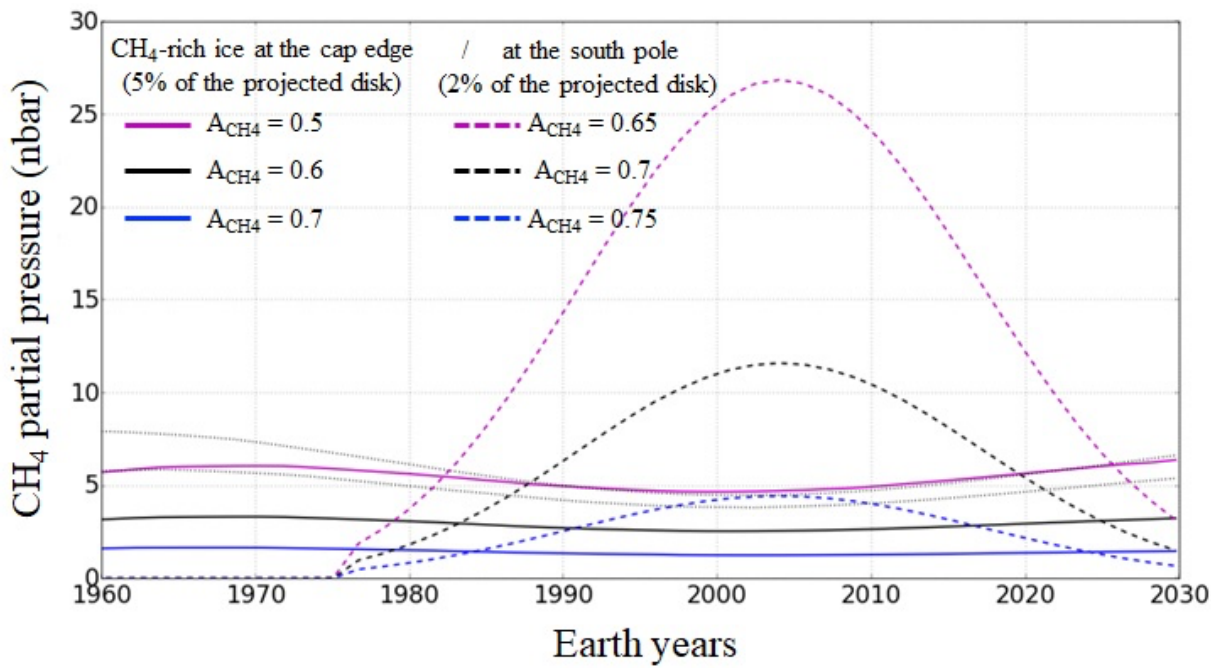

*Figure 24: Global mean $CH_4$ partial pressure (nbar, or x10^{-4} Pa) over time, as obtained for the simulation TI=2000 SI, R=450 m, $A_{bed}$=0.6, with $CH_4$-rich ice at the cap edge (fractional area of 5% of the visible disk in 2010, solid lines) and at the south pole (2% of the visible disk, dashed lines), for different $CH_4$ ice albedos. The two grey curves are for a simulation with $CH_4$-rich ice at the cap edge (and $A_{CH4}$=0.6, as for the solid black curve) but covering 9% and 12% of Triton's visible surface area, respectively.*

## 8. Discussion

### 8.1. On the formation of a northern (polar) cap

All long-term simulations performed in Section 5 with North-South asymmetries in internal heat flux, topography, and $N_2$ ice albedo, along with a global $N_2$ ice reservoir of 300 m, predict the formation of a permanent $N_2$ (polar) cap in the northern hemisphere (>100 m thick). Most of the best-case simulations (Class #1 to #3) performed in Section 7 also predict a permanent $N_2$ cap in the northern hemisphere. The northern cap is much smaller and thinner in the best-case simulations performed with an internal heat flux of 30 mW m^{-2}, due to the warmer subsurface. A few results with intermediate thermal inertia and moderate global $N_2$ reservoir (TI=1000 SI, R=250-350 m, $A_{bed}$ = 0.6) predict no permanent cap. The northern cap is not permanent if a smaller global $N_2$ ice reservoir (< 250 m) is used along with large North-South asymmetries (see Figure 13, R=200 m, and $\Delta A_{N2}$ = 0.1). However, these low-reservoir simulations are not consistent with the available observations.

The formation of a permanent $N_2$ northern cap is primarily the result of two effects:

1. The northern polar night is long enough so that surface temperatures of the bedrock drop below 34 K for a bedrock surface albedo of 0.6. Consequently, $N_2$ condensation is easily triggered at the north pole if the surface pressure becomes higher than the pressure at solid-gas equilibrium (e.g., ~1 Pa at 37 K).

2. In the simulations, $N_2$ ice flows (slowly, on timescales longer than the seasonal timescales) from the southern cap towards the warmer sublimation-dominated mid-to-low latitudes, thus maintaining large amounts of ice in these regions, supplying $N_2$ for further sublimation and allowing the surface pressure to exceed that at solid-gas equilibrium at the north pole.

Note that all best-case simulations explored predict that $N_2$ condensation has been occurring in the northern hemisphere since 1980. They show that at least a seasonal deposit of a few mm of $N_2$ ice was covering the northern polar latitudes in 1989 and should cover the northern latitudes down to 30°N-45°N in the period 2010-2030 (see Figure 22). In general, the current extent of the northern cap is related to the extent of the southern cap by the surface pressure: a large extent of the northern cap down to 30°N-45°N (i.e., a large condensation-dominated area) is only possible if the southern cap remains extended to at least 15°S (i.e., a large sublimation-dominated area) during this period, and a smaller southern cap would imply little to no $N_2$ deposits in the northern hemisphere, otherwise northern condensation would largely dominate southern sublimation and the surface pressure would have significantly dropped in 2017 to levels inconsistent with observations. The scenario of a large permanent southern cap and the seasonal appearance of $N_2$ ice deposits in the northern hemisphere below the terminator could also explain the increase in $N_2$ band absorption observed with SpeX/IRTF during the period 2002-2020 (Holler et al., 2016, 2020).

## 8.2. On the evolution of surface pressure during the period 1989-2017

All best-case simulations show the same trend for the evolution of surface pressure, with an increase from 1920 to 2005-2010 (shortly after the southern summer solstice) followed by a decrease back to the 1920 levels in 2080.

Our results do not suggest a strong surge in surface pressure, as it has been reported for the period 1995-1997 (e.g., Elliot et al., 2000). During the 1989-2005 period, our best-case class #3 simulations suggest an increase in surface pressure by a factor of 1.1-1.2 only, with a peak around 1.5-1.7 Pa. Some of the Class #1 simulations show a peak in surface pressure reaching 2.3 Pa, i.e. a factor of 1.6 compared to the Voyager 2 epoch (these simulations typically predict a 2017 surface pressure that is above the 1.41+0.2 Pa constraint, not shown). These values are consistent with the 1995 occultation event and the derived pressure reported by Olkin et al. (1997), and also eventually with the 1997 occultation event and the derived pressure reported by Marques Oliveira et al. (2021), although the simulations only match this latter value at the 1σ significance level. However, they are inconsistent with the higher values reported in 1997 (Elliot et al., 2000, 2003). We note that a reanalysis of some of these data indicates that this pressure increase was not significant at the 3-σ level (Marques Oliveira et al., 2021). Such a strong increase in surface pressure could still have occurred if processes not taken into account in our model play an important role on Triton, such as strong surface albedo feedback for instance.

## 8.3. On the presence of volatile ice at the south pole

As detailed in Section 2.3, ground-based spectroscopy of Triton's surface performed with IRTF/SpeX in 2002-2014 and VLT/SINFONI in 2010-2013 showed little longitudinal variability for $CO_2$ and $H_2O$ ice, unlike the three volatile ices $N_2$, $CH_4$, CO (Grundy et al. 2010, Holler et al., 2016, Merlin et al., 2018). To explain these observations, it has been suggested that $CO_2$ and $H_2O$ ices are exposed at the south pole (the very southernmost latitudes, roughly 90°S-60°S), which would be bare of volatile ice (Grundy et al. 2010, Holler et al., 2016).

However, our volatile transport model challenges this scenario since it always predicts a permanent cap at the south pole, except if an extremely low $N_2$ reservoir, i.e. only seasonal frosts, is assumed, but these simulation cases typically present (at least for moderate to high thermal inertias) a too low surface pressure in 2017 (< 1.41 Pa, Figure 16), a low volatile fractional area (< 40%, Figure 15) and a southern cap extent poleward of 45°S in 1989 (Figure 14). Our model, and its current modeled physical processes, cannot predict such a "hole" in volatile ice at the south pole (90°S-60°S). Physical processes not taken into account in the model may play an important role to mask the detection or completely remove $N_2$ ice at the south pole (see model uncertainties in Section 8.7). For instance, the predicted sublimation of 30-50 cm of $N_2$ ice at the south pole during the period 1970-2000 (Figure 11, Figure 22) could have led to an accumulation of $CH_4$-rich ice at the south pole could mask the detection of $N_2$ ice and be a possible scenario, although it would imply a relatively large $CH_4$-rich-covered area that would remain to be consistent with other observations.

The near-IR VLT/SINFONI observations and subsequent modeling of Merlin et al. (2018) showed that $CO_2$ ice is actually present in the form of very small grains and has two components (see their Table 9), one in which it is mixed into the $N_2$:$CH_4$:CO matrix (unit 1) and another in which it is mixed into $H_2O$ ice (unit 2). These results suggest that $CO_2$ ice is present almost everywhere (with an equivalent area of 50% and very small grains). In addition, their model suggests that $H_2O$ ice is present in unit 2 only ($H_2O$+$CO_2$+$CH_4$-rich) and does not need to be invoked in unit 1 ($N_2$:$CH_4$:CO). Based on these results, Merlin et al. (2018) imagined two scenarios that could reconcile the presence of both $N_2$ and $CO_2$ ices at the south pole: (1) A bedrock surface ($CO_2$ ice) covered by small $CO_2$ grains (regolith) covered by a transparent $N_2$-based matrix (allowing $CO_2$ to stay detectable), and (2) Small $CO_2$ particles on top or inside the $N_2$-based matrix, as the result of endogenous processes (e.g. active geysers tearing particles of $CO_2$ from the bedrock, transport and spatial redistribution of these particles by the winds, and deposition onto the surface).

Our model results are consistent with these two scenarios, since they do not prevent $N_2$ ice to be present at the south pole. However, in general, it remains difficult to be confident on the composition of Triton's south pole, given the limited amount of observational datasets, the relative simplicity of the VTM, and the large number of model parameters in both the VTM and the near-IR spectra analysis models.

## 8.4. On the different terrains observed at the cap edge by Voyager 2

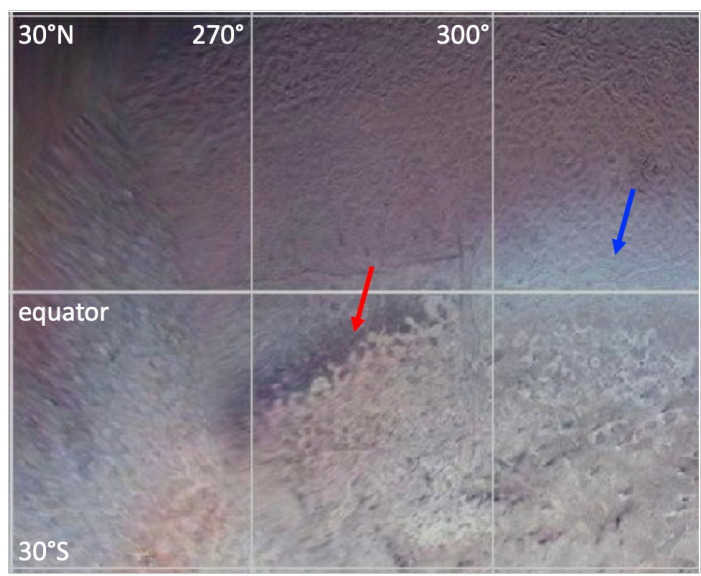

*Figure 25: Equatorial regions on Triton at longitudes 240°E-330°E displaying alternation of relatively bright blue frosts (blue arrow, 240°E-270°E and 300°E-330°E) and relatively dark terrains without blue frost (red arrow, 270°E-300°E) at the edge of the southern cap.*

Figure 25 highlights two different terrains observed in the equatorial regions at the cap edge: (1) a relatively bright blue (or less red) surface, with photometric properties consistent with a freshly deposited frost (McEwen, 1990); this terrain is seen at almost all longitudes and forms a fringe at the cap edge (Figure 2), and (2) a relatively dark surface, without much blue frost, around 270°E-300°E.

We hypothesize that this dark terrain is the result of a high concentration of dark materials (e.g. tholins-like haze deposits and/or irradiated ices) that had been mixed in $N_2$ ice and have now been revealed at the surface as the $N_2$ ice sublimates away. The composition of this unit may be similar to some degree to the dark $N_2$ ice plains located at the northern edge of Sputnik Planitia on Pluto, which are enriched in dark materials and in $CH_4$ ice due to intense $N_2$ ice sublimation (White et al., 2017). The fact that this unit on Triton may be enriched in $CH_4$ ice is consistent with IRTF/SpeX observations, which show a peak absorption for $CH_4$ near ~290°E and also a maximum of shift in the position of the $CH_4$ bands indicating an increased amount of $CH_4$-rich ice. However, these observations are for the period 2002-2014 only (Grundy et al., 2010, Holler et al., 2016). We also note that the formation of this dark terrain on Figure 25 could be related to topography (see Section 8.5).

This observed longitudinal asymmetry in $CH_4$ ice distribution could suggest the presence of $CH_4$-rich deposits at the cap edge, rather than at the south pole. Our simulations with $CH_4$-rich ice at the cap edge could explain the fourfold increase in $CH_4$ partial pressure from 1989 to 2009 reported by Lellouch et al. (2010) if the $CH_4$ ice albedo significantly lowered or if the $CH_4$ ice surface coverage significantly increased during this period. The latter would also be consistent with the observed increase in $CH_4$ band absorption during the period 2002-2020 (Holler et al., 2016, 2020). It should be noted that such a scenario of a variable-width ring of segregated less-volatile ice surrounding retreating volatile seasonal deposits has already been observed on Mars in the analogue case of water ice trapped in seasonal $CO_2$ ice deposits (Appéré et al. 2011). On Mars, this moving ring also triggers a strong seasonal increase in water vapor during its sublimation (e.g. Pankine et al., 2009, 2010).

Our simulations with $CH_4$-rich ice at the south pole also explain the fourfold increase in $CH_4$ partial pressure (suggested for the period 1989-2009) as being driven by insolation changes. However, it remains unclear how $CH_4$-rich ice would form there in the first place. The surface composition of the south pole on Triton in 1989 could also be comparable to the northern dark plains of Sputnik Planitia on Pluto in 2015, since these plains correspond to thick $N_2$ ice deposits experiencing the constant insolation of the polar day, as Triton's south pole.  In Sputnik Planitia, the plains display darker $N_2$-rich ice deposits enriched in $CH_4$ ice. The permanent and cold $N_2$ ice layer located there may buffer the increase of surface temperatures and limit the formation of $CH_4$-rich deposits on top of the ice layer.  On Pluto, $CH_4$-rich ice is mostly detected outside Sputnik Planitia, with more spatial variability in the $CH_4$:$N_2$ mixture at mid latitudes, where thinner seasonal deposits are found (Schmitt et al., 2017).  By analogy, on Triton, $CH_4$-rich ice may tend to form primarily at mid-to-equatorial latitudes where volatile ice deposits are thinner, and at the cap edge where the deposits are seasonal.

The bright "blue" surface may be freshly deposited $N_2$ frost (relatively clean ice, hence the less red color). Our model suggests that $N_2$ recondenses in the equatorial regions after 1980 (and until 2020, Figure 22). This happens only if the exposed "bedrock" locally has a high surface albedo or if $N_2$ ice is already present or was recently present in these regions; in the latter case the exposed "bedrock" surface would have remained cold enough to permit $N_2$ recondensation. The locations of this hypothetical blue $N_2$ frost are relatively consistent with IRTF/SpeX observations, which show a minimal (respectively maximal) $N_2$ absorption roughly where the combined extent of the cap and of the blue fringe is minimal (respectively maximal). The formation and expansion of this frost would also be consistent with the observed increase in $N_2$ band absorption during the period 2002-2020 (Holler et al., 2016, 2020).

Alternatively, the bright "blue" surface could be made of freshly-condensed $CH_4$-rich ice. As detailed in Section 2.3, the amount of $CH_4$ ice that is not diluted in $N_2$ ice is estimated to cover a relatively small area on Triton's surface: 2–3% of the surface projected on the visible disk in 2010 (Merlin et al., 2018) and a maximum of 10% in 1995. The blue fringe roughly covers a latitudinal band of 15° at the equator, which accounts for ~10% of the disk-projected surface, so exposed $CH_4$-rich frost at the location of the blue fringe would remain in the range of possible values (although close the highest possible value) consistent with surface spectroscopic observations. Would there be enough gaseous $CH_4$ in the atmosphere to permit $CH_4$ condensation onto the surface in the equatorial regions? We can estimate the saturation vapour pressure volume mixing ratio of $CH_4$ (above a volatile-free surface) by using equation 14 in Forget et al. (2017), derived from the thermodynamic relations computed by Fray and Schmitt (2009). For a surface temperature of 38 K, 39 K, 40 K and 41 K (i.e. slightly warmer than $N_2$ ice in 1989), we find a $CH_4$ saturation volume mixing ratio of 0.017%, 0.036%, 0.075% and 0.153% respectively. Consequently, a volume mixing ratio of $CH_4$ of ~0.03% (measured by Voyager 2 in 1989) would permit condensation onto a surface colder than ~39 K.

However, the hypothesis of $CH_4$-rich frost for the blue fringe faces three main issues: (1) In our model, the bedrock surface temperatures in 1989 in the equatorial regions at the cap edge are relatively warm at ~42-44 K and become colder than 39 K in the terrains poleward of 30°N (besttsurf1989), which is not consistent with a blue fringe made of freshly-condensed $CH_4$. We note that $CH_4$ condensation could still be favored at the cap edge by locally enhanced winds or if the bedrock surface is locally colder than modeled (e.g. with a higher albedo). (2) A $CH_4$ ice composition of that region does not match well the longitudinal trends seen in the near-IR spectra. (3) The scenario of a $CH_4$ ice deposit left after $N_2$ sublimation is not likely, as the remaining deposit would also contain a certain amount of tholins-like dark materials and therefore would not appear less red (i.e. cleaner) than the surrounding terrains.

Finally, on Pluto, it has been shown that albedo and composition positive feedback could further increase local contrasts in ice sublimation/condensation rates, which could explain the transition from bright to dark plains in Sputnik Planitia (Earle et al., 2018, Bertrand et al., 2020a). Such processes may also be operating on Triton and could explain the diversity of terrains and colors observed at the cap edge. For instance, $N_2$ sublimation and the subsequent darkening of the surface would lead to an amplifying positive feedback by increasing the absorption of incoming radiation and thus the sublimation rate, whereas $N_2$ condensation and the subsequent brightening of the surface would lead to further condensation.

## 8.5. On the topography and $N_2$ ice thickness in the southern hemisphere

Our best-case simulations have been obtained with a relatively large global reservoir of $N_2$ ice (>200 m), and suggest a thick southern cap (> 1 km at the south pole, > 100 m at the mid-latitudes and > 1 m at the equator, balanced by glacial flow). Interestingly, some simulations performed with a low bedrock thermal inertia of 200 SI have been able to produce a southern cap extended to the equatorial regions in 1989 with a lower $N_2$ global reservoir in the range 20 m - 200 m, leading to $N_2$ ice thickness of 0.1-1 m in the equatorial to mid-latitude regions and >100m at the south pole (Figure 14, and see solid orange line on Figure 23). In these simulations, the equatorial to mid-latitude deposits are not balanced by glacial flow (i.e., they are not connected to the polar deposits) but remain permanent due to the relatively cold bedrock, although in some of these cases a thin mid-latitude band of $N_2$ ice entirely sublimates in the period 2030-2050. However, these simulations with low bedrock thermal inertia are not among our best-case simulations (they are only of Class #1 at best), see Figure 20), mostly because they predict $N_2$ ice deposits in the northern hemisphere extending from 90°N to at least 30°N in 1989 (due to the colder surface induced by lower thermal inertia, Figure 14), and consequently a surface pressure in 2017 much lower than the reported value of 1.41 Pa.

In general, the relatively coarse spatial resolution of Voyager 2 images and the lack of useful topography data above the southern cap do not allow us to determine with confidence the order of magnitude of the thickness of the bright southern deposits, which makes it difficult to conclude on the nature (less than 1 m thick seasonal vs. more than 1 m thick permanent) of these large-scale deposits. The spatial resolution also does not allow us to determine if glacial flow has occurred or not. However, the sharp dichotomy between the southern and northern terrains suggests that the northern bright deposit boundary is controlled by topography, possibly by a topographic barrier (e.g., scarps) of ~100 m to several 100 m (Figure 26), which reinforces our best-case model results suggesting thick and permanent volatile ice deposits in these regions.

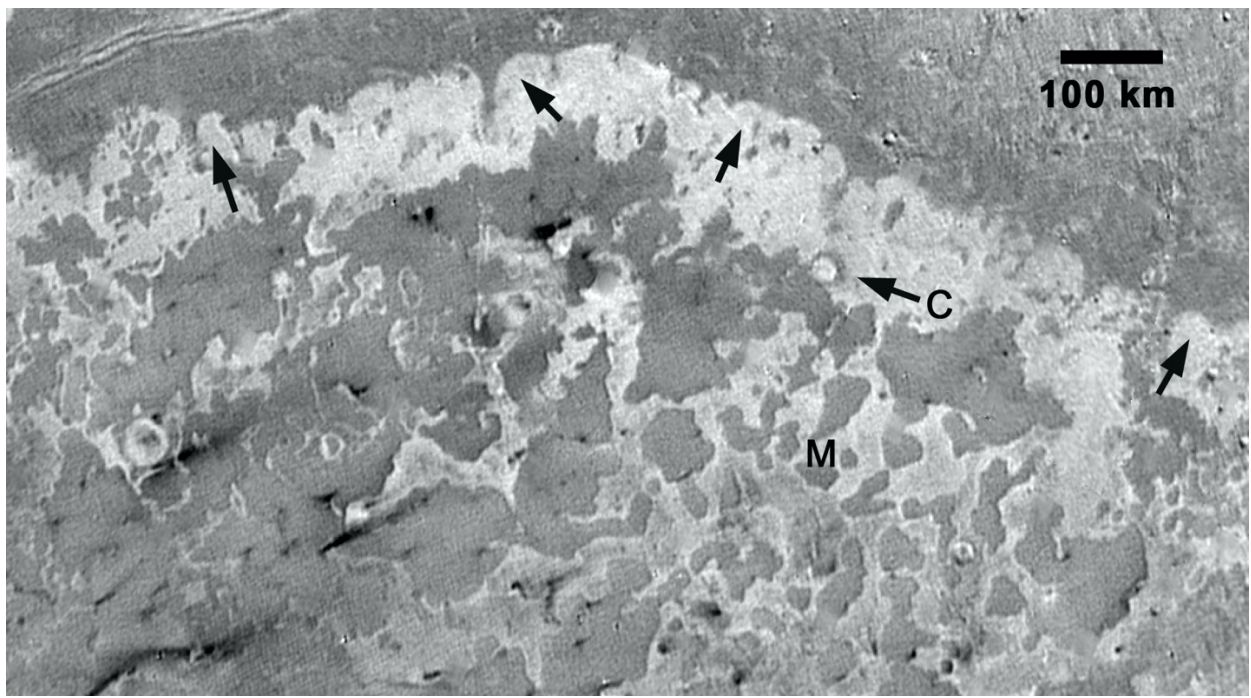

*Figure 26: Enlargement of global cylindrical map showing macular southern terrains near the equator of Triton. Possible example of a topographic barrier in the form of scarps (arrows), possibly ~100 m to several 100 m in height, along the sharp dichotomy between the bright southern and northern terrains seen by Voyager 2. "M" denotes a region of mesas south of the main boundary. "C" shows a possible, partially buried crater. Image center ~15°S, 0°E. North is up.*

Finally, the sigmoidal contact line of the edge of the dark terrain in Figure 25 seems to be related to topography (Paul Schenk, personal communication). This region forms the southwestern extent of the known region of cantaloupe terrain (Croft et al., 1995), which is characterized by closed topographic cells 20-50 km across and up to 500 m deep (Schenk et al., 2021). The dark material appears to occur mainly on the elevated boundaries between the cells, suggesting that these elevated terrains prevent volatile ice condensation or accelerate sublimation. These color and topography contrasts between the northern and southern deposits at the cap edge could indicate a gradual decrease in elevation to the south.

## 8.6. Model predictions for the next decades

### 8.6.1. General climate predictions

**Surface pressure and stellar occultations**

Our best-case simulations that best match the observed surface pressure of the 2017 event suggest a slow decrease in surface pressure over the next decades, at least until 2080 when the subsolar point will be above 30°N and when northern sublimation should dominate over southern condensation (Figure 21). These simulations suggest a surface pressure of ~1-1.5 Pa in the period 2020-2040 and ~0.5-1 Pa by 2060. This is large enough for Triton's atmosphere to remain global during this period. In the next decades, stellar occultation events will be key to confirming or disproving the trend suggested by our model. The next Triton occultation will be on Oct. 6, 2022, and should indicate a surface pressure close to the 2017 value.

**The southern and northern volatile ice deposits**

Our results suggest that most of the $N_2$ deposits in the southern hemisphere are permanent, and the southern cap extent should therefore not change much over seasonal timescales (Figure 22, Figure 27.B). Our simulations performed with a low global $N_2$ reservoir (< 200 m) tend to predict a retreat of the southern cap, but they fail to meet the observational constraints on the northern cap extent in 1989 and on the surface pressure in 2017. In particular, the southern cap should not have retreated by more than 30° latitude since 1989 from its ~15°S extent at that time, otherwise the surface pressure would have dramatically collapsed since 1989, and would be inconsistent with the 2017 occultation (see Figure best_pres.B, Figure profile.B).

We obtained a large diversity of results regarding the thickness and the nature (perennial, seasonal, or non-existent) of the northern deposits. Our best-case simulations suggest that the $N_2$ northern cap will extend to at least 60°N in 2040 (and possibly down to 30°N, Figure 22, Figure 27). The extent of the seasonal northern $N_2$ deposits should be maximal during the period 2010-2030 (Figure 22). As the subsolar latitude currently increases with time, northern latitudes that were hidden in the polar night during the Voyager 2 flyby in 1989 start to be revealed. In the following decades, the northern cap at 60°N (at least its seasonal deposits), if existing, should therefore become visible below the northern limb in ~2025 (subsolar latitude ~30°S). Future observations of Triton with HST, JWST, and the Extremely Large Telescope (ELT) would be able to provide a new and rich dataset of Triton's surface (e.g., surface composition, rotational light-curves, and albedo maps at different wavelengths), which should give strong constraints on the extent of the northern and southern caps as well as on the surface properties of the different terrains (see examples in Section 8.6.2).

In our model, $N_2$ ice sublimation is more intense at the south pole than at the cap edge during the current season, and will continue until 2025-2030. The south pole may thus appear darker and be enriched in $CH_4$ ice in the coming years. In 2030, the south pole should start to be dominated by $N_2$ condensation (Figure 22), which should reverse this trend. Finally, equatorial $N_2$ condensation is predicted until ~2025, possibly forming bright $N_2$-rich frosts. After 2025, $N_2$ sublimation should dominate in the equatorial regions, and any equatorial seasonal $N_2$ frost should start to disappear.

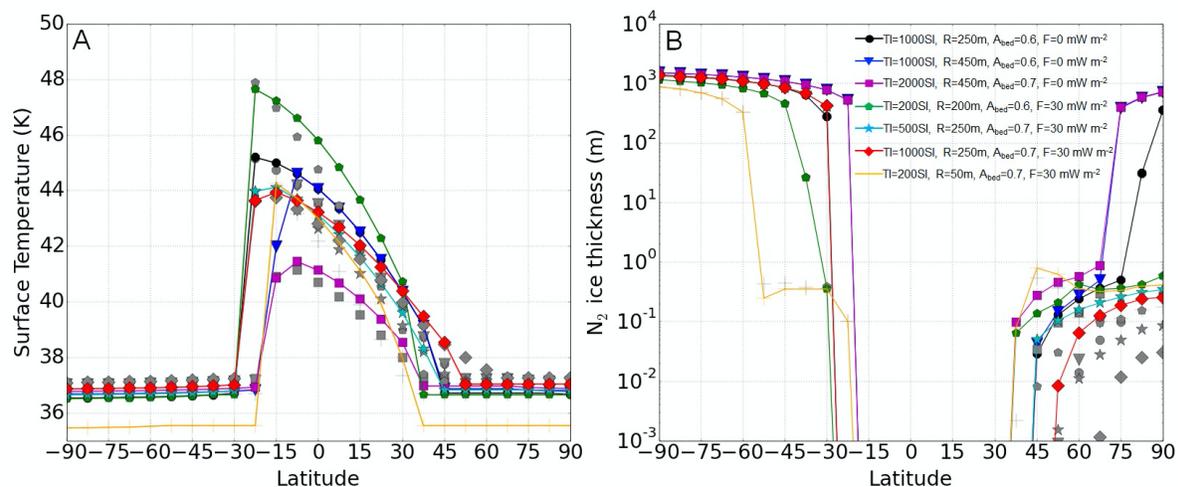

*Figure 27: A. Surface temperature vs latitude for the best-case simulations in 1989 (grey symbols, as in Figure 23) and in 2030 (solid line with colored symbols). Without internal heat flux : TI=1000 SI - R=250 m - $A_{bed}$=0.6 (black circle), TI=1000 SI - R=450 m - $A_{bed}$=0.6 (blue triangle), TI=2000 SI - R=450 m - $A_{bed}$=0.7 (purple square). With internal heat flux: TI=200 SI - R=200 m - $A_{bed}$=0.6 (green polygon), TI=500 SI - R=250 m - $A_{bed}$=0.7 (cyan star), TI=1000 SI - R=250 m - $A_{bed}$=0.7 (red losange). The orange solid line shows the cases of a low reservoir and low thermal inertia simulation (TI=200 SI - R=50 m - $A_{bed}$=0.7 - F=30 W m$^{-2}$) that is consistent with observed caps extent and volatile fractional area but inconsistent with the observed pressure in 2017 (too low, see text in Section 8.3). Right. Same as A but showing the surface $N_2$ ice thickness.*

**Atmospheric abundances of CO and CH$_4$**

According to our model and assuming a constant and uniform CO ice mixing ratio into $N_2$ ice of 0.04%-0.08%, the CO gas volume mixing ratio should remain relatively constant with time, with values around 0.005%-0.01% in 2040 (see Section 7.4.1). The evolution of the CH$_4$ atmospheric mixing ratio in the next decades remains uncertain (and very sensitive to the surface area covered by CH$_4$-rich deposits, see Section 7.4.2). It would tend to increase as the surface pressure decreases, in particular if new CH$_4$-rich deposits form in the mid-to-low latitudes at the edge of the southern cap. However, if most of the CH$_4$-rich deposits formed at the south pole, the CH$_4$ atmospheric mixing ratio would decrease in the next decades (see Figure 24) as the south pole will approach polar winter. CH$_4$ gas is relatively easy to observe from the ground (Lellouch et al. 2010) and could be further monitored with e.g. CRIRES+ at the VLT, and subsequently with HIRES on ELT.

### 8.6.2. Thermal lightcurve predictions for JWST

In this section, we model thermal lightcurves of Triton for different climate scenarios in 2022. Thermal lightcurves of Triton have never been observed before due to the inability of previous facilities (e.g., Spitzer, Herschel) to separate Triton from Neptune, given the maximum ~15" angular distance. Although ALMA can easily resolve Triton from Neptune, thermal radiation at mm wavelength is actually more sensitive to emissivity effects than to thermal inertia. JWST would be able to measure Triton's thermal lightcurve for the first time with the Mid-Infrared Instrument (MIRI, imaging mode), at 21 and 25.5 μm. We show here that future observations of Triton with JWST/MIRI would provide us with a one-of-a-kind dataset of Triton's surface that would strongly constrain the ice distribution and the surface properties (temperature, emissivity, roughness…) of the volatile-covered vs. bedrock terrains. In addition, these observations would be able to test our model predictions and to discriminate between the climate scenarios presented in this section.

We designed and ran three simulations of the current seasonal cycle of Triton based on some of our best-case simulations (TI=1000 SI, no internal heat flux, R=200 and 650m, A=0.6 and 0.7). In these three simulations, we artificially prescribed the longitudinal asymmetry of the southern cap seen in the Voyager 2 images and inferred from the IRTF/SpeX spectra, with the bright deposits roughly extending to the equator on the sub-Neptunian hemisphere and to ~30°S in the opposite hemisphere (peak-to-peak magnitude of ±15° in latitude), and we assumed that the longitudinal variation of the spectrum has not changed since 1989. The three simulations are realistic in the sense that they are consistent with (1) a projected surface area of the volatile ice (given by NIR spectra) of ~60-70% in the period 1995-2010 (2) a southern cap extending to 15°S ±15° and no northern deposits south of 45°N in 1989 (consistent with Voyager 2 images), (3) a disk-averaged brightness temperature of 38-41 K at 45 µm in 1989 (consistent with the values retrieved from Voyager 2/IRIS), (4) relatively high Bond albedos for all terrains (0.6-0.8), and (5) a surface pressure of ~1.4 ± 0.2 Pa in 1989 and ~1.41 ± 0.4 Pa in 2017 (see Figure 29.A).

However, the three simulations differ in terms of bedrock surface albedo (A=0.6 vs. A=0.7) and of the extent of the southern cap in 2022 (large permanent cap to 7.5°S vs small cap to 37.5°S) as shown by Figure 28, which leads to three different climate scenarios (relatively cold, relatively warm, and intermediate):

**(A) Warm scenario:** Based on the simulation with TI=1000 SI, A=0.6, R=200m. In this scenario, we artificially set the thickness of $N_2$ ice to a few centimeters in the mid-to-equatorial southern regions so that these deposits become seasonal, undergo sublimation since 1989, and lead to a retreat of the southern cap from the equator in 1989 to 37.5°S in 2022. We also assume a warm bedrock surface (A=0.6) north of these deposits, which leads to a relatively warm global mean surface temperature.

**(B) Intermediate scenario:** Based on the simulation with TI=1000 SI, A=0.6, R=650m. This scenario assumes a large permanent $N_2$ southern cap (extending to 7.5°S) and a warm bedrock surface (A=0.6).

**(C) Cold scenario:** Based on the simulation with TI=1000 SI, A=0.7, R=650m. This scenario assumes a large permanent $N_2$ southern cap (extending to 7.5°S), a cold bedrock surface (A=0.7), and more extended $N_2$ ice in the northern hemisphere, which leads to a relatively cold global mean surface temperature.

We also performed similar simulations with no longitudinal asymmetry for the southern cap or with a longitudinal asymmetry amplified to a magnitude of ±25° latitude. The boundary of the cap edge in these cases is illustrated by the dashed lines on Figure 28.B.

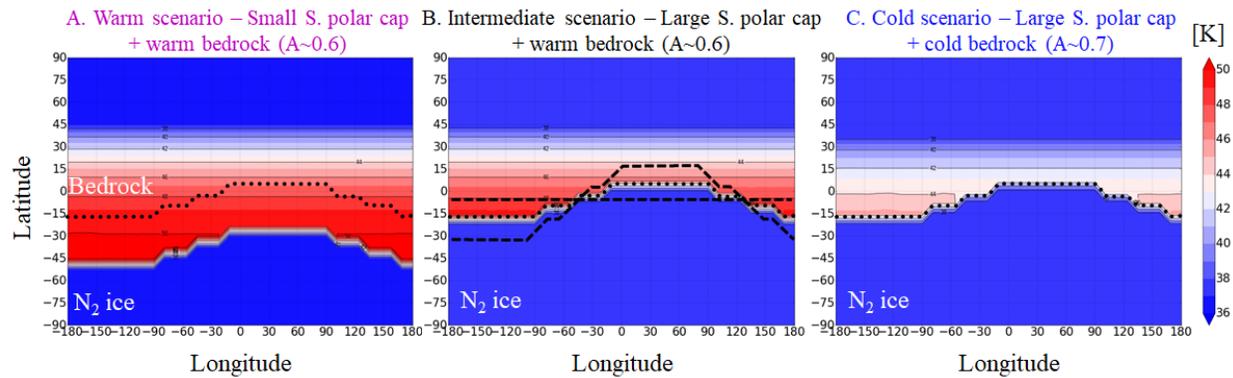

Figure 28: *Modeled surface temperatures on Triton in 2022 (at local noon) for three climate scenarios with different extents for the southern cap and bedrock surface albedo. A. Warm scenario: reduced cap extent and warm bedrock (A=0.6). B. Intermediate scenario: large cap extent and warm bedrock (A=0.6). C. Cold scenario: large cap extent, cold bedrock (A=0.7), and more extended $N_2$ ice in the northern hemisphere. Dotted lines indicate the modeled cap extent in 1989 (the cap retreated only in scenario A). Dashed lines in panel B illustrate the cap extent in alternative simulations in which the longitudinal asymmetry disappeared or amplified to a magnitude of ±25° latitude (±15° is the reference case and roughly corresponds to the asymmetry seen by Voyager 2 in 1989).*

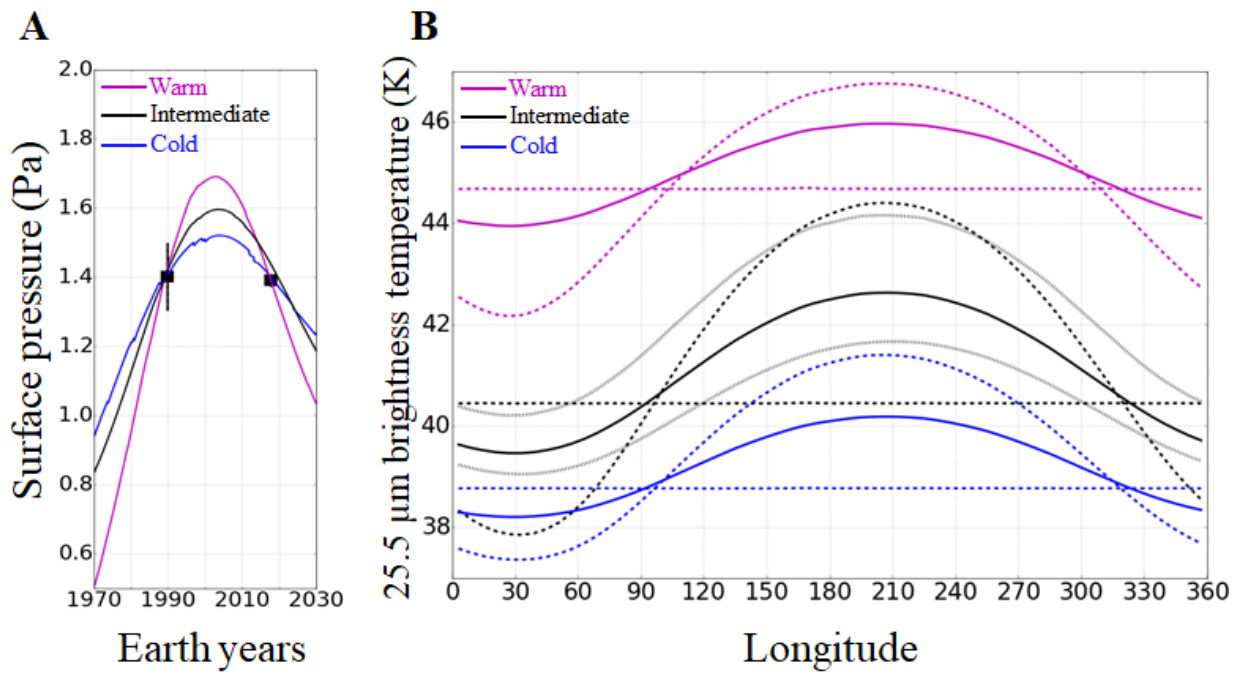

Figure 29: *Triton climate model results. A. Surface pressure evolution for the 3 scenarios presented on Figure 28. Black squares indicate the Voyager 2 and 2017 occultation observations (Gurrola, 1995, Marques Oliveira et al., 2021). B. Triton 25.5 μm lightcurve for the 3 climate scenarios (with a 15° longitudinal asymmetry), with a diurnal thermal inertia of 20 SI (solid line). Dashed lines show the light curve for the alternative simulations with no or with ±25° longitudinal asymmetry. Grey lines for the intermediate scenario are for TI=10 (warmer) and TI=40 SI (colder).*

Figure 28 shows that the $N_2$ southern cap surface temperature in 2022 is ~37 K (at solid-gas equilibrium) for the 3 cases while the equatorial volatile-free bedrock surface temperature (at local noon) is ~46 K for the warm and intermediate scenarios (A=0.6) and ~44 K for the cold scenario (A=0.7). In the case of a significant cap retreat since 1989, the volatile-free surface temperature in the mid-southern latitudes may reach ~50 K in 2022 (Figure 28.A). We note that locally, even warmer (resp. colder) patches could exist on Triton's surface due to subsurface heating activity (hot spots) or if the bedrock surface Bond albedo is significantly lower than 0.6 (resp. higher than 0.7). JWST/MIRI will be able to measure the thermal lightcurve of Triton (as seen from the Earth, i.e. of the projected disk) at 25.5 µm (and also at 21 µm) with its F2550W filter. Figure 29.B shows the modeled thermal lightcurves at 25.5 µm for the 3 climate scenarios in 2022. The thermal rotational variation on Triton is anti-correlated to the (subdued) optical light curve, with the warmest brightness temperatures being associated with the darkest regions (note that $N_2$-rich regions have a uniform temperature, ~37K, controlled by the surface pressure and the global albedo but not the local albedo). The thermal lightcurve variation and overall flux level therefore sensitively depend on ice distribution, mainly the presence/absence of volatile ice at the south pole, the southern cap extent, and the longitudinal variability between 45°S-20°N. Although $N_2$ ice deposits could cover the latitudes 90°N-30°N in 2022 and thus be seen by Earth-based telescopes in a small crescent below the terminator (56°N), they will have a weak quasi-negligible contribution to the lightcurves due to their cold surface and their small contribution (~ 3% in 2022, see Appendix A in Holler et al., 2016) to the total projected area relative to Triton's visible disk.

The thermal lightcurve is also sensitive to the thermal and energetic properties of the $N_2$-free terrains (thermal inertia, surface roughness, bolometric and spectral emissivity, Lellouch et al., 2011, 2016). In particular, the measured mean brightness temperature can discriminate between warm or cold scenarios (Figure 29.B). We expect generally low diurnal thermal inertias (TI=10-30 SI) as on Pluto (Lellouch et al., 2016), but this can be confirmed or disproved by thermal lightcurve measurements as the amplitude, general flux level and shape of the light curve will decrease for higher thermal inertia (see grey lines on Figure 29.B).

## 8.7. Uncertainties and future work

Although we explored the volatile cycles on Triton by using a large range of model parameters, other combinations may exist that also reproduce the observations. In particular, in the volatile transport simulations of this paper (Section 7), we fixed the diurnal and seasonal thermal inertia of $N_2$ ice to 20 SI and 800 SI respectively, and we only explored a north-south asymmetry in $N_2$ ice albedo ($\Delta A_{N2}$=0.1). Other values for the thermal inertia are possible, although they should not impact the results to first order if they remain in the range of what has been suggested for Pluto (10-40 SI and 500-2000 SI, respectively). Other north-south or local asymmetries in albedo, emissivity, ice composition or internal heat flux are possible and may significantly impact the results. In particular, albedos and ice composition feedback probably have a significant role as geysers could deposit dark material (resp. bright ice grain) on top of the ice, and further darken (resp. brighten) the ice. Haze particle deposition and volatile ice sublimation could also contribute to darken the ice locally. We also consider that Triton's atmosphere remains always global, even when it hits a non-global limit in the past, with $N_2$ ice temperatures varying over the body. These processes are not taken into account in our model.

In addition, in our model, when both $CH_4$ and $N_2$ ices are present on the surface and $CH_4$ is sublimating, we assume that $CH_4$ is diluted in a solid solution $N_2$:$CH_4$ with a mole fraction of $CH_4$ in $N_2$ ice of 0.05% (Quirico et al. 1999, Merlin et al., 2018). We apply Raoult's law to compute the $CH_4$ atmospheric mixing ratio at saturation, as in Forget et al. (2017). We do not take into account possible variations and deviations from equilibrium. Tan and Kargel (2018) showed that these volatile ices should not exhibit such ideal behavior and form solid solutions whose phases follow ternary phase equilibria ($N_2$, CO, $CH_4$). We note that sophisticated equations of state exist for the $N_2$-CO-$CH_4$ systems under Triton surface conditions (CRYOCHEM, Tan and Kargel, 2018). As these equations of state have not been coded for use in a climate model or been applied to the specific distribution of ices and temperatures seen on Triton, we have substituted the alternative of relying on Raoult's law.

Future work involving laboratory experiments, spectroscopic analyses, thermodynamic models, and global circulation models (GCMs) is strongly needed to improve the models, constrain the timescales for ice relaxation toward thermodynamic equilibrium, and explore in detail the effect of the ternary phase equilibrium on Triton and Pluto (and on other Trans-Neptunian objects). Lastly, coupling the volatile transport model with models of internal structure (including cryovolcanic activity, obliquity tides) and outgassing could also help us better understand the formation of geysers on Triton and atmospheric escape.

## Conclusions

We simulated the long-term and seasonal volatile cycles of Triton, exploring a large range of model parameters (thermal inertia, bedrock surface albedo, global reservoir of $N_2$ ice, internal heat flux) and comparing with available observations (Voyager 2 images and surface pressure, infrared surface emission measurements, albedo maps, volatile fractional area from Earth-based near-infrared spectra and surface pressure evolution from stellar occultations) to better constrain these parameters. In particular, we use an extremely high-quality occultation dataset obtained in 2017 to define our best-case simulations (Marques Oliveira et al., 2021). The following results were obtained:

1. Permanent volatile ice caps form at the poles and extend to low latitudes through glacial flow or through the formation of thinner seasonal deposits. North-South asymmetries in surface properties can favor the development of one cap over the other, as previously evidenced. A difference in topography has little impact on the North-South $N_2$ ice asymmetry.

2. Best-case simulations are obtained for a bedrock surface albedo of 0.6-0.7, a global reservoir of $N_2$ ice thicker than 200 m, and a bedrock thermal inertia larger than 500 SI or smaller but with a large internal heat flux (>30 mW m$^{-2}$). The large $N_2$ ice reservoir implies a permanent $N_2$ southern cap (several 100 m thick, and up to 1.5 km thick) extending to the equatorial regions with higher amounts of volatile ice at the south pole, which is not inconsistent with Voyager 2 images but does not reconcile well with observed full-disk near-infrared spectra. In particular, the lack of variability of the non-volatile ices ($H_2O$, $CO_2$, and ethane) with longitude (Holler et al., 2016) is not explained by our model.

3. Our results suggest that a small permanent polar cap should exist in the northern hemisphere, in particular if the internal heat flux remains relatively low (e.g. radiogenic, < 3 mW m$^{-2}$) and if the bedrock albedo is >0.7. A non-permanent northern polar cap was only obtained for simulations with high internal heat flux (30 mW m$^{-2}$). The northern cap will possibly extend to 30°N in the next decade, thus becoming visible by Earth-based telescopes. The southern cap should not significantly retreat in the next decades compared to what has been observed in 1989.

4. Our model predicts $N_2$ condensation at the northern edge of the southern cap in the period 1980-2020, which could explain the bright equatorial fringe observed by Voyager 2 in 1989. The model predicts that over the last 30 years, the southern polar cap lost ~0.3 m of $N_2$ ice by sublimation, while ~0.1 m of ice deposited in the equatorial regions and mid northern latitudes.

5. Best-case surface pressures are consistent with a moderate increase of pressure during the 1990-2000 period. The model suggests that the peak of surface pressure occurred between 2000-2010 and did not exceed ~2 Pa (although albedo feedback may have increase the peak amplitude).

6. According to our model, the atmosphere of Triton never collapses. The surface pressure should slowly decrease but remain larger than 0.5 Pa by 2060. Simulations performed with low ice emissivity ($\varepsilon_{N2}$=0.3-0.5) show that the surface pressure could always remain larger than 0.5 Pa. Seasonal thin deposits play a significant role in the evolution of the surface pressure. The internal heat flux does not significantly impact the surface pressure curve since the ice albedo is adjusted to match the Voyager 2 pressure constraint (it impacts locally the sublimation-condensation rates but not the net mass balance of $N_2$).

7. In our model, CO tends to follow $N_2$ and condenses where $N_2$-rich deposits are already present. By using a CO/$N_2$ ice mixing ratio of 0.04%-0.08%, we obtain an atmospheric CO gas volume mixing ratio of 0.006%-0.012% during the period 2000-2020, and slightly slower values for the next decades (0.005%-0.01%).

8. Unlike for Pluto, no permanent $CH_4$-rich deposit form in our Triton model. We tested two scenarios with small $CH_4$-rich ice patches to explain the observed $CH_4$ atmospheric mixing ratio. Simulations with $CH_4$-rich ice at the cap edge (equatorial regions) produce a $CH_4$ partial pressure of 1-5x10$^{-4}$ Pa that remains relatively constant during the period 1980-2030. Simulations with $CH_4$-rich ice at the south pole produce a strong increase in $CH_4$ partial pressure until 2005, followed by a decrease as the south pole exits polar day. Both "cap edge" and "south pole" scenarios could reconcile the 2.45x10$^{-4}$ Pa partial pressure observed by Voyager 2 and the increase in the $CH_4$ partial pressure (factor of 4) observed from 1989 to 2009 (Lellouch et al., 2010), although the "cap edge" scenario would require a decrease in $CH_4$ ice albedo or increase in $CH_4$ ice coverage with time to lead to a significant increase in $CH_4$ partial pressure.

9. JWST/MIRI will be able to measure the thermal lightcurve of Triton at 25.5, 21µm, which will help in discriminating several possible climate scenarios, as well as in providing strong constraints on the thermal and energetic properties of the $N_2$-free terrains.

Triton remains relatively unexplored. In particular, we note that there is a strong need for more and continuing Earth-based observations (JWST, HST, VLT, Keck, ELT) of Triton's surface and atmosphere in order to provide a temporal context and understand Triton's seasonal evolution. Surface ice distribution has never been mapped, only inferred from disk integrated rotational curves of ice band depths, and except for very limited observations, surface temperatures have not been measured. This manifests as a substantial gap in our understanding of Triton's climate evolution. Finally, an orbiter mission such as Neptune Odyssey or a flyby mission to Triton such as Trident would provide a rich science-breakthrough dataset with which to understand the physical and dynamical processes at play on Triton and their seasonal evolution (by comparison with existing Earth-based observations), which would be insightful to further understand this class of volatile-rich object, includingPluto, Eris, and Makemake.

# Acknowledgments

T. B. was supported for this research by an appointment to the National Aeronautics and Space Administration (NASA) Postdoctoral Program at the Ames Research Center administered by Universities Space Research Association (USRA) through a contract with NASA. Part of the work leading to these results has received funding from the European Research Council under the European Community's H2020 2014-2021 ERC Grant Agreement n°669416 "Lucky Star". Under contract with NASA, part of this work was conducted at the Jet Propulsion Laboratory, California Institute of Technology.

# References

– Agnor, C.B., Hamilton, D.P., 2006. Neptune's capture of its moon Triton in a binary-planet gravitational encounter. Nature 441, 192-194.

– Augé, B. and 12 colleagues, 2016. Irradiation of nitrogen-rich ices by swift heavy ions. Clues for the formation of ultracarbonaceous micrometeorites. Astronomy and Astrophysics 592. doi:10.1051/0004-6361/201527650.

– Bauer, J.M., Buratti, B.J., Li, J.-Y., Mosher, J.A., Hicks, M.D., Schmidt, B.E., Goguen, J.D., 2010. Direct detection of seasonal changes on Triton with Hubble Space Telescope. ApJL 723, L49-L52.

– Bertrand, T., & Forget, F., 2016. Observed glacier and volatile distribution on Pluto from atmosphere-topography processes. Nature, 540(7631). https://doi.org/10.1038/nature19337.

– Bertrand, T., Forget, F., Umurhan, O. M., Grundy, W. M., Schmitt, B., Protopapa, S., Zangari, A. M., White, O. L., Schenk, P. M., Singer, K. N., Stern, A., Weaver, H. A., Young, L. A., Ennico, K., & Olkin, C. B., 2018. The nitrogen cycles on Pluto over seasonal and astronomical timescales. Icarus, 309, 277–296. https://doi.org/10.1016/j.icarus.2018.03.012.

– Bertrand, T., Forget, F., Umurhan, O. M., Moore, J. M., Young, L. A., Protopapa, S., Grundy, W. M., Schmitt, B., Dhingra, R. D., Binzel, R. P., Earle, A. M., Cruikshank, D. P., Stern, S. A., Weaver, H. A., Ennico, K., & Olkin, C. B., 2019. The $CH_4$ cycles on Pluto over seasonal and astronomical timescales. Icarus. https://doi.org/10.1016/j.icarus.2019.02.007.

– Bertrand, T., Forget, F., White, O., Schmitt, B., Stern, S. A., Weaver, H. A., Young, L. A., Ennico, K., & Olkin, C. B., 2020a. Pluto's Beating Heart Regulates the Atmospheric Circulation: Results From High-Resolution and Multiyear Numerical Climate Simulations. Journal of Geophysical Research: Planets, 125(2), 1–24. https://doi.org/10.1029/2019JE006120.

– Bertrand, T., Forget, F., Schmitt, B., White, O. L., & Grundy, W. M., 2020b. Equatorial mountains on Pluto are covered by methane frosts resulting from a unique atmospheric process. Nature Communications, 11, 5056.

– Broadfoot, A. L., et al., 1989. Ultraviolet spectrometer observations of Neptune and Triton. Science 246, 1459-1466.

– Brown, R. H., Johnson, T. V., Goguen, J.D., Schubert, G., Ross, M. N., 1991. Triton's Global Heat Budget. Science 251, 1465–1467. doi:10.1126/science.251.5000.1465.

– Brown, R. H., Kirk, R. L., 1994. Coupling of volatile transport and internal heat flow on Triton. Journal of Geophysical Research 99, 1965–1982. doi:10.1029/93JE02618.

–  Brown, R. H., Cruikshank, D. P., Veverka, J., Helfenstein, P., Eluszkiewicz, J., 1995. Surface composition and photometric properties of Triton. Neptune and Triton, 991–1030.

–  Buratti, B. J., Goguen, J. D., Gibson, J., Mosher, J., 1994. Historical photometric evidence for volatile migration on Triton. Icarus 110, 303-314.

–  Buratti, B. J., Bauer, J. M., Hicks, M. D., Hillier, J. K., Verbiscer, A., Hammel, H., Schmidt, B., Cobb, B., Herbert, B., Garsky, M., Ward, J., & Foust, J., 2011. Photometry of Triton 1992-2004: Surface volatile transport and discovery of a remarkable opposition surge. Icarus, 212(2), 835–846. https://doi.org/10.1016/j.icarus.2011.01.012.

–  Chen, E. M. A., Nimmo, F., Glatzmaier, G. A., 2014. Tidal heating in icy satellite oceans. Icarus 229, 11–30. doi:10.1016/j.icarus.2013.10.024.

–  Conrath, B. and 15 colleagues, 1989. Infrared Observations of the Neptunian System. Science 246, 1454–1459. doi:10.1126/science.246.4936.1454.

–  Correia, A. C. M., 2009. Secular evolution of a satellite by tidal effect: Application to triton. Astrophysical Journal. https://doi.org/10.1088/0004-637X/704/1/L1.

–  Croft, S. K., Kargel, J. S., Kirk, R. L., Moore, J. M., Schenk, P. M., Strom, R. G., 1995. The geology of Triton. In: Cruikshank (Ed.), Neptune and Triton. University of Arizona Press, Tucson, pp. 879–947.

–  Cruikshank, D. P. and 7 colleagues, 1991. Tentative Detection of CO and CO2 Ices on Triton. Bulletin of the American Astronomical Society.

–  Cruikshank, D. P., Roush, T. L., Owen, T. C., Geballe, T. R., de Bergh, C., Schmitt, B., Brown, R. H., Bartholomew, M. J., 1993. Ices on the surface of Triton. Science 261, 742-745.

–  Cruikshank, D. P., Schmitt, B., Roush, T. L., Owen, T. C., Quirico, E., Geballe, T. R., de Bergh, C., Bartholomew, M. J., Dalle Ore, C. M., Douté, S., Meier, R., 2000. Water ice on Triton. Icarus 147, 309-316.

–  DeMeo, F. E., Dumas, C., de Bergh, C., Protopapa, S., Cruikshank, D. P., Geballe, T.R., Alvarez-Candal, A., Merlin, F., Barucci, M. A., 2010. A search for ethane on Pluto and Triton. Icarus 208, 412-424.

–  Douté, S., B. Schmitt, E. Quirico, T. C. Owen, D. P. Cruikshank, C. de Bergh, T. R. Geballe, and T.L. Roush, 1999. Evidence for methane segregation at the surface of Pluto. Icarus, 142, 421-444.

–  Dubois, D. , Patthoff, D.A, Pappalardo, R.T., 2017. Diurnal, Nonsynchronous Rotation and Obliquity Tidal Effects on Triton Using a Viscoelastic Model: SatStressGUI. Implications for Ridge and Cycloid Formation. LPSC, LPI #1964, 2897.

–  Duxbury, N. S., Brown, R. H., 1993. The Phase Composition of Triton's Polar Caps. Science 261, 748–751. doi:10.1126/science.261.5122.748.

–  Earle, A. M., Binzel, R. P., Young, L. A., Stern, S. A., Ennico, K., Grundy, W., Olkin, C. B., Weaver, H. A. New Horizons Surface Composition Theme, 2018. Albedo matters: understanding runaway albedo variations on Pluto. Icarus 303, 1–9. doi: 10.1016/j.icarus.2017.12.015.

–  Elliot, J. L., et al., 1998. Global warming on Triton. Nature 393, 765-767.

– Elliot, J. L., Strobel, D. F., Zhu, X., Stansberry, J. A., Wasserman, L. H., & Franz, O. G., 2000. The thermal structure of triton's middle atmosphere. Icarus, 143(2), 425–428. https://doi.org/10.1006/icar.1999.6312.

– Elliot, J. L., Person, M. J., Qu, S., 2003. Analysis of Stellar Occultation Data. II. Inversion, with Application to Pluto and Triton.\ The Astronomical Journal 126, 1041–1079. doi:10.1086/375546

– Eluszkiewicz, J., 1991. On the microphysical state of the surface of Triton. J. Geophys. Res. 96, 19217-19229.

– Ferrari, C. & Lucas, A., 2016. Low thermal inertias of icy planetary surfaces. Evidence for amorphous ice? Astronomy and Astrophysics, 588, A133.

– Forget, F., Bertrand, T., Vangvichith, M., Leconte, J., Millour, E., & Lellouch, E., 2017. A post-New Horizons global climate model of Pluto including the $N_2$, $CH_4$ and CO cycles. Icarus, 287, 54–71. https://doi.org/10.1016/j.icarus.2016.11.038.

– Fray, N., Schmitt, B., 2009. Sublimation of ices of astrophysical interest: A bibliographic review. P&SS 57, 2053-2080.

– Gaeman, J., Hier-Majumder, S., Roberts, J. H., 2012. Sustainability of a subsurface ocean within Triton's interior. Icarus 220, 339–347. doi:10.1016/j.icarus.2012.05.006.

– Grundy, W. M., Fink, U., 1991. A new spectrum of Triton near the time of the Voyager encounter. Icarus 93, 379–385. doi:10.1016/0019-1035(91)90220-N.

– Grundy, W. M., & Young, L. A., 2004. Near-infrared spectral monitoring of Triton with IRTF/SpeX I: Establishing a baseline for rotational variability. Icarus, 172(2), 455–465. https://doi.org/10.1016/j.icarus.2004.07.013

– Grundy, W. M., Young, L. A., Stansberry, J. A., Buie, M. W., Olkin, C. B., Young, E. F., 2010. Near-infrared spectral monitoring of Triton with IRTF/SpeX II: Spatial distribution and evolution of ices. Icarus 205, 594-604.

– Grundy, W. M., Binzel, R. P., Buratti, B. J., Cook, J. C., Cruikshank, D. P., Dalle Ore, C. M., Earle, A. M., Ennico, K., Howett, C. J. A., Lunsford, A. W., Olkin, C. B., Parker, A. H., Philippe, S., Protopapa, S., Quirico, E., Reuter, D. C., Schmitt, B., Singer, K. N., Verbiscer, A. J., … Young, L. A., 2016. Surface compositions across Pluto and Charon. Science, 351(6279). https://doi.org/10.1126/science.aad9189.

– Gurrola, E. M., 1995. Interpretation of Radar Data from the Icy Galilean Satellites and Triton. Ph.D. thesis, Stanford University.

– Gurwell, M., Lellouch, E., Butler, B., Moreno, R., Moullet, A., Strobel, D., 2019. The Atmosphere of Triton Observed With ALMA. EPSC-DPS Joint Meeting 2019.

– Hansen, C.J., Paige, D.A., 1992. A thermal model for the seasonal nitrogen cycle on Triton. Icarus 99, 273-288.

– Herbert, F., Sandel, B.R., 1991. $CH_4$ and haze in Triton's lower atmosphere. JGR Supp. 96, 19241-19252.

– Hillier, J., Helfenstein, P., Verbiscer, A., & Veverka, J., 1991. Voyager photometry of Triton: Haze and surface photometric properties. Journal of Geophysical Research, 96(S01), 19203. https://doi.org/10.1029/91ja01736.

- Hillier, J., Veverka, J., Helfenstein, P., Lee, P., 1994. Photometric Diversity of Terrains on Triton. Icarus 109, 296–312. doi:10.1006/icar.1994.1095.

- Holler, B. J., Young, L. A., Grundy, W. M., & Olkin, C. B., 2016. On the surface composition of Triton's southern latitudes. Icarus, 267, 255–266. https://doi.org/10.1016/j.icarus.2015.12.027.

- Howard, A. D., Moore, J. M., White, O. L., Umurhan, O. M., Schenk, P. M., Grundy, W. M., Schmitt, B., Philippe, S., McKinnon, W. B., Spencer, J. R., Beyer, R. A., Stern, S. A., Ennico, K., Olkin, C. B., Weaver, H. A., & Young, L. A., 2017. Pluto: Pits and mantles on uplands north and east of Sputnik Planitia. Icarus, 293, 218–230. https://doi.org/10.1016/j.icarus.2017.02.027.

- Hussmann, H., Sohl, F., Spohn, T., 2006. Subsurface oceans and deep interiors of medium-sized outer planet satellites and large trans-neptunian objects. Icarus 185, 258–273. doi:10.1016/j.icarus.2006.06.005.

- Krasnopolsky, V. A., Cruikshank, D. P., 1999. Photochemistry of Pluto's atmosphere and ionosphere near perihelion. J. Geophys. Res. 104, 21979–21996. http://dx.doi.org/10.1029/1999JE001038.

- Johnson, P. E. and 8 colleagues, 2021. Modeling Pluto's minimum pressure: Implications for haze production. Icarus 356. doi:10.1016/j.icarus.2020.114070.

- Johnson, R.E., Oza, A. Young, L.A., Volkov, A. N., Schmidt, C. , 2015. Volatile Loss and Classification of Kuiper Belt Objects. ApJ, 809, 1, 43. doi:10.1088/0004-637X/809/1/43

- Jovanovic, L., Gautier, T., Vuitton, V., Wolters, C., Bourgalais, J., Buch, A., Orthous-Daunay, F., Vettier, L., Flandinet, L., Carrasco, N., 2020. Chemical composition of Pluto aerosol analogues. Icarus, 346, 113774, 10.1016/j.icarus.2020.113774.

- Laskar J, Robutel P, 1993. The chaotic obliquity of the planets. Nature 361:608–612.

- Lee, P., Helfenstein, P., Veverka, J., & McCarthy, D., 1992. Anomalous-scattering region on Triton. Icarus, 99(1), 82–97. https://doi.org/10.1016/0019-1035(92)90173-5.

- Lellouch, E., de Bergh, C., Sicardy, B., Ferron, S., Käufl, H.-U., 2010. Detection of CO in Triton's atmosphere and the nature of surface-atmosphere interactions. A&A 512, L8-L13.

- Lellouch, E., de Bergh, C., Sicardy, B., Käufl, H. U., Smette, A., 2011. High resolution spectroscopy of Pluto's atmosphere: detection of the 2.3 μm $CH_4$ bands and evidence for carbon monoxide. Astron. Astrophys. 530, L4. http://dx.doi.org/10.1051/0004-6361/201116954.

- Lellouch, E. and 11 colleagues, 2016. The long-wavelength thermal emission of the Pluto-Charon system from Herschel observations. Evidence for emissivity effects. Astronomy and Astrophysics 588. doi:10.1051/0004-6361/201527675.

- Lellouch, E., Gurwell, M., Butler, B., Fouchet, T., Lavvas, P., Strobel, D. F., Sicardy, B., Moullet, A., Moreno, R., Bockelée-Morvan, D., Biver, N., Young, L., Lis, D., Stansberry, J., Stern, A., Weaver, H., Young, E., Zhu, X., & Boissier, J., 2017. Detection of CO and HCN in Pluto's atmosphere with ALMA. Icarus, 286, 289–307. https://doi.org/10.1016/j.icarus.2016.10.013.

- Lewis, B. L. and 12 colleagues, 2021. Distribution and energy balance of Pluto's nitrogen ice, as seen by New Horizons in 2015. Icarus 356. doi:10.1016/j.icarus.2020.113633.

- Li, D., Johansen, A., Mustill, A. J., Davies, M. B., Christou, A. A., 2020. Capture of satellites during planetary encounters. A case study of the Neptunian moons Triton and Nereid. Astronomy and Astrophysics 638. doi:10.1051/0004-6361/201936672.

- Lunine, J, E., and Stevenson, D. J., 1985. Physical state of volatiles on the surface of Triton. Nature 317, 238-240.

- Mangold, N., 2011. Ice sublimation as a geomorphic process: A planetary perspective. Geomorphology 126, 1–17. doi:10.1016/j.geomorph.2010.11.009.

- Marques Oliveira, J. et al. 2021. Structure and evolution of Triton's atmosphere from the 5 October 2017 stellar occultation and previous observations. Submitted to A&A.

- Martin-Herrero, A., Romeo, I., & Ruiz, J., 2018. Heat flow in Triton: Implications for heat sources powering recent geologic activity. Planetary and Space Science, 160, 19–25. https://doi.org/10.1016/j.pss.2018.03.010.

- Materese, C. K., Cruikshank, D. P., Sandford, S. A., Imanaka, H., Nuevo, M. 2015. Ice Chemistry on Outer Solar System Bodies: Electron Radiolysis of $N_2$-, $CH_4$-, and CO-Containing Ices. The Astrophysical Journal 812. doi:10.1088/0004-637X/812/2/150.

- McCord, T. B., 1966. Dynamical evolution of the Neptunian system. Astron. J. 71, 585. http://dx.doi.org/10.1086/109967.

- McEwen, A. S., 1990. Global color and albedo variations on Triton. Geophys Res Lett 17, 1765–1768. http://dx.doi.org/10.1029/GL017i010p01765.

- McKinnon, W. B., 1984. On the origin of Triton and Pluto. Nature 311, 355–358. http://dx.doi.org/10.1038/311355a0.

- McKinnon, W. B., Lunine, J. I., Banfield, D., 1995. Origin and evolution of Triton. In: Cruikshank (Ed.), Neptune and Triton. University of Arizona Press, Tucson, pp. 807–877.

- McKinnon W. B., Nimmo F., Wong T., et al., 2016. Convection in a volatile nitrogen-ice-rich layer drives Pluto's geological vigour. Nature, 534, 82-85.

- Merlin, F., 2015. New constraints on the surface of Pluto. Astron. Astrophys. 582, A39. http://dx.doi.org/10.1051/0004-6361/201526721.

- Merlin, F., E. Lellouch, E. Quirico, and B. Schmitt, 2018. Triton's surface ices: Distribution, temperature and mixing state from VLT/SINFONI observations. Icarus 314, 274-293.

- Meza, E., Sicardy, B., Assafin, M., Ortiz, J. L., Bertrand, T., Lellouch, E., Desmars, J., Forget, F., Bérard, D., Doressoundiram, A., Lecacheux, J., Oliveira, J. M., Roques, F., Widemann, T., Colas, F., Vachier, F., Renner, S., Leiva, R., Braga-Ribas, F., Benedetti-Rossi, G., Camargo, J. I. B., Dias-Oliveira, A., Morgado, B., Gomes-Júnior, A. R., Vieira-Martins, R., Behrend, R., Tirado, A. C., Duffard, R., Morales, N., Santos-Sanz, P., Jelínek, M., Cunniffe, R., Querel, R., Harnisch, M., Jansen, R., Pennell, A., Todd, S., Ivanov, V. D., Opitom, C., Gillon, M., Jehin, E., Manfroid, J., Pollock, J., Reichart, D. E., Haislip, J. B., Ivarsen, K. M., LaCluyze, A. P., Maury, A., Gil-Hutton, R., Dhillon, V., Littlefair, S., Marsh, T., Veillet, C., Bath, K.-L., Beisker, W., Bode, H.-J., Kretlow, M., Herald, D., Gault, D., Kerr, S., Pavlov, H., Faragó, O., Klös, O., Frappa, E., Lavayssière, M., Cole, A. A., Giles, A. B., Greenhill, J. G., Hill, K. M., Buie, M. W., Olkin, C. B., Young, E. F., Young, L. A., Wasserman, L. H., Devogèle, M., French, R. G., Bianco, F. B., Marchis, F., Brosch, N., Kaspi, S., Polishook, D., Manulis, I., Ait Moulay Larbi, M., Benkhaldoun, Z., Daassou, A., El Azhari, Y., Moulane, Y., Broughton, J., Milner, J., Dobosz, T., Bolt, G., Lade, B., Gilmore, A., Kilmartin, P., Allen, W. H., Graham, P. B., Loader, B., McKay, G., Talbot, J., Parker, S., Abe,

L., Bendjoya, P., Rivet, J.-P., Vernet, D., Di Fabrizio, L., Lorenzi, V., Magazzú, A., Molinari, E., Gazeas, K., Tzouganatos, L., Carbognani, A., Bonnoli, G., Marchini, A., Leto, G., Sanchez, R. Z., Mancini, L., Kattentidt, B., Dohrmann, M., Guhl, K., Rothe, W., Walzel, K., Wortmann, G., Eberle, A., Hampf, D., Ohlert, J., Krannich, G., Murawsky, G., Gährken, B., Gloistein, D., Alonso, S., Román, A., Communal, J.-E., Jabet, F., deVisscher, S., Sérot, J., Janik, T., Moravec, Z., Machado, P., Selva, A., Perelló, C., Rovira, J., Conti, M., Papini, R., Salvaggio, F., Noschese, A., Tsamis, V., Tigani, K., Barroy, P., Irzyk, M., Neel, D., Godard, J. P., Lanoiselée, D., Sogorb, P., Vérilhac, D., Bretton, M., Signoret, F., Ciabattari, F., Naves, R., Boutet, M., De Queiroz, J., Lindner, P., Lindner, K., Enskonatus, P., Dangl, G., Tordai, T., Eichler, H., Hattenbach, J., Peterson, C., Molnar, L. A., & Howell, R. R., 2019. Lower atmosphere and pressure evolution on Pluto from ground-based stellar occultations, 1988-2016. Astronomy and Astrophysics, 625, A42.

–   Moore, J. M., Spencer, J. R., 1990. Koyaanismuuyaw - The hypothesis of a perenially dichotomous Triton. GRL 17, 1757-1760.

–   Moore, M. H., Hudson, R. L., 2003. Infrared study of ion-irradiated $N_2$-dominated ices relevant to Triton and Pluto: formation of HCN and HNC. Icarus 161, 486–500. http://dx.doi.org/10.1016/S0019-1035(02)00037-4.

–   Nimmo, F., & Spencer, J. R., 2015. Powering Triton's recent geological activity by obliquity tides: Implications for Pluto geology. Icarus, 246(C), 2–10. https://doi.org/10.1016/j.icarus.2014.01.044.

–   Nogueira, E., Brasser, R., & Gomes, R., 2011. Reassessing the origin of Triton. Icarus. https://doi.org/10.1016/j.icarus.2011.05.003.

–   Olkin, C. B., Elliot, J. L., Hammel, H. B., Cooray, A. R., McDonald, S. W., Foust, J. A., Bosh, A. S., Buie, M. W., Millis, R. L., Wasserman, L. H., Dunham, E. W., Young, L. A., Howell, R. R., Hubbard, W. B., Hill, R., Marcialis, R. L., McDonald, J. S., Rank, D. M., Holbrook, J. C., & Reitsema, H. J., 1997. The thermal structure of Triton's atmosphere: Results from the 1993 and 1995 occultations. Icarus, 129(1), 178–201. https://doi.org/10.1006/icar.1997.5757.

–   Paige, D. A., 1985. The Annual Heat Balance of the Martian Polar Caps from Viking Observations. Ph.D. thesis, Calif. Inst. of Technol., Pasadena.

–   Pankine, A. A., Tamppari, L. K., Smith, M. D., 2009. Water vapor variability in the north polar region of Mars from Viking MAWD and MGS TES datasets. Icarus 204, 87–102. doi:10.1016/j.icarus.2009.06.009.

–   Pankine, A. A., Tamppari, L. K., Smith, M. D., 2010. MGS TES observations of the water vapor above the seasonal and perennial ice caps during northern spring and summer. Icarus 210, 58–71. doi:10.1016/j.icarus.2010.06.043.

–   Prokhvatilov, A.I., Yantsevich, L.D., 1983. X-ray investigations of the equilibrium phase diagram of CH–N, solid mixtures. Sov. J. Low Temp. Phys. 9, 94–97.

–   Protopapa, S., Grundy, W. M., Reuter, D. C., Hamilton, D. P., Dalle Ore, C. M., Cook, J. C., Cruikshank, D. P., Schmitt, B., Philippe, S., Quirico, E., Binzel, R. P., Earle, A. M., Ennico, K., Howett, C. J. A., Lunsford, A. W., Olkin, C. B., Parker, A., Singer, K. N., Stern, A., Young, L. A., 2017. Pluto's global surface composition through pixel-by-pixel Hapke modeling of New Horizons Ralph/LEISA data. Icarus, 287, 218–228. https://doi.org/10.1016/j.icarus.2016.11.028.

- Quirico, E., S. Douté, B. Schmitt, C. de Bergh, D.P. Cruikshank, T.C. Owen, T.R. Geballe, and T.L. Roush, 1999. Composition, physical state and distribution of ices at the surface of Triton. Icarus, 139, 159-178.

- Ruiz, J., 2003. Heat flow and depth to a possible internal ocean on Triton. Icarus 166, 436–439.

- Schenk, P. M., & Zahnle, K., 2007. On the negligible surface age of Triton. Icarus. https://doi.org/10.1016/j.icarus.2007.07.004.

- Schenk P. M., Beyer R. A., McKinnon W. B., et al., 2018. Basins, fractures and volcanoes: Global cartography and topography of Pluto from New Horizons. Icarus, 314, 400-433.

- Schenk, P. M., et al., 2021. Triton : Topography and Geology of a (Probable) Ocean World with Comparison to Pluto and Charon. Submitted.

- Schmitt, B., Philippe, S., Grundy, W. M., Reuter, D. C., Côte, R., Quirico, E., Protopapa, S., Young, L. A., Binzel, R. P., Cook, J. C., Cruikshank, D. P., Dalle Ore, C. M., Earle, A. M., Ennico, K., Howett, C. J. A., Jennings, D. E., Linscott, I. R., Lunsford, A. W., Olkin, C. B., … Weaver, H. A., 2017. Physical state and distribution of materials at the surface of Pluto from New Horizons LEISA imaging spectrometer. Icarus, 287, 229–260. https://doi.org/10.1016/j.icarus.2016.12.025.

- Smith, B. A., Soderblom, L. A., Banfield, D., Barnet, C., Basilevsky, A. T., Beebe, R. F., Bollinger, K., Boyce, J. M., Brahic, A., Briggs, G. A., Brown, R. H., Chyba, C., Collins, S. A., Colvin, T., Cook II, A. F., Crisp, D., Croft, S. K., Cruikshank, D., Cuzzi, J. N., Danielson, G. E., Davies, M. E., De Jong, E., Dones, L., Godfrey, D., Goguen, J., Grenier, I., Haemmerle, V. R., Hammel, H., Hansen, C. J., Helfenstein, C. P., Howell, C., Hunt, G. E., Ingersoll, A. P., Johnson, T. V., Kargel, J., Kirk, R., Kuehn, D.I., Limaye, S., Masursky, H., McEwen, A., Morrison, D., Owen, T., Owen, W., Pollack, J. B., Porco, C. C., Rages, K., Rogers, P., Rudy,D., Sagan, C.,Schwartz, J., Shoemaker, E. M., Showalter, M., Sicardy,B., Simonelli, D., Spencer, J.,Sromovsky, L. A., Stoker,C., Strom, R. G., Suomi,V.E., Synott, S. P., Terrile, R. J., Thomas, P., Thompson, W. R., Verbiscer, A., Veverka, J., 1989. Voyager 2 at Neptune – imaging science results. Science 246, 1422–1449.

- Soderblom, L. A., Kieffer, S. W., Becker, T. L., Brown, R. H., Cook II, A. F., Hansen, C. J., Johnson, T. V., Kirk, R. L., Shoemaker, E. M., 1990. Triton's geyser-like plumes: discovery and basic characterization. Science 250, 410–415.

- Spencer, J. R., 1990. NITROGEN FROST MIGRATION ON TRITON: A historical model. Geophysical Research Letters 17, 1769–1772. doi:10.1029/GL017i010p01769.

- Spencer, J. R., & Moore, J. M., 1992. The influence of thermal inertia on temperatures and frost stability on Triton. Icarus, 99(2), 261–272. https://doi.org/10.1016/0019-1035(92)90145-W.

- Stansberry, J. A., Lunine, J. I., Porco, C. C., McEwen, A. S., 1990. Zonally averaged thermal balance and stability models for nitrogen polar caps on Triton. Geophysical Research Letters 17, 1773–1776. doi:10.1029/GL017i010p01773.

- Stansberry, J. A., Yelle, R. V., Lunine, J. I., McEwen, A. S., 1992. Triton's surface-atmosphere energy balance. Icarus 99, 242–260. doi:10.1016/0019-1035(92)90144-V.

- Stansberry, J. A., Pisano, D. J., Yelle, R. V., 1996a. The emissivity of volatile ices on Triton and Pluto. Planetary and Space Science 44, 945–955. doi:10.1016/0032-0633(96)00001-3.

– Stansberry, J. A., Spencer, J. R., Schmitt, B., Benchkoura, A.-I., Yelle, R. V., & Lunine, J. I., 1996b. A model for the overabundance of methane in the atmospheres of Pluto and Triton. Planetary and Space Science, 44, 1051.

– Stern, S. A., McKinnon, W. B., 2000. Triton's Surface Age and Impactor Population Revisited in Light of Kuiper Belt Fluxes: Evidence for Small Kuiper Belt Objects and Recent Geological Activity. The Astronomical Journal 119, 945–952. doi:10.1086/301207.

– Stern, S. A., Bagenal, F., Ennico, K., Gladstone, G. R., Grundy, W. M., Mckinnon, W. B., Moore, J. M., Olkin, C. B., Spencer, J. R., Weaver, H. A., Young, L. A., Andert, T., Andrews, J., Banks, M., Bauer, B., Bauman, J., Barnouin, O. S., Bedini, P., Beisser, K., Versteeg, M. H., 2015. The Pluto system: Initial results from its exploration by New Horizons.

– Stone, E. C., Miner, E. D., 1989. The Voyager 2 encounter with the Neptunian system. Science 246, 1417-1421.

– Strobel, D. F., Summers, M. E., 1995. Triton's upper atmosphere and ionosphere. Neptune and Triton, 1107–1148.

– Tan, S. P., & Kargel, J. S., 2018. Solid-phase equilibria on Pluto's surface. Monthly Notices of the Royal Astronomical Society, 474(3), 4254–4263. https://doi.org/10.1093/mnras/stx3036.

– Tegler, S. C. and 13 colleagues, 2019. A New Two-molecule Combination Band as a Diagnostic of Carbon Monoxide Diluted in Nitrogen Ice on Triton. The Astronomical Journal 158. doi:10.3847/1538-3881/ab199f.

– Trafton, L., 1984. Large seasonal variations in Triton's atmosphere. Icarus 58, 312-324.

– Trafton, L. M., D. L. Matson, and J. A. Stansberry, 1998. Surface/atmosphere Interactions and Volatile Transport (triton, Pluto and Io). Solar System Ices 227, 773.

– Trafton, L. M., Stansberry, J. A., 2015. On the Departure from Isothermality of Pluto's Volatile Ice due to Local Insolation and Topography. AAS/Division for Planetary Sciences Meeting Abstracts #47.

– Trafton, L. M., 2015. On the state of methane and nitrogen ice on Pluto and Triton: Implications of the binary phase diagram. Icarus, 246(C), 197–205. https://doi.org/10.1016/j.icarus.2014.05.022.

– Tryka, K. A., Brown, R. H., Anicich, V., Cruikshank, D. P., Owen, T. C., 1993. Spectroscopic determination of the phase composition and temperature of nitrogen ice on Triton. Science 261, 751-754.

– Tryka, K. A., Brown, R. H., Chruikshank, D. P., Owen, T. C., Geballe, T. R., de Bergh, C., 1994. Temperature of nitrogen ice on Pluto and its implications for flux measurements. Icarus 112, 513–527. http://dx.doi.org/10.1006/icar.1994.1202.

– Tyler, G. L., Sweetnam, D. N., Anderson, J. D., Borutzki, S. E., Campbell, J. K., Kursinski, E. R., Levy, G. S., Lindal, G. F., Lyons, J. R., Wood, G. E., 1989. Voyager radio science observations of Neptune and Triton. Science 246, 1466-1473.

– Umurhan, O. M., Howard, A. D., Moore, J. M., Earle, A. M., White, O. L., Schenk, P. M., Binzel, R. P., Stern, S. A., Beyer, R. A., Nimmo, F., McKinnon, W. B., Ennico, K., Olkin, C. B., Weaver, H. A., & Young, L. A., 2017. Modeling glacial flow on and onto Pluto's Sputnik Planitia. Icarus, 287, 301–319. https://doi.org/10.1016/j.icarus.2017.01.017.

– Vetter, M., Jodl, H.-J., Brodyanski, A., 2007. From optical spectra to phase diagrams–the binary mixture $N_2$-CO. Low. Temp. Phys. 33, 1052-1060.

– White, O. L., Moore, J. M., McKinnon, W. B., Spencer, J. R., Howard, A. D., Schenk, P. M., Beyer, R. A., Nimmo, F., Singer, K. N., Umurhan, O. M., Stern, S. A., Ennico, K., Olkin, C. B., Weaver, H. A., Young, L. A., Cheng, A. F., Bertrand, T., Binzel, R. P., Earle, A. M., … Schmitt, B., 2017. Geological mapping of Sputnik Planitia on Pluto. Icarus, 287, 261–286. https://doi.org/10.1016/j.icarus.2017.01.011.

– Wood, S. E., and Paige, D. A., 1992. Modeling the martian seasonal CO2 cycle: Fitting the Viking lander pressure curves. Icarus 99, 1–14.

– Young, L. A., Bertrand, T., Trafton, L. M., Forget, F., Earle, A. M. and Sicardy, B., 2021. Pluto's Volatile and Climate Cycles on Short and Long Timescales. In *The Pluto System After New Horizons (S. A. Stern, J. M. Moore, W. M. Grundy, L. A. Young, and R. P. Binzel, eds)*, p.321-361. Univ. of Arizona, Tucson.

– Young, L. A., 2017. Volatile transport on inhomogeneous surfaces: II. Numerical calculations (VT3D). Icarus 284, 443–476. doi:10.1016/j.icarus.2016.07.021.

– Zent, A., et al., 1989, Grain size metamorphism in polar nitrogen ice on Triton. Geophys. Res. Lett. 16, 965-968.

# Appendix - Calculation of the temporal variations of Triton's subsolar latitude

In our volatile transport model, the calculation of Triton's subsolar latitude is performed by using a trigonometric equation corresponding to a fit of a solution obtained by a dynamic model of Triton's motion.

The solution of the dynamic model was obtained over the period 06/09/-807 to 01/01/2100 using the *ephemph* application of the EPROC project (Berthier, 1998, Forget, 2000). The method, described by Le Guyader, 1993, includes the disturbances caused by Neptune's flatness at the poles (Peters, 1981), by the Sun and by the 8 other planets of the solar system, with their revised coordinates (Davies, 1996).

The trigonometric fit of this solution is based on the trigonometric equation proposed by Harris (1984) and Trafton (1984), from a suggestion by Dobrovolskis (1980). Changes in subsolar latitude $\lambda$ are expressed as a sinus function, resulting from the superposition of 3 harmonics of different amplitudes and frequencies, as :

$$\sin\big(\lambda(t)\big) = A\cos(\alpha_0 - \alpha_1\mu) + B\sin(\alpha_0 - \beta_0 + (\beta_1 - \alpha_1)\mu) + C\sin(\alpha_0 + \beta_0 - (\alpha_1 + \beta_1)\mu)$$

With:

$$\mu = \frac{365.25\ t\ L_s}{D}$$

Where t is the time in Earth year from year 0 (i.e., t = 2000 corresponds to year 2000 A.D) $L_s$ is the length of one Triton day in seconds (507773 s) and $L_d$ is the length of one Triton day in Earth days (5.877 days).

Since the work of Harris (1984) and Trafton (1984), the angles and coefficients have been revised (Berthier, 1998, Forget, 2000) and the values are:

A = 0.429870

B = 0.370543

C = 0.0225091

D=31557595

$\alpha_0$ = 241.52577°

$\alpha_1$ = 0.038142 rad/year

$\beta_0$ = -100.79473°

$\beta_1$ = 0.009131 rad/year

The first harmonic evolves at the frequency $\alpha_1$= 0.0381 rad/year, which corresponds to the sideral period of Neptune (165 Earth years). The second harmonic has a frequency equal to ($\beta_1$-$\alpha_1$) with $\beta_1$ the frequency corresponding to the precession of Triton's orbit around Neptune's pole (period

= 650 Earth years). The period of this harmonic is ~221 Earth years. The third harmonic has a low amplitude compared to the others, and its frequency is equal to $(\alpha_1+\beta_1)$, which corresponds to a period of 131.5 Earth years.

The error made on the subsolar latitude by using the trigonometric equation (compared to the dynamical solution) is always smaller than 0.5°, which is largely sufficient for the purpose of this paper (the latitudinal resolution of our model is 7.5°). In fact, the errors are periodic and their analysis in frequency (Laskar, 1999) show that the frequency peaks at 82.2, 93.4 and 683.4 years with an amplitude of the order of 0.1°, while other peaks have an amplitude at least 10 times smaller.

In conclusion, the trigonometric equation provides an excellent approximation of Triton's subsolar latitude that can be extrapolated in time. In our model, we compute the past climates of Triton by using these equations with t typically ranging from $t=-9 \times 10^6$ (9 million years ago) to t=2100 (end of the current century). Note that over time, Triton's inclination is slowly increasing (getting closer to perfectly retrograde) and its precession period is slowly getting shorter as its orbit shrinks, but these are *very* long-term effects that are not significant over the timescale used in this paper (~10 Myrs). The obliquity of Neptune does not change significantly over this period either (it is suggested to be primordial, Laskar and Robutel, 1993).

− Berthier, J., 1998. Définitions relatives aux éphémérides pour l' observation physique des corps du système solaire. Notes scientifiques et techniques du Bureau des longitudes, S061.

− Davies, M.E., Abalakin, V.K., Bursa, M., Lieske, J.H., Morando, B., Morison, D., Seidelmann, P.K., Sinclair, A.T., Yallop, B., Tjuflin, Y.S., 1996. Report of the TAU/IAG/COSPAR working group on cartographic coordinates and rotational elements of the planets and satellites: 1994. Celes. Mech 63, 127-148.

− Forget, F., Decamp, N., Hourdin, F., 1999. A 3D general circulation model of Triton's atmosphere and surface. In "Piuto and Triton, comparisons and evolution over time", Lowell Observatory's fourth annual worshop, Flagstaff, Arizona.

− Laskar, J., 1999. Introduction to frequency map analysis. NATO ASI Hamiltonian Systems with Three or More Degrees of Freedom. C. Simd Ed. Kluwer, Dordrecht. 134-150.

− Le Guyader, C., 1993. Solution of the N-body problem expanded into Taylor series of high orders. Applications to the solar system over large time range. Astron. Astrophys. 272, 687-694.

− Harris, A. W., 1984. Physical characteristics of Neptune and Triton inferred from the orbital motion of Triton. Paper presented to I.A.U Colloquium 77, July 5-9, 1983 at Cornell Univ.

− Peters, C. F., 1981. Numerical integration of the satellites of the outer planets. Astron. Astrophys. 104, 37-41.